\documentstyle[epsfig,onecolumn]{mn}

\title{Reprocessed emission from warped accretion discs 
with application to X-ray iron line profiles}
 
\author[Sean A. Hartnoll and Eric G. Blackman]
{{\textbf{Sean A. Hartnoll$^{1,2}$ and Eric G. Blackman$^{1,3}$} } \\ {
 1. Theoretical Astrophysics, California Institute of Technology, Pasadena CA 91125, USA;} \\  { 2. St. John's College, University of Cambridge,
Cambridge CB2 1TP, UK;} \\ { 3. Department of Physics and Astronomy, University of Rochester, Rochester NY 14627, USA;} \\ {e-mails:} \\ {sah40@hermes.cam.ac.uk } \\ {blackman@tapir.caltech.edu}}

\date{Submitted to MNRAS}

\pubyear{2000}

\begin{document}

\maketitle

\begin{abstract} 
Flourescent iron line profiles currently provide the best 
diagnostic for active galactic nuclei (AGN) engine geometries.
Here we construct a method for calculating the relativistic iron line 
profile from an arbitrarily warped accretion disc, illuminated from above
and below by hard X-ray sources. 
This substantially generalises previous calculations of 
reprocessing by accretion discs by including non-axisymmetric 
effects. We include a relativistic 
treatment of shadowing by ray-tracing photon paths 
along Schwarzchild geodesics. 
We apply this method to two classes of warped discs, and
generate a selection of resulting line profiles.
New profile features include the possibility of sharper red, and
softer blue fall-offs, a time varying line
profile if the warp precesses about the disc, and
some differences between `twisted' and `twist-free'
warps.  We discuss some qualitative 
implications of the line profiles in the context of  
Type I and II Seyfert AGN.
\end{abstract}


\section{Introduction}

Emission from active galactic nuclei (AGN) and some X-ray binaries is thought to result from accretion onto a central massive black hole (e.g. Pringle 1981; Rees 1984). This paradigm allows spectra of Seyferts to be modelled as a combination of direct emission and reprocessed emission from a cold, optically thick accretion disc (see Reynolds 1999 for a review). The reprocessed component includes the prominent broad iron K$\alpha$ flourescence line of rest energy 6.4 keV. Since X-rays originate from the innermost regions of the accretion flow, the X-ray emission carries information about geometry and dynamics of the gas very near the black hole. ASCA has observed iron lines in 18 Seyfert Is (Nandra \emph{et al.} 1997). The best studied iron line is that of MCG-6-30-15 (e.g. Tanaka \emph{et al.} 1995; Iwasawa \emph{et al.} 1996, 1999). The line profiles observed are consistent with reprocessing of an X-ray illuminated accretion disc under the influence of strong gravity.

Iron line profiles for accreting engines around black holes are sensitive to the disc illumination law, the inclination of the disc to the line of sight, the inner and outer radii, and the disc geometry. 
(Fabian \emph{et al.} (1989); Matt \emph{et al.} (1993) 
consider the effect of extreme inclination angles; Reynolds and Begelman (1997) discuss the illumination law). Most work on reprocessing in AGN
has considered flat discs, 
although Pariev \& Bromley (1998) considered 
some effects of finite disc thickness, 
Blackman (1999) considered concave accretion discs, 
and Bachev (1999) has a non-relativistic treatment of broad line 
H $\beta$ profiles from warped discs.

There are several motivations for considering iron line profiles from warped accretion discs. The first is to provide a predictive signature for 
theoretical models of warping (e.g. radiative warping (Pringle 1996),  tidal warping (Terquem and Bertout 1993)). Observational evidence for warping of discs in AGN already
includes the water maser emission of NGC 4258 at  
0.1pc from the central engine (on larger scales than the inner accretion disc) 
which traces a disc warp 
(eg. Miyoshi \emph{et al.} 1995; Herrnstein \emph{et al.} 1996). There is also indirect evidence on these larger scales from
some observations of Seyfert Is which 
suggest that the BLRs are not coplanar with the inner disc 
(Nishiura \emph{et al.} 1998). There have also been 
suggestions that the dusty tori of 
Seyfert unification paradigms might be warped discs (Maloney 1999; 
Phinney 1989). The second motivation is that some observational features of
iron lines directly tempt consideration of non-flat or warped accretion 
discs with an  inclusion of shadowing effects:
(i) The sharper red than blue fall-off in some Seyfert Is
(e.g. Turner \emph{et al.} 1997) and possibly soft blue fall-offs
in Seyfert Is (Nandra \emph{et al.} 1997). 
(ii) Profile time variation between observed 
epochs, including the `deep minimum'(Iwasawa \emph{et al.} 1996, 1999; Sulentic \emph{et al.} 1998; Wijers and Pringle 1999; Weaver and Yaqoob 1998). 
(iii) The excessive reprocessed fraction of ultrasoft narrow-line 
Seyferts (Brandt, private communication 1999). 
A disc with a concave surface may help with (iii) (Blackman 1999), but
the other features require shadowing and non-axisymmetric effects.
Finally, while some AGN line profiles such as MCG-6-30-15 may be 
quite consistent with flat discs (Iwasawa \emph{et al.} 1996;  Lee \emph{et al.} 1999), 
the data at present are somewhat limited and many objects remain to be 
studied. Consideration of a wider range of plausible disc geometries is 
necessary so that geometric effects can be disentangled from other effects on 
the line profiles. 

In section 2 we relate two mathematical descriptions of a warped disc. 
This facilitates applying various analytic studies of 
accretion disc warping (e.g. Maloney \emph{et al.} 1996, Pringle 1997) 
to the calculation of line profiles. We then  
describe `twisted' and `twist-free' warps (see Figs. 1 and 2) for later
use. Section 3 gives a detailed procedure for calculating the 
iron line profile from a general warped accretion disc 
about a Schwarzchild black hole. 
This constitutes a substantial generalisation of the currently 
standard procedure described by Fabian \emph{et al.} (1989) for flat discs. 
The consideration of shadowing requires ray-tracing of the 
photon paths in a Schwarzchild geometry. This calculation can be 
straightforwardly adapted to calculate the iron line profile for any 
shape of warp.
Section 4 begins with a discussion of the values of parameters used, in particular the outer radius and illumination law. 
We then give frequency emission maps of a few discs, which illustrate the effects of shadowing and warping. Finally, we comment on a range of line profiles, relating the profile features to the form of the warp and to observed phenomena. Section 5 is the conclusion.

Our calculations of emission line profiles are generically 
applicable to any reprocessing line from a 
warped disc which is illuminated from above and below.

\section{Warped discs}

\subsection{Relationship between different formalisms for warped discs}

Geometrically thin warped accretion discs can be constucted from
a series of concentric rings of increasing radii. The rings have varying
inclination defined by the Eulerian angles $\beta_{eul}(R,t)$ and $\gamma_{eul}(R,t)$, where $\beta_{eul}$
is the angle between the normal to the disc and the normal to the equatorial
plane and $\gamma_{eul}$ is the angle of the node with respect to some fixed axis
on the equatorial plane. Here $R$ is the radius of the ring and $t$ allows for time dependence of the warp (for an illustration see Bachev 1999).

Although
this is a conceptually and mathematically useful formalism that emphasises the idea of Keplerian orbits about the central object, for flux-related calculations it can be more useful to have the disc described 
as a surface with height as a function of radius and azimuthal angle from 
some fixed equatorial axis, $h(r,\phi)$.  
This cylindrical formalism is also closer to those
used for flat discs in flux calculations, so it is easier to appreciate 
the required generalistations. 
Such a description of warped discs has been used by Terquem 
\& Bartout (1993, 1996).

The two formalisms described above can be related by the formula
\begin{equation}
h(r,\phi,t)=r\tan [\beta_{eul}(\sqrt{r^{2}+h^{2}(r,\phi,t)},t)] \sin [\phi + \gamma_{eul}(\sqrt{r^{2}+h^{2}(r,\phi,t)},t)].
\end{equation}
Given the difficulty of solving the above for $h(r,\phi,t)$ we will 
employ the approximations ${\rm tan}\beta_{eul}\sim \beta_{eul}$ 
and $\sqrt{r^{2}+h^{2}}\sim r$.

\subsection{Disc forms considered}

We consider two specific classes of warp, which are motivated and
described below. 
Throughout this paper, distances such as $h$ and $r$ 
are taken in units of $R_{g}\equiv GM/c^2$. 
Negative values of $h$ denote a height
below the equatorial plane containing the black hole.
The discs are taken to extend from the innermost stable orbit of the 
Schwarzchild solution ($r_{in}=6$) to some outer radius $r_{out}$, 
the value of which will be discussed later.   Points on the disc thus satisfy 
\begin{equation}
r_{in} \leq \sqrt{r^{2}+h^{2}} \leq r_{out}
\end{equation}
Note that, for an accretion disc, it is $\sqrt{r^2+h^2}$ which is bounded,
not $r$.
This is necessary to preserve the use of Keplerian orbits
needed for calculating Doppler shifts. It appears that Terquem and Bertout (1993) bound r.

Note that we consider here only Schwarzchild black holes. Kerr solutions have a smaller inner radius and also a critical radius within which there is no warping, due to the Bardeen-Peterson effect; in the presence
of a sufficiently visicous accretion disc however, this
is critical radius is not particularly large (Kumar \& Pringle 1985).

\subsubsection{Twist-free form with power law r dependence}

This form is given by (Fig. 1)
\begin{equation} h(r,\phi) = a_{1} r_{out} 2^{(b-1)/2} \left(\frac{r}{r_{out}}\right)^{b}
\cos(\phi-\omega t), \end{equation}
where $a_{1}$ 
measures the magnitude of the warp 
and is chosen so that 
$h(r_{out}/\sqrt{2},\omega t)= a_{1} r_{out}/\sqrt{2}$. 
We will consider $a_{1}=0.25$ and $a_{1}=1$. The $\omega t$ term allows for possible precession with angular frequency $\omega$, or consideration of discs aligned at different azimuthal angles. 
This is mathematically equivalent to changing the azimuthal angle of the observer, but computationally more convenient. The quantity $b$ gives the 
curvature of the disc. We consider $b=2.0$ and $b=4.0$.
\begin{figure}
\epsfig{file=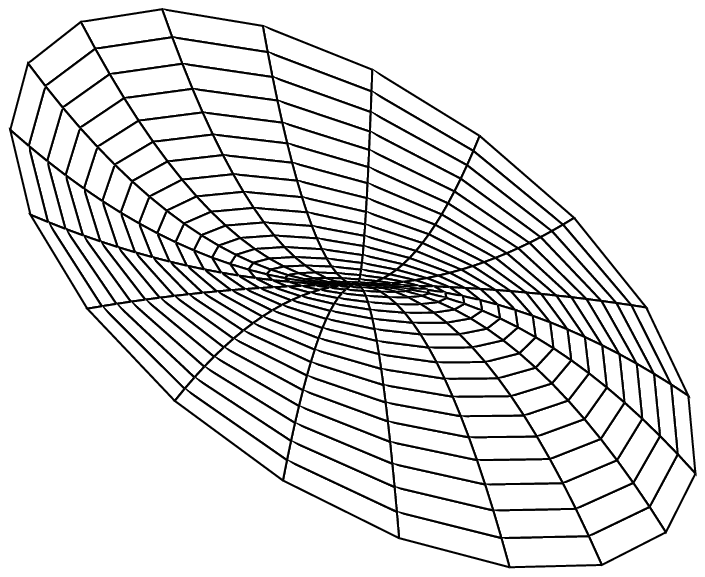} \hspace{1cm}
\epsfig{file=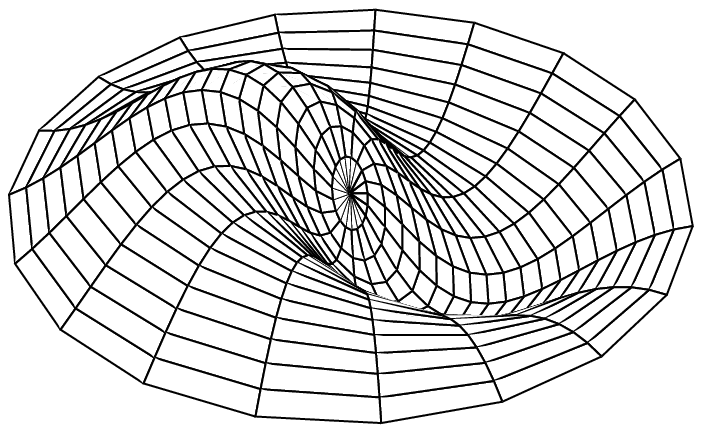}

\noindent\textbf{Figure 1:} Twist-free warped disc. $b=2$, $a_{1}=1$. \hspace{1in}
\noindent\textbf{Figure 2:} Twisted warped disc. $a_{2}=2$.

\noindent In both figures, a large magnitude of warp is chosen for illustration.
\end{figure}
This form is simple, but
contains the important features of warped discs that we wish to
consider -- non-axisymmetric curvature, shadowing, possiblity of
emission from the lower surface. We call the disc of eqn. (3) 
twist-free because when it is related to its Eulerian angles through (1), 
$\gamma_{eul}$ is independent of $r$ so the line of nodes is straight. 
Also, $\beta$ is indepedent of time.
Within the context of radiative warping models, (3)
is essentially the same form as the generalised `constant-torque' 
solution for certain parameter choices, as 
described in Maloney \emph{et al.} (1996).

Outside of the context of AGN, this form has arisen in warps of T Tauri accretion discs (Terquem and Bertout 1993, 1996) where it is the stable state under tidally induced warping. It is also the form of the warp of the galaxy NGC 660 (Arnaboldi and Galletta 1992), which has been imaged.

\subsubsection{Twisted radiatively-induced warp}

The second form we consider is (Fig. 2)
\begin{equation}
h(r,\phi) = a_{2} \sqrt{\frac{r_{out} r}{10}}\sin\left( \sqrt{\frac{10r}{r_{out}}}\right) \cos \left( \phi-\sqrt{\frac{10r}{r_{out}}}-\omega t \right),
\end{equation}
where $a_{2}$ allows the absolute magnitude of the warp to be fixed. We consider $a_{2}=1$ and $a_{2}=2$.
This form  has arisen in both numerical and analytic calculations of radiatively-induced warping (for example, Pringle 1997; Maloney \emph{et al.} 1996; Wijers and Pringle 1999). Again, $\omega$ is the precession frequency. Note that $\omega$ can be negative to give precession in the opposite direction.

We have rescaled the steady state solution in Maloney \emph{et al.} (1996) to the dimensions arising in Pringle (1997) so that the first radial zero (i.e. smallest radius for which $h=0$) coincides with the outer radius. 
An alternative would have been to use the steady state solution in its initial scaling, which would have produced a multiply peaked disc (see Maloney \emph{et al.} 1996, Fig. 2). For the geometry of the warp,
we have used $\tan\beta_{eul}\sim \beta_{eul}$ and  $r\sim \sqrt{r^2+h^2}$, 
as described above. These approximations do not significantly 
change the qualitatitve form of the disc or line profiles.

\section{Flux calculation: reprocessing from warped discs}

\subsection{Incident flux}

We consider a hard X-ray point 
source at a height $H_{e} = 10$ directly above and below the
black hole. 
Note that for the twisted disc there is a finite tilt at the origin. 
The K$\alpha$ iron line is produced by reprocessing
(absorbtion and fluorescent emission) in the disc (cf. Reynolds 1999).

We define the position of the sources as 
${\mathbf{s}} = {\pm s_{x}} {\mathbf{i}} \pm {s_{y}} {\mathbf{j}} \pm {s_{z}} 
\mathbf{k}$. The cartesian coordinate system is the usual cartsian system associated with the cylindrical polars of the disc (i.e. x-axis at $\phi=0)$. For the twisted disc
\begin{equation}
{\mathbf{s}} = \pm \frac{H_{e}}{\sqrt{1+a^{2}_{2}}}(-
a_{2}\cos\omega t {\mathbf{i}}-a_{2}\sin\omega t{\mathbf{j}}
+{\mathbf{k}}).
\end{equation}
Define the (normalised) position vector
\begin{equation}
{\mathbf{\hat{r}}}=\frac{r\cos\phi{\mathbf{i}}
+r\sin\phi{\mathbf{j}}
+h{\mathbf{k}}}{\sqrt{r^{2}+h^{2}}},
\end{equation}
ww will use $\mathbf{r}$ for the actual position vector.
We take the illumination law for flux impinging on an infinitessimal area dA of the disc to be
\begin{equation} dF =
\frac{\mathbf{n \cdot n_{d}}
\mathrm{V_{1}}(r,\phi) dA}
{ |{\mathbf{s}}-{\mathbf{r}}|^q } =
\frac{\mathbf{n \cdot n_{d}}
\mathrm{V_{1}}(r,\phi)
\sqrt{1+\left(\frac{\partial h}{\partial r}\right)^{2}+\frac{1}{r^{2}}
\left(\frac{\partial h}{\partial \phi}\right)^{2}} r dr d\phi}{ |\mathbf{s}-\mathbf{r}|^q},
\end{equation}
where $h(r,\phi)$ is the vertical height of the disc, with $(r,\phi)$ the
usual cylindrical coordinates; 
$q$ is the power law index;
$\mathrm{V_{1}}(r,\phi)$ is the first visibility
function with value 1 if dA is visible to the source under consideration and 0 if dA is shadowed. 
This is determined using a ray-tracing technique. ${\rm V_{1}}(r,\phi)$ will not in general be the same for the upper and lower sources.
Finally,  $\mathbf{n \cdot n_{d}}$ is the cosine of
the angle between the incident photon and the normal to the disc, with 
$\mathbf{n}$ and $\mathbf{n_{d}}$ given by
\begin{equation}
{\mathbf{n_{d}}} = \pm
\frac{(-\frac{\partial h}{\partial r}\cos\phi+
\frac{\partial h}{\partial \phi}\frac{\sin\phi}{r})
{\mathbf{i}}-
(\frac{\partial h}{\partial r}\sin\phi+
\frac{\partial h}{\partial \phi}\frac{\cos\phi}{r}){\mathbf{j}}+
{\mathbf{k}}}{\sqrt{1+\left(\frac{\partial h}{\partial r}\right)^{2}+
\frac{1}{r^{2}} \left(\frac{\partial h}{\partial \phi}\right)^{2}}},
\end{equation}
and
\begin{equation} \mathbf{n} =
\frac{{\mathbf{s}}-{\mathbf{r}}}{|{\mathbf{s}}-{\mathbf{r}}|}
\end{equation}
respectively.
Note that $\mathbf{n}$ is directed from the disc to the source and that
$\mathbf{n_{d}}$ can be the upward or downward normal, as appropriate.
The orientation is defined by sign of the z-component.
We will refer to the disc faces as ``upper'' and ``lower'' respectively. 
If the dot product of (8) and (9) is negative, 
then the point is shadowed from the source and there is no contribution 
to the flux.  This accounts for most of the shadowing but not all, 
hence the need for the visibility function above.

The formula used above for source flux does not account for the bending 
of photon paths due to the Schwarzchild geometry. 
However, this is not a serious problem because (i) 
not much is known about the X-ray illumination law and the general uncertainty swamps any changes that light bending to the disc might cause, 
and (ii) for $H_{e}\geq 10$ the effect is not significant 
(Reynolds and Begelman 1998).
We do consider light bending for the reprocessed emission below. 

\subsection{Emitted flux}

Define the normalised vector towards the obsever as
\begin{equation}
{\mathbf{i_{obs}}} = {\sin} a_{i} {\mathbf{j}} 
+ {\cos} a_{i} {\mathbf{k}},
\end{equation}
where $0 \le a_{i} \le \frac{\pi}{2}$ 
is the inclination angle of the observer with respect to the
z-axis.  The discs are non-axisymmetric, so 
this range of $a_{i}$ must be combined with 
the full $2\pi$ range in azimuthal angle 
to cover all viewing angles of the disc.
We must be careful to allow for the fact that the observer will generally
see regions of both the upper and lower sides of the disc.

We now compute the observed flux and energy, 
generalising formulae from Fabian \emph{et
al.} (1989) to warped geometry.  The geometric approach is close to
that in Terquem and Bertout (1993, 1996). 
For a flat disc, the equations reduce appropriately. 
The energy received by the observer is 
\begin{equation}
E_{obs} = \frac{E_{em}}{(1+z)},
\end{equation}
where $E_{em}=6.4 {\rm keV}$ for the iron line, and
\begin{equation}
(1+z) = \frac{1+
\frac{\cos \beta_{ot}}{\left( \sqrt{r^{2}+h^{2}}(1+\tan^{2}{\xi})-2 \right)^
{1/2}}}
{\left( 1-\frac{3}{\sqrt{r^{2}+h^{2}}} \right) ^{1/2}}.
\end{equation}
At each point on the disc, $\beta_{ot}$ is defined as the angle between the
plane of the Keplerian orbit at that point and the photon trajectory plane, defined by the black
hole, the emitting point and the direction to the observer. (In the literature $\beta$ has aquired several meanings, here we use $\beta_{eul}$ for the Eulerian angle and $\beta_{ot}$ for the observer-trajectory angle).  Thus
\begin{equation}
\cos\beta_{ot} = \mathbf{n_{o} \cdot n_{t}},
\end{equation}
where $\mathbf{n_{t}}$ is the normal to the
trajectory plane and is given by
\begin{equation}
{\mathbf{n_{t}}} = \frac{\mathbf{\hat{r} \times i_{obs}}}{|{\mathbf{\hat{r} \times i_{obs}}}|} =
\frac{(r\sin\phi\cos a_{i} - h\sin a_{i}){\mathbf{i}}-r\cos\phi\cos a_{i}{\mathbf{j}}
+r\cos\phi\sin a_{i} {\mathbf{k}}}
{\sqrt{r^{2}({\cos}^{2}a_{i}+{\sin}^{2}a_{i}{\cos}^{2}\phi)+h^{2}\sin^{2}a_{i}-2r h \sin\phi\cos a_{i}\sin a_{i}}},
\end{equation}
$\mathbf{n_{o}}$ is the normal to the orbit plane given by
\begin{equation}
\mathbf{n_{o}} = \frac{\mathbf{ e_{\gamma} \times \hat{r}}}{\left| \mathbf{e_{\gamma}} \times \mathbf{\hat{r}} \right|}
\end{equation}
with $\mathbf{e_{\gamma}}$ 
in the x-y plane, making an angle 
$\gamma$ with the x-axis such that $h(\sqrt{r^{2}+h^{2}},\gamma) = 0$ 
(this is a vector we know lies in the orbit plane, see Fig. 3), 
and chosen 
so that $\mathbf{n_{o}}$ has a positive z-component. 
For the twist-free disc this is
\begin{equation}
{\mathbf{e_{\gamma}}} = \pm \sin\omega t {\mathbf{i}} \mp \cos\omega t {\mathbf{j}},
\end{equation}
and for the twisted disc
\begin{equation}
{\mathbf{e_{\gamma}}} = \pm \sin\left( \omega t + \sqrt{\frac{10}{r_{out}}}\sqrt[4]{r^{2}+h^{2}} \right) {\mathbf{i}} \mp \cos\left( \omega t + \sqrt{\frac{10}{r_{out}}}\sqrt[4]{r^{2}+h^{2}} \right) {\mathbf{j}}.
\end{equation}
The normal to the orbit plane is undefined in this formalism for the single line at which 
$\mathbf{e_{\gamma}}$ coincides with $\mathbf{\hat{r}}$, but this represents a negligible number of points.
\begin{figure}
\epsfig{file=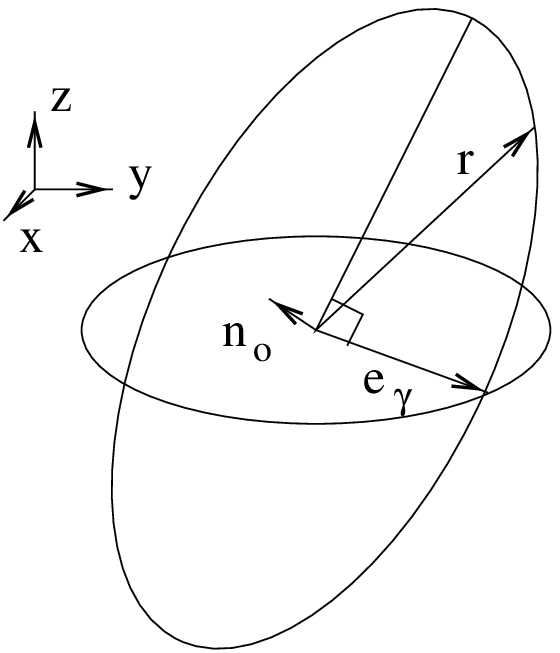}

\noindent\textbf{Figure 3:} The orbit plane is shown with its defining vectors ${\mathbf r}$ and ${\mathbf e_{\gamma}}$ (not normalised).
\end{figure}
Define $\xi$ as the emission angle on the trajectory plane (see (25) and Fig. 4), so $\xi+\frac{\pi}{2}$ is the angle between
the direction of emission of the photon and the position vector of the point
relative to the black hole. We calculate this by first calculating $\tan\xi^{\prime}$, the Euclidean case, in which the direction of emission is the direction to the observer, so that
\begin{equation}
\cos(\xi^{\prime}+\frac{\pi}{2}) = \mathbf{- \hat{r} \cdot i_{obs}}, 
\end{equation}
and
\begin{equation}
\tan\xi^{\prime} = \frac{\sin\xi^{\prime}}{\cos\xi^{\prime}} =
\frac{-\cos(\xi^{\prime}+\frac{\pi}{2})}{\sqrt{1-{\cos}^{2}(\xi^{\prime}+\frac{\pi}{2})}}.
\end{equation}
We then find $\tan\xi$ from $\tan\xi^{\prime}$ using properties of photon paths near Schwarzchild black holes. 
These formulae are only strictly correct for $-\pi/2 < \xi^{\prime} < \pi/2$ (which is all that is needed for a flat disc). However, the symmetry of the emitted photon paths about the plane of the Keplerian orbit means we can integrate the paths initially underneath the orbit plane as paths above the orbit plane (which have $\xi^{\prime}$ in the range required), although we have to take this into account when considering shadowing.
We can now write down the expression for the observed flux as 
\begin{equation}
dF_{obs} = \frac{\mathrm{V_{2}}(r,\phi) {\mathbf{R}}({\mathbf{i_{obs}}}){\mathbf{\cdot n_{d}}} dF}{(1+z)^{3}}, 
\end{equation}
where $\mathrm{V_{2}}(r,\phi)$ is the second visibility function and is 1 if
dA is visible to the observer and 0 if it is shadowed. 
In calculating ${\rm V}_2$ we ray-trace along the bent photon paths. 
The vector $\mathbf{R}(\mathbf{i_{obs}})$ is $\mathbf{i_{obs}}$ rotated by $\xi^{\prime}-\xi$ in the photon trajectory plane, and is given by
\begin{equation}
\mathbf{R}(\mathbf{i_{obs}}) = \sin\xi \mathbf{\hat{r}} + \cos\xi \mathbf {n_t \times \hat{r}}
\end{equation}
The dot product will always be positive, providing we use the correct normal to the disc.

From any given emission site on the disc, 
the observer can only see flux emitted from either 
the upper face or the lower face of the disc, because the disc is assumed to be optically thick. Light bending can potentially allow a point to be visible from both faces; we do not consider this small effect, and similarly, (22) below, is not corrected. We determine which face is visible by noting that the upper face will be visible, subject to possible shadowing, 
if
\begin{equation}
\frac{\partial h(r,\phi)}{\partial y} \leq \tan \left( \frac{\pi}{2} - a_{i} \right)
\end{equation}
Whilst the lower face is visible if the inequality is reversed.
Finally, note that we 
have assumed that the emitted flux is proportional to the incident flux, 
and so we do not take into account effects due to photoionisation
(cf. Reynolds \& Begelman 1998).

\subsection{Calculation of photon paths in Schwarzchild geometry}

The potential for shadowing to the observer 
requires that each photon path be 
checked for possibile intersection with the disc.
We explain this calculation now.
In the following $(r_{em},\phi_{em})$ are the coordinates of the emitting point on the disc, $\theta$ is the azimuthal angle in the plane of the photon trajectory and  $\rho$ is the radial distance in this plane (see Fig. 4).
\begin{figure}
\epsfig{file=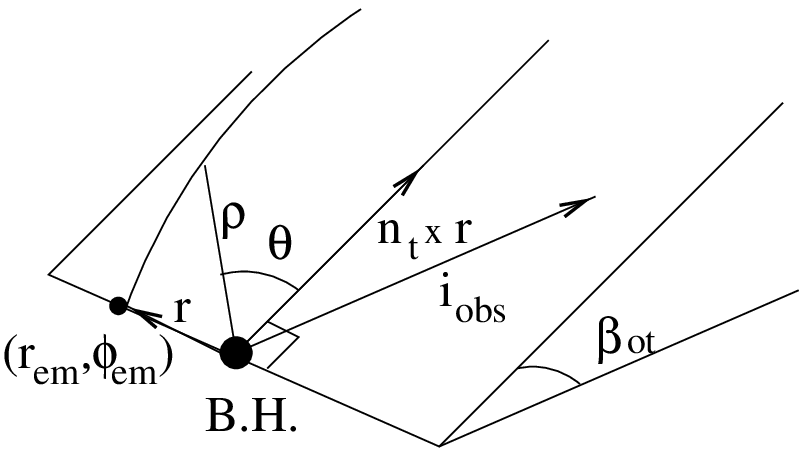}

\noindent\textbf{Figure 4:} Schematic picture of the plane of the photon trajectory, showing the photon path, the meaning of $(\rho,\theta)$, the positions of the black hole (B.H.) and the emitting point $(r_{em},\phi_{em})$. $\beta_{ot}$ is the angle from the plane of the photon trajectory to the plane defined by the orbit of the emitting point on the disc. Also shown are several relevant vectors (note that $\mathbf{n_{t}}$ is not in general parallel to ${\mathbf i_{obs}}$, not shown).
\end{figure}
The paths are calculated by a numerical integration of the Schwarzchild geometry null geodesic equation (for example, Misner, Thorne and Wheeler 1973),

\begin{equation}
\frac{d^{2}u}{d \theta^{2}}=3u^2 - u
\end{equation}
with $u=\frac{1}{\rho}$.
The integration is done using a fourth order Runge-Kutta routine.
The coordinates on the trajectory plane are related to the 
cartesian coordinates of the disc by (see Fig. 4)
\begin{equation}
{\mathbf{x}} = x{\mathbf{i}}+y{\mathbf{j}}+z{\mathbf{k}} = 
\rho \sin\theta \mathbf{\hat{r}}+\rho \cos\theta \mathbf{n_{t} \times \hat{r}},
\end{equation}
where $\mathbf{n_{t}} \perp \mathbf{\hat{r}}$. 
We use this relation in our  
check for photon path intersection with the disc. 
Here $\mathbf{n_{t}}$, $\mathbf{\hat{r}}$ and the black hole define the photon trajectory. 
This coordinate change is correct for paths above and below the disc.

We must determine the initial conditions for these paths,
noting that modifications to angles of emission also
affect the flux calculation of the previous section.
The initial conditions are (note the difference with Fabian \emph{et al.}(1989))
\begin{equation}
\begin{array}{l}
u_{em} = \frac{1}{\sqrt{r_{em}^{2}+h(r_{em})^{2}}} \\
\frac{du}{d\theta}_{em} = -u_{em}\tan{\xi}.
\end{array}
\end{equation}
Here $\xi$, to be found, must be such that the \emph{photon path is asympotically parallel to the direction to the observer}. This requires
\begin{equation}
\theta_{\infty}(\xi) = \xi^{\prime}
\end{equation}
where $\theta_{\infty}(\xi)$ is the asymptotic angle of the photon path starting with angle $\xi$ and $\xi^{\prime}$ is the angle of the Euclidean path to the observer, as in the flux calculation above. This is solved numerically, with $\theta_{\infty}(\xi)$ calculated by 
integrating the differential equation for the trajectory up to a 
large distance. It is sufficient to consider $-\pi/2 < \xi < \pi/2$ for upper side trajectories. The photon makes no contribution 
to the flux if its path takes it inside the event horizon (at $r=2$).

\section{Line profiles and discussion}

\subsection{Discussion of parameter values}

There is a wide range in the values of $r_{out}$ used in iron line models. Frequently very small values are used
(e.g. Tanaka \emph{et al.} (1995) have $r_{out} \approx 20$), however, 
much larger values have also been employed (e.g. Nandra \emph{et al.} (1997) use $r_{out}=10^{3}$). For the illumination laws considered, 
the use of a large $r_{out}$ effectively sets
the radial cutoff at infinity for flat discs.
A significant effect of disc curvature is to increase the contribution to 
the line profile from photons originating from large radii (Blackman 1999), so it can be instructive to consider even larger radii. 
We consider $r_{out}=10^{2}$, $r_{out}=10^{3}$ and $r_{out}=10^{4}$.

The meaning of $q$ is slightly different than in the model described in the original Fabian \emph{et al.} (1989) paper (and used subsequently by, for example, Matt \emph{et al.} 1993; Tanaka \emph{et al.} 1995; Iwasawa \emph{et al.} 1999; Nandra \emph{et al.} 1997) because they have the emitting point in
the plane of the disc and consequently cannot include the cosine factor, $\mathbf{n \cdot n_{d}}$. 
Their use of a single power law does 
not take into account the changing angle of incident flux onto the disc.
Hence, $q=2$ for us is only roughly comparable to $q=3$ for the Fabian 
\emph{et al.} model. This issue is discussed in Laor (1991). 
Several authors who do not actually employ the power law change, acknowledge the issue (e.g. Fabian \emph{et al.} 1989; Tanaka \emph{et al.} 1995; Nandra \emph{et al.} 1997) and a few others do formally implement
a source displaced above the disc 
(Laor 1991; Matt \emph{et al.} 1992 but not Matt \emph{et al.} 1993; Reynolds and Begelman 1998; Young \emph{et al.} 1998).

The magnitude of the warp in the innermost regions ($\leq 10^{4}$) of the accretion disc is unknown. For each of the two warps, we initially consider two possibilities, one in which the magnitude of the warp is ``large'' ($h(r)$ of order $r$ at the maximimally warped points) and one in which it is significantly smaller.
\begin{figure}

3 images here (in separate files ter0.gif, ter1.gif, ter3.gif).

(a)\epsfig{file=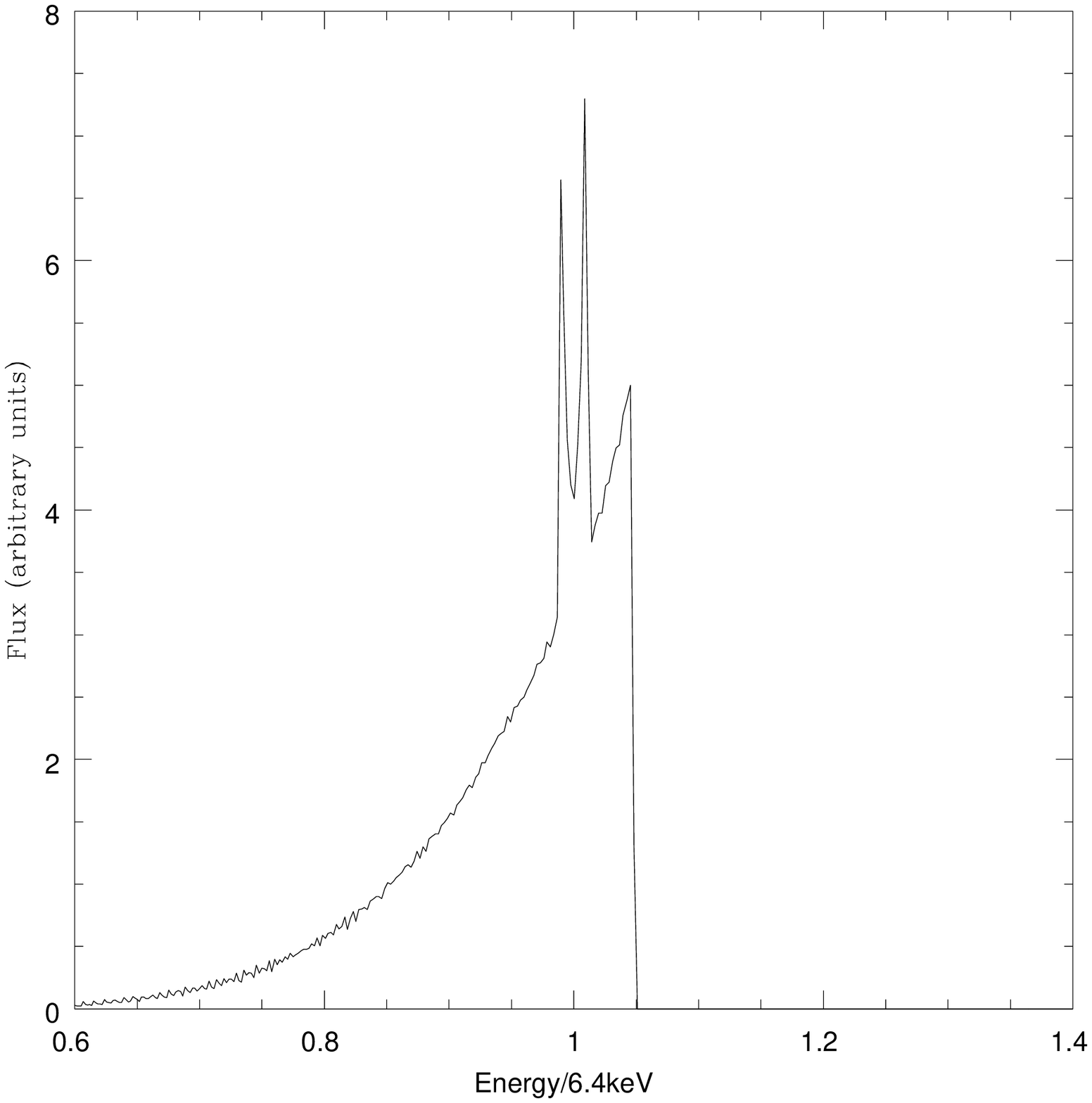,height=2in}
(b)\epsfig{file=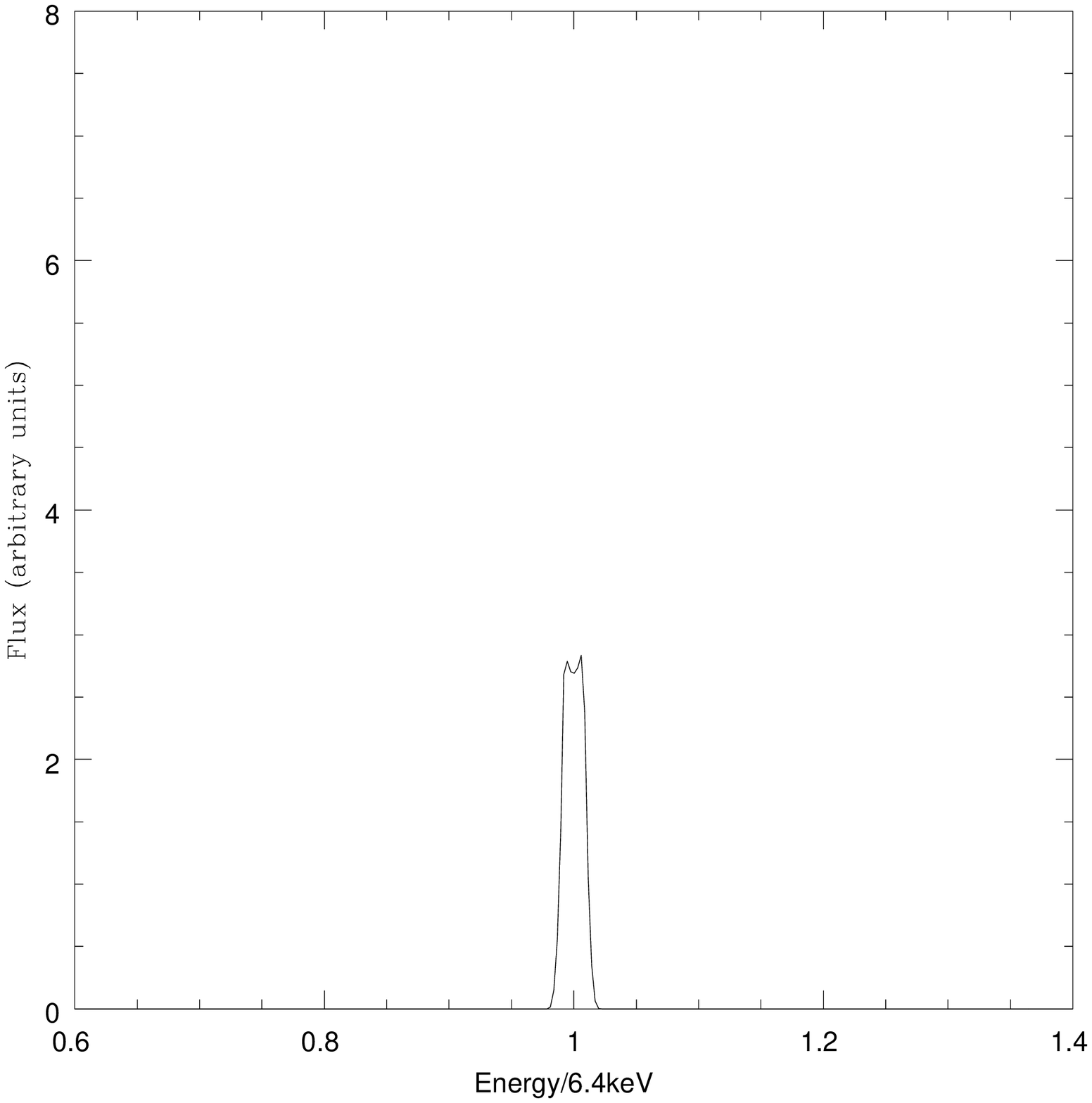,height=2in}
(c)\epsfig{file=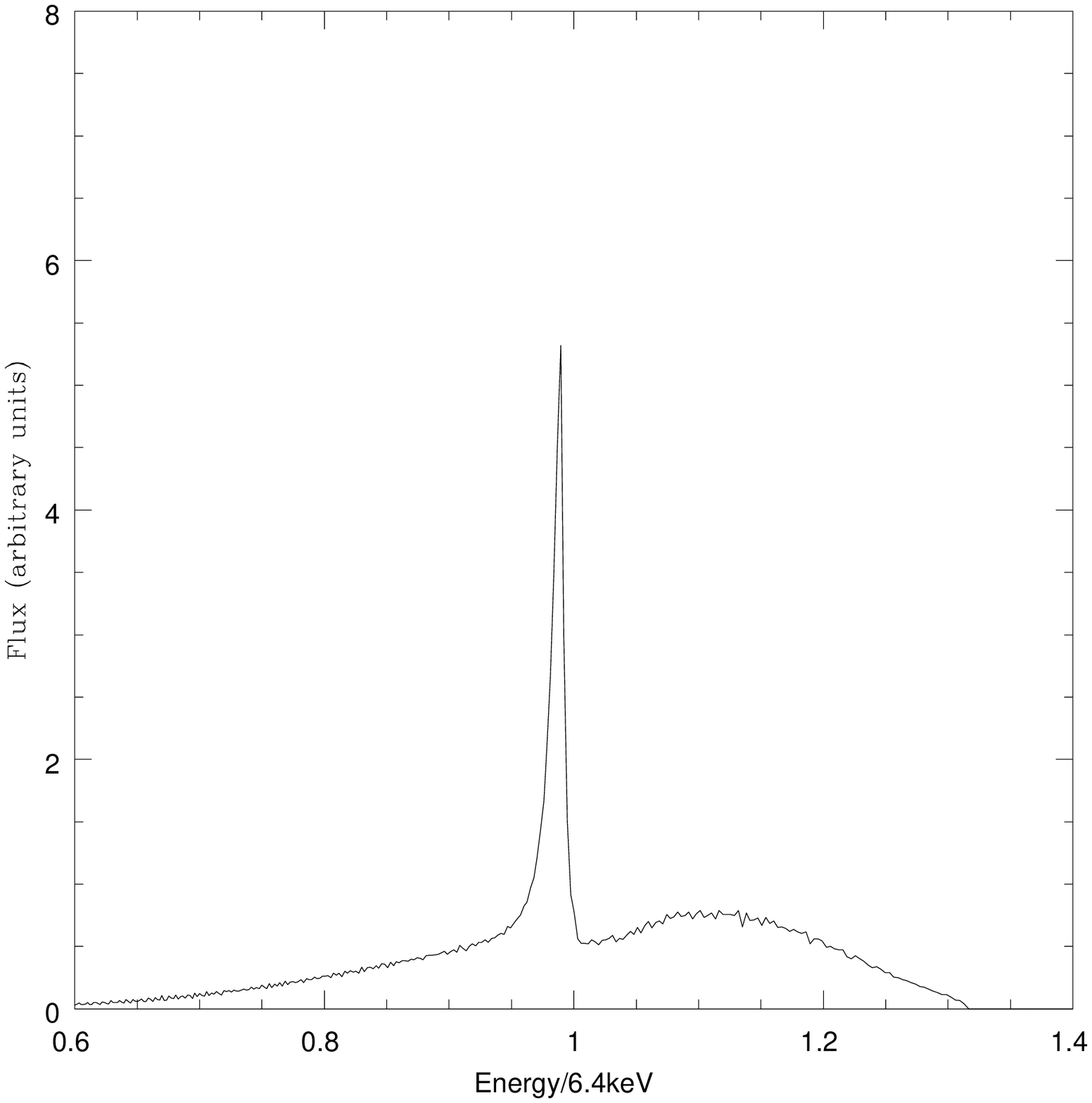,height=2in}

\noindent\textbf{Figure 5:} Frequency maps for three twist-free discs with the resulting line profiles. Black regions show shadowing and darker tone indicates greater redshift. The observer is viewing the disc from the top of the page. The oval shape is due to projection of the tilted disc. All these discs have $r_{out}=10^{4}$. (a) This disc is viewed at $a_{i}=30^{o}$ and has $b=4$ and $\omega t = \pi/2$, so the height of the disc is increasing towards the observer. The dark lower part of the figure represents shadowing from the source whilst the dark region in the top half of the figure represents shadowing of the disc to the observer. (b) This disc is being viewed at $a_{i}=70^{o}$ with $b=2$ and $\omega t = \pi/2$, as before. Only photons from the underside of the disc reach the observer at this angle. The dark upper region is due to 
shadowing from the underside source whilst the lower dark region and part 
of the upper dark region (the central part) represents 
shadowing of the disc to the observer. (c) This disc also has $a_{i}=70^{o}$ and $b=2$ but is rotated, $wt=0$, so it is being viewed ``side on'' (i.e. along the line of nodes). The dark region on the left represents shadowing from the source, and the dark region on the right represents shadowing of the disc to the observer.
\end{figure}
\begin{figure}

2 images here (in separate files pri1.gif, pri3.gif).

(a)\epsfig{file=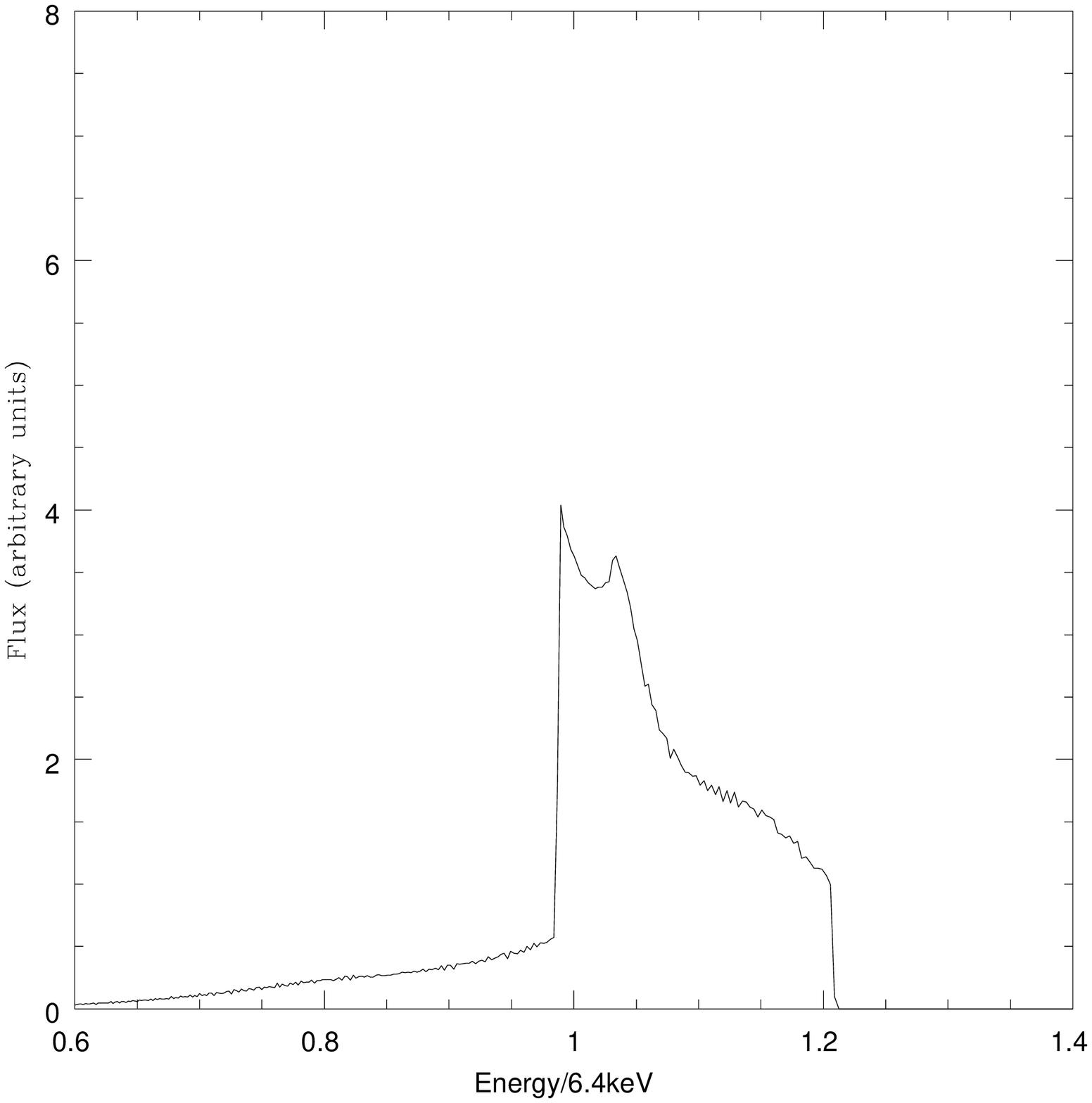,height=2in}
(b)\epsfig{file=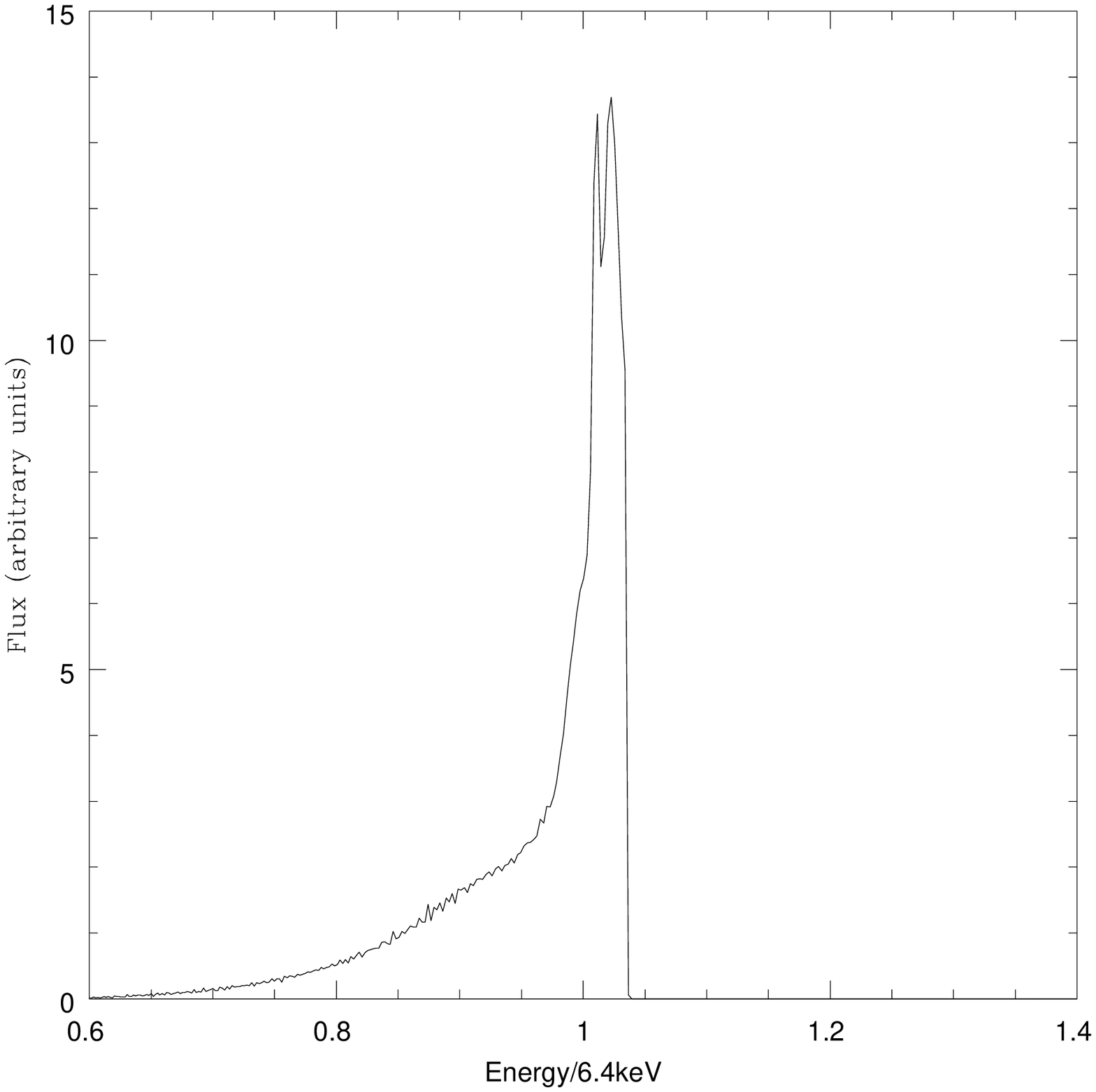,height=2in}

\noindent\textbf{Figure 6:} Frequency maps for two twisted discs with the resulting line profiles. Black regions show shadowing and darker tone indicates greater redshift. The observer is viewing the disc from the top of the page. (a) 
This disc has $r_{out}=10^{3}$, $a_{2}=1$, $\omega t = 0$ and is being viewed at $a_{i}=30^{o}$. Here the dark region is due to 
shadowing from the source. (b) This  disc has $r_{out}=10^{4}$, $a_{2}=2$, $\omega t = 3 \pi / 2$ and is being viewed at $a_{i}=40^{o}$. We see a small additional dark region, which results from shadowing of the disc to the observer.
\end{figure}
\begin{figure}

1 image here (in separate file log1.gif).

\noindent\textbf{Figure 7:} The first disc (disc (a)) from Fig. 5 is shown with the radius plotted logarithmically, thus emphasising the innermost regions. Darker tone indicates greater redshift in the frequency of emmision, as shown in the colourbar (units are keV). The central dark region is inside $r_{in}$ and is oval due to projection. The contours show the effect of curvature in the central regions (cf. Fig. 2, although Fig. 2 is not logarithmic so some care is required).
\end{figure}
\subsection{Frequency maps of warped discs}

Figs. 5 and 6 show frequency maps for a selection of warped discs above 
the corresponding iron line profiles. This helps to visualise the effect of shadowing and the correlation between geometry and line profile. Figs. 1 and 2 are useful for interpreting these maps. The parameters for these discs (i.e. high inclination angle of the observer, large outer radius, large magnitude of warp) were chosen for illustration of the various types of shadowing. Fig. 7 shows one of the twisted discs with the radius plotted logarithmically. This clearly illustrates the effect of the warp on the frequency shifts in the inner regions of the disc.
\begin{figure}
\epsfig{file=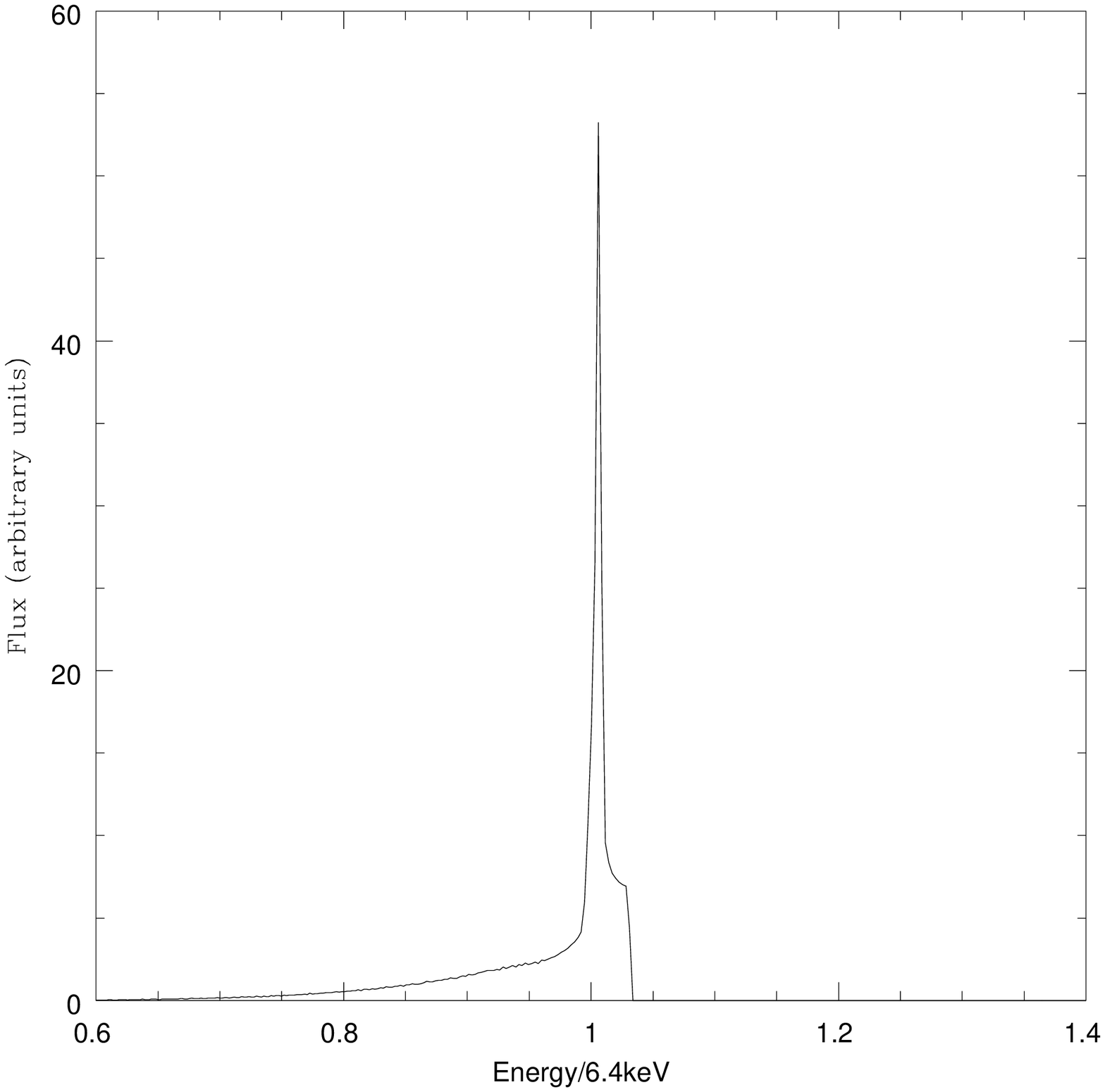,height=2in}

\noindent\textbf{Figure 8:} Profile from a twisted disc with $a_{2}=1$,$r_{out}=10^{4}$, $\omega t = 1.25 \pi$ and $a_{i}=30^{o}$, illustrating the phenomenon of a large, thin peak due to the large outer radius.
\end{figure}
\subsection{Large outer radius and large magnitude of warp}

As mentioned above, an important effect of curvature is to increase the flux from the outer regions of the disc, relative to the flat disc case, because the angle of incident flux from the source is more favourable. If the outer radius is large ($10^{4}$, or $10^3$ with a large magnitude of warp), then this effect can dominate the line profile. This happens despite the falloff of the illumination law as $|{\mathbf{s}}-{\mathbf{r}}|^{-q}$ due to the large ammount of disc surface involved (of order $r^{2}$). The result is a very sharp, thin peak close to the rest energy (the outer parts of the disc are not very Doppler shifted because the Keplerian velocity falls off as $r^{-1/2}$). We include an example of a resulting line profile (Fig. 8), as such lines may be relevant to other cases of line emission or even to better resolved or as yet unobserved iron lines, but we consider smaller $r_{out}$ in the following section.

\begin{figure}
(a)\epsfig{file=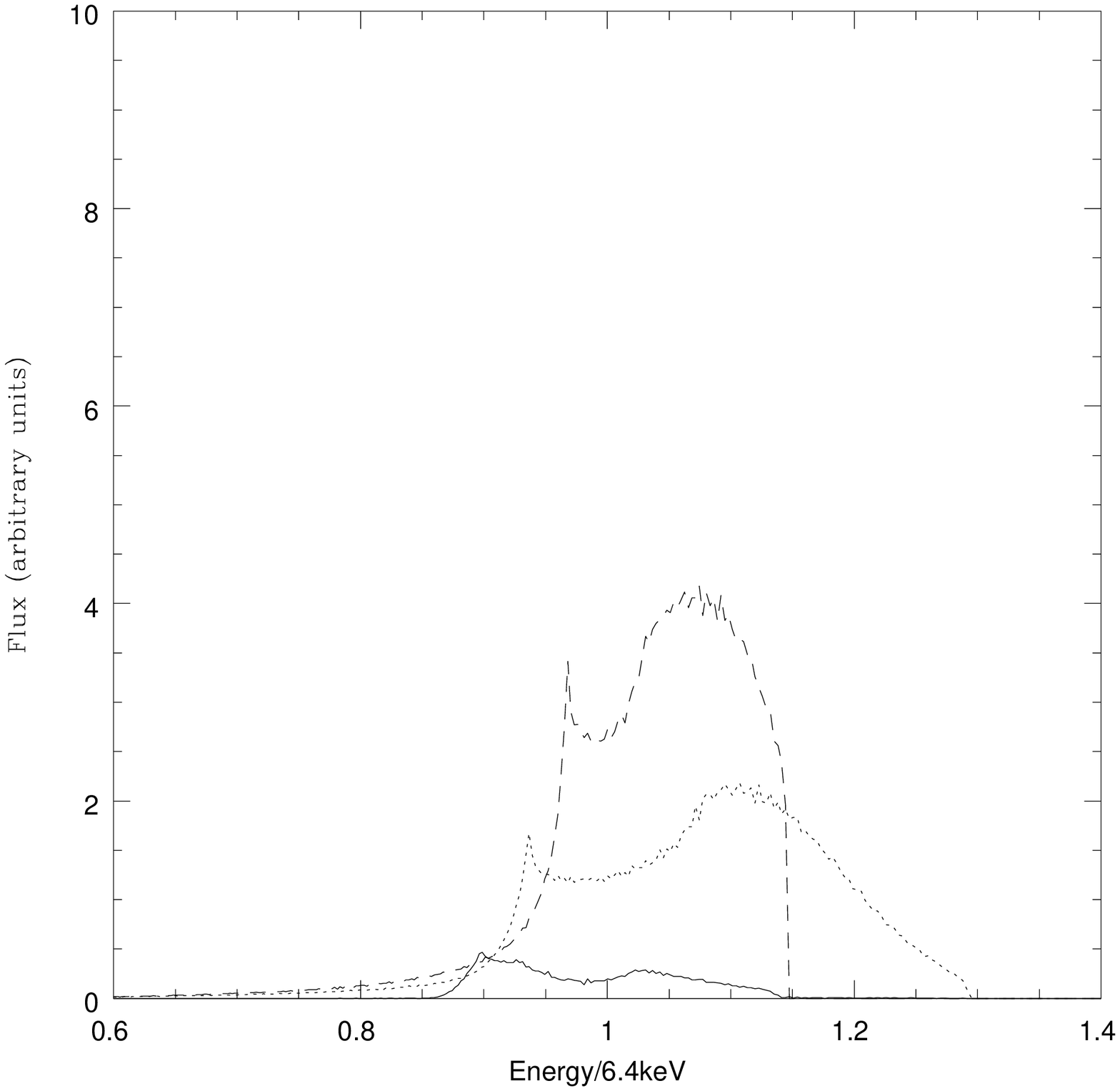,height=2in}
(b)\epsfig{file=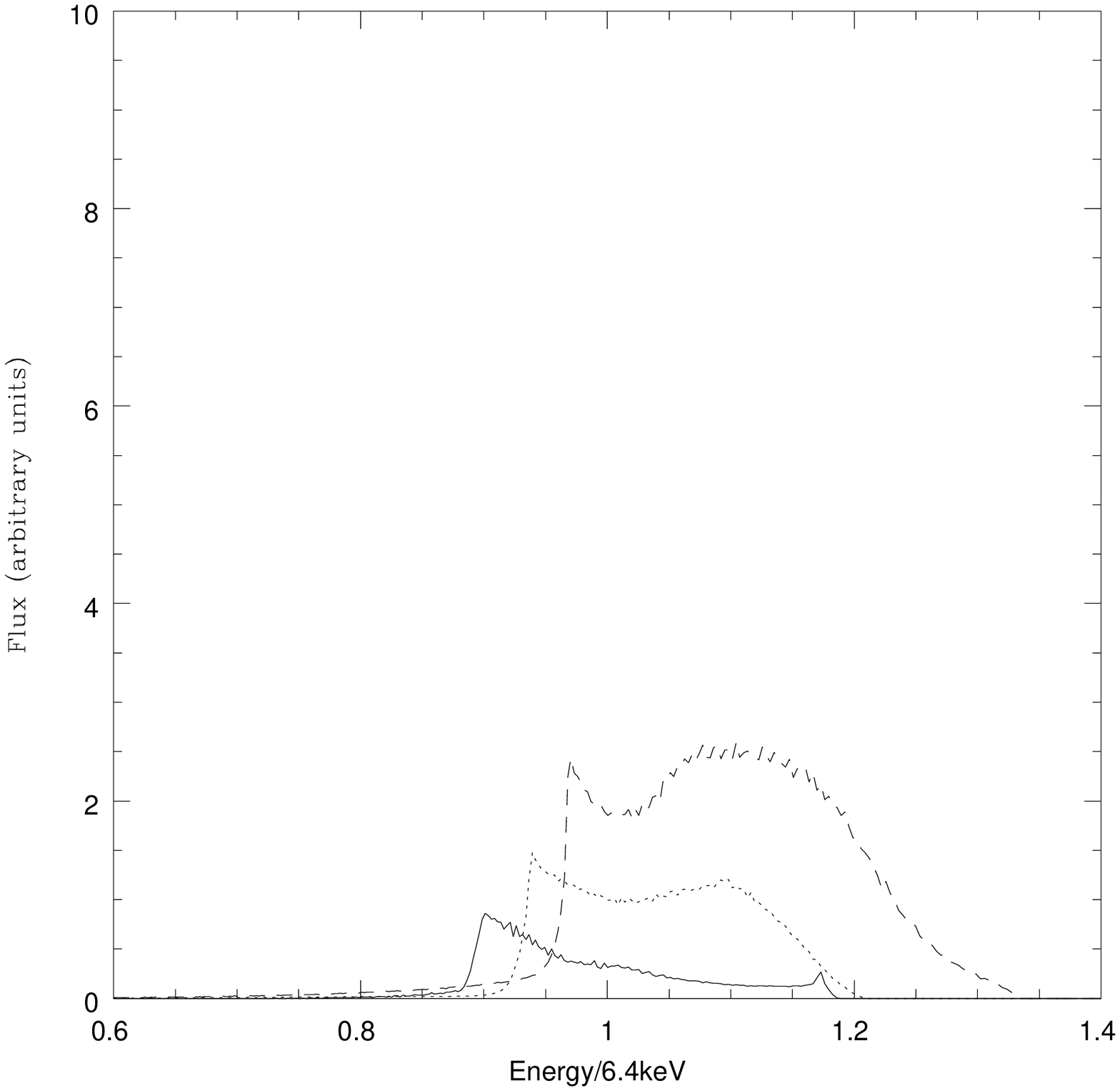,height=2in}
(c)\epsfig{file=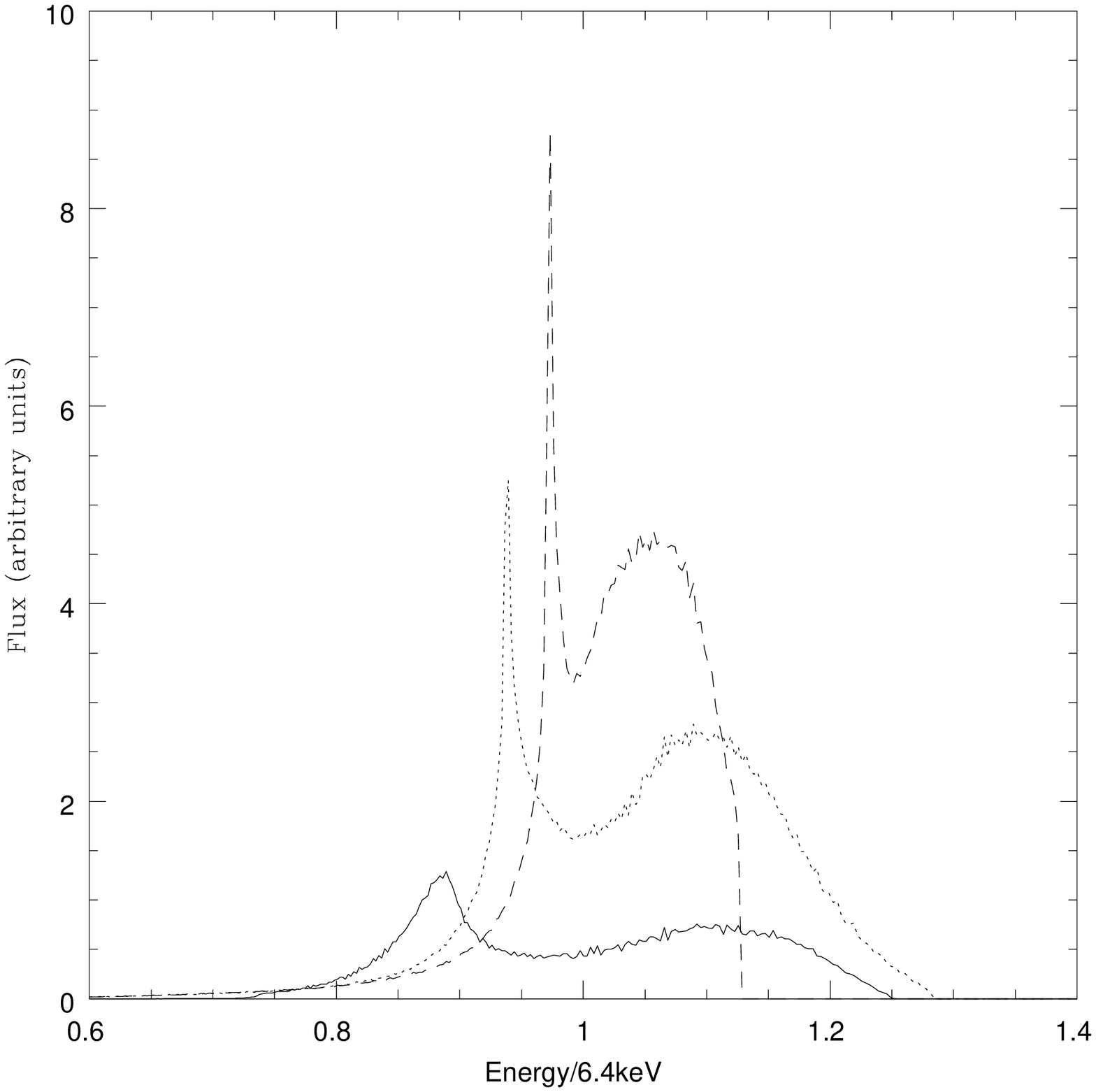,height=2in}

(d)\epsfig{file=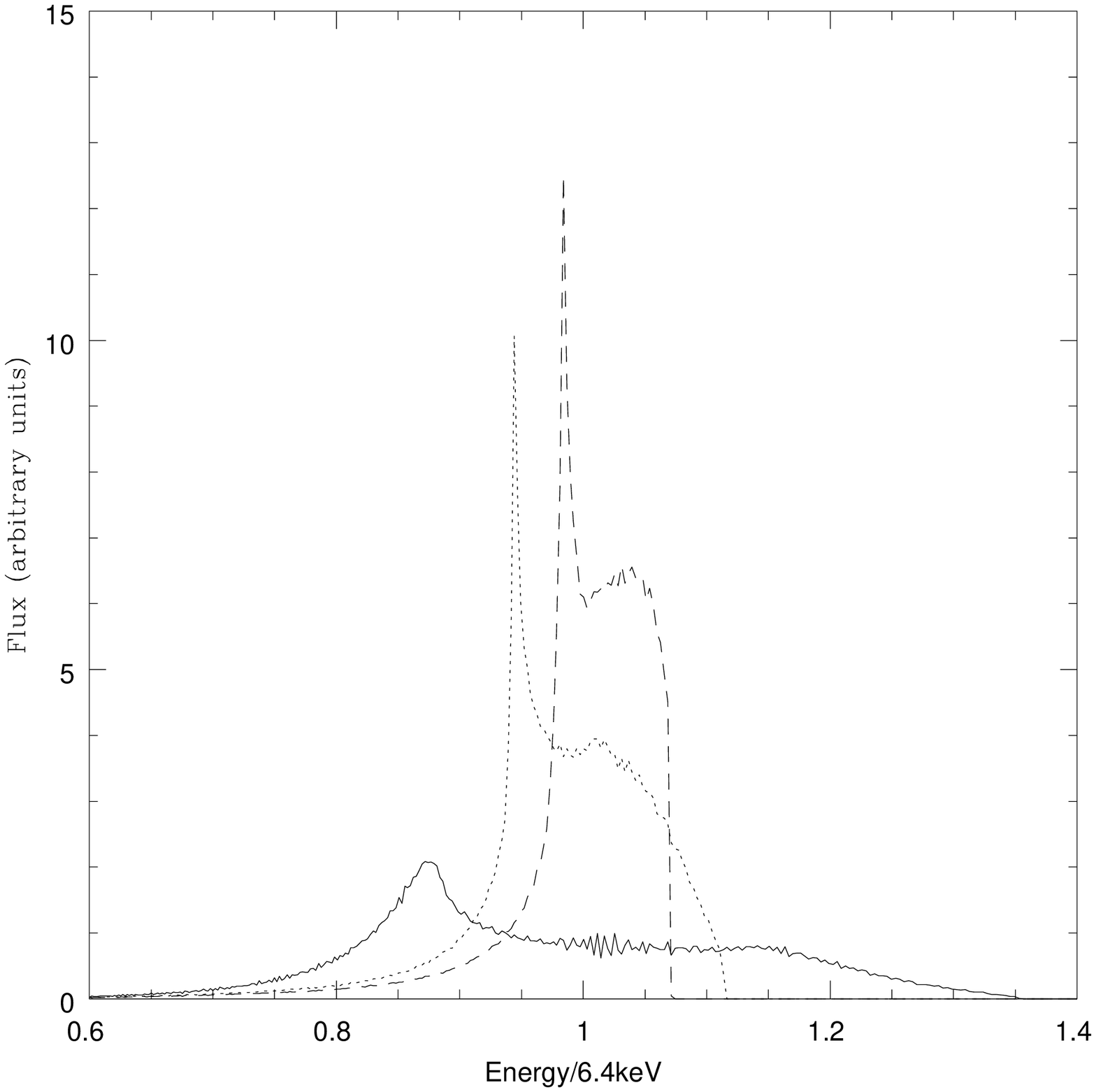,height=2in}
(e)\epsfig{file=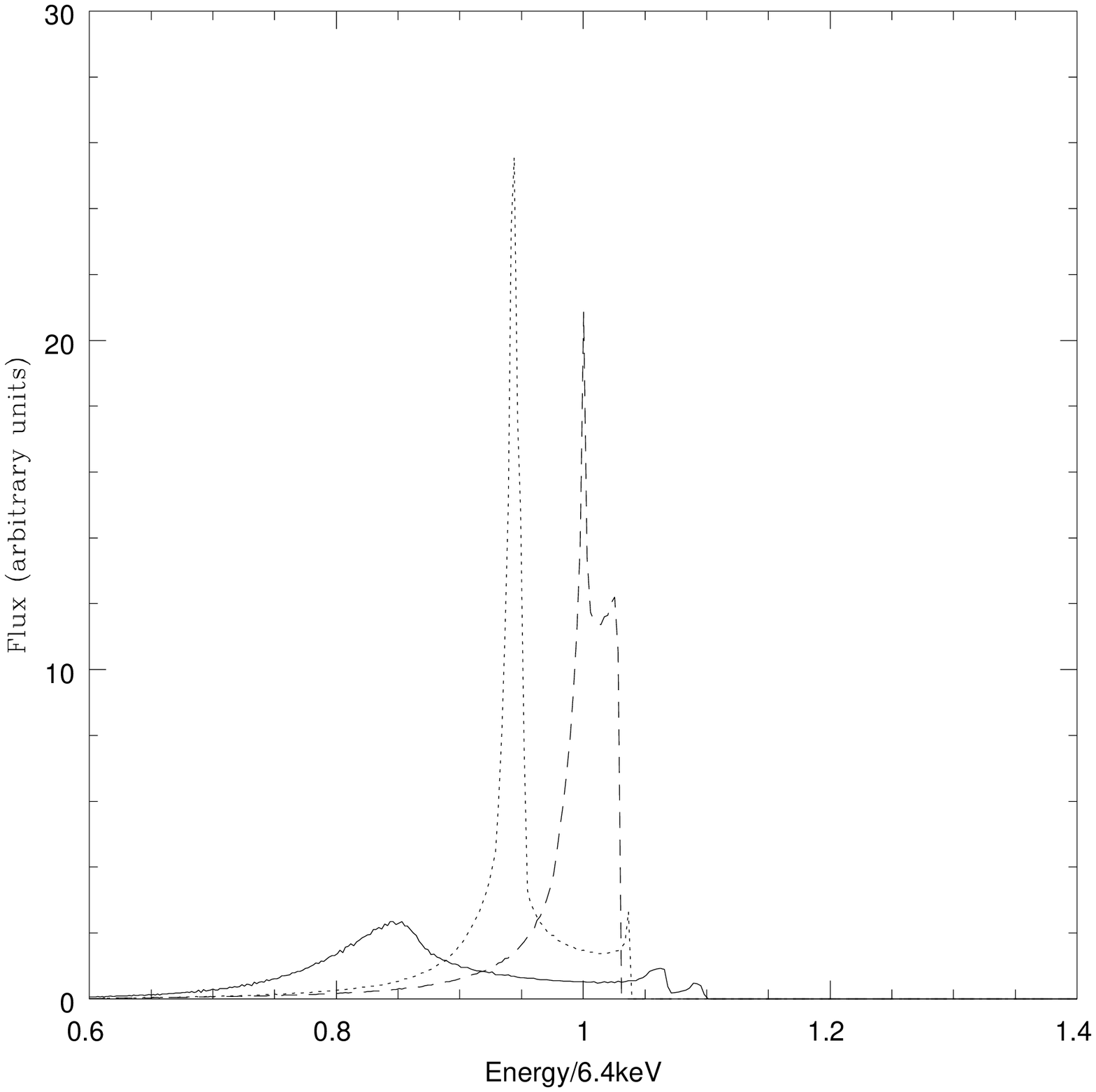,height=2in}
(f)\epsfig{file=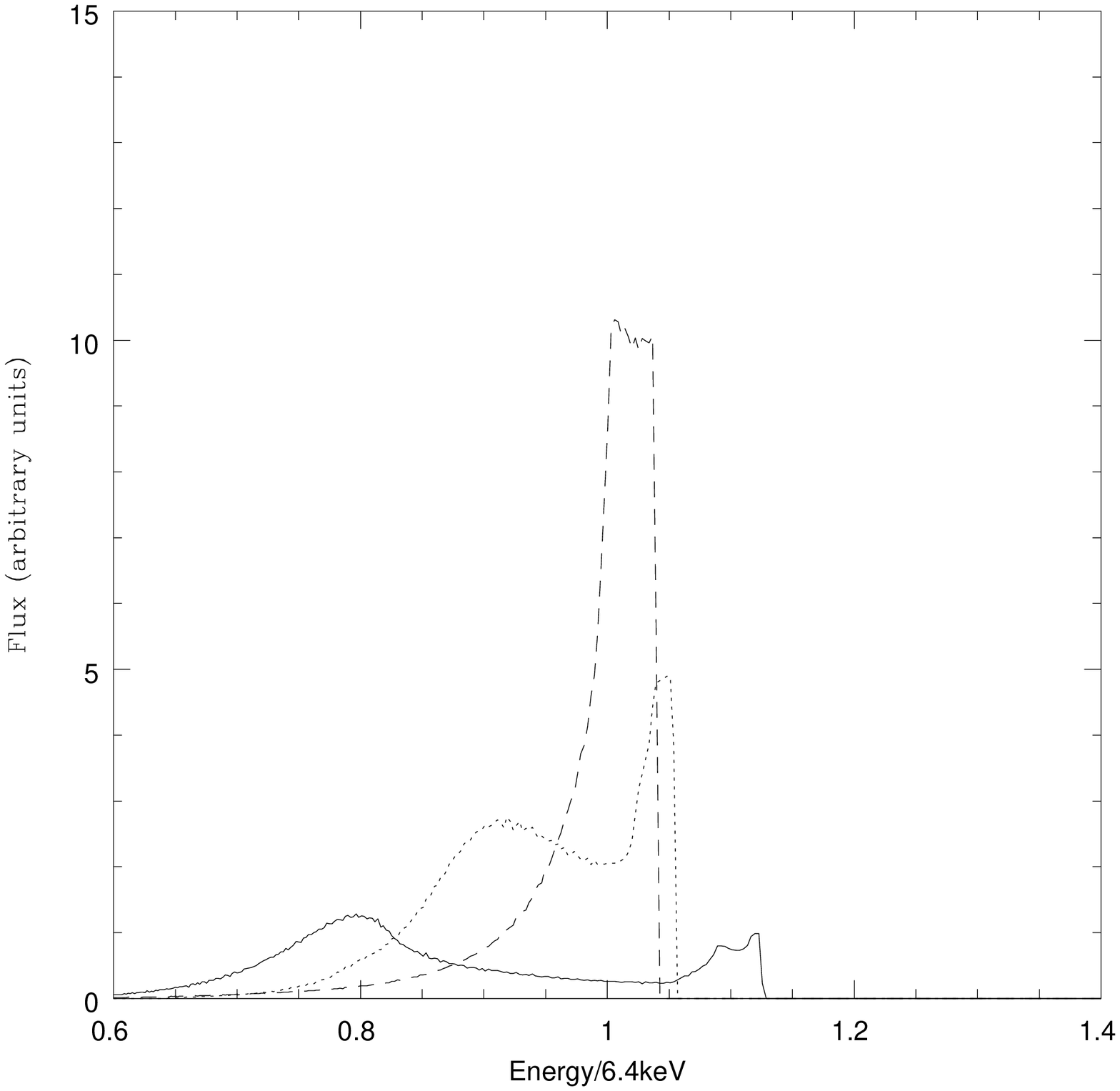,height=2in}

(g)\epsfig{file=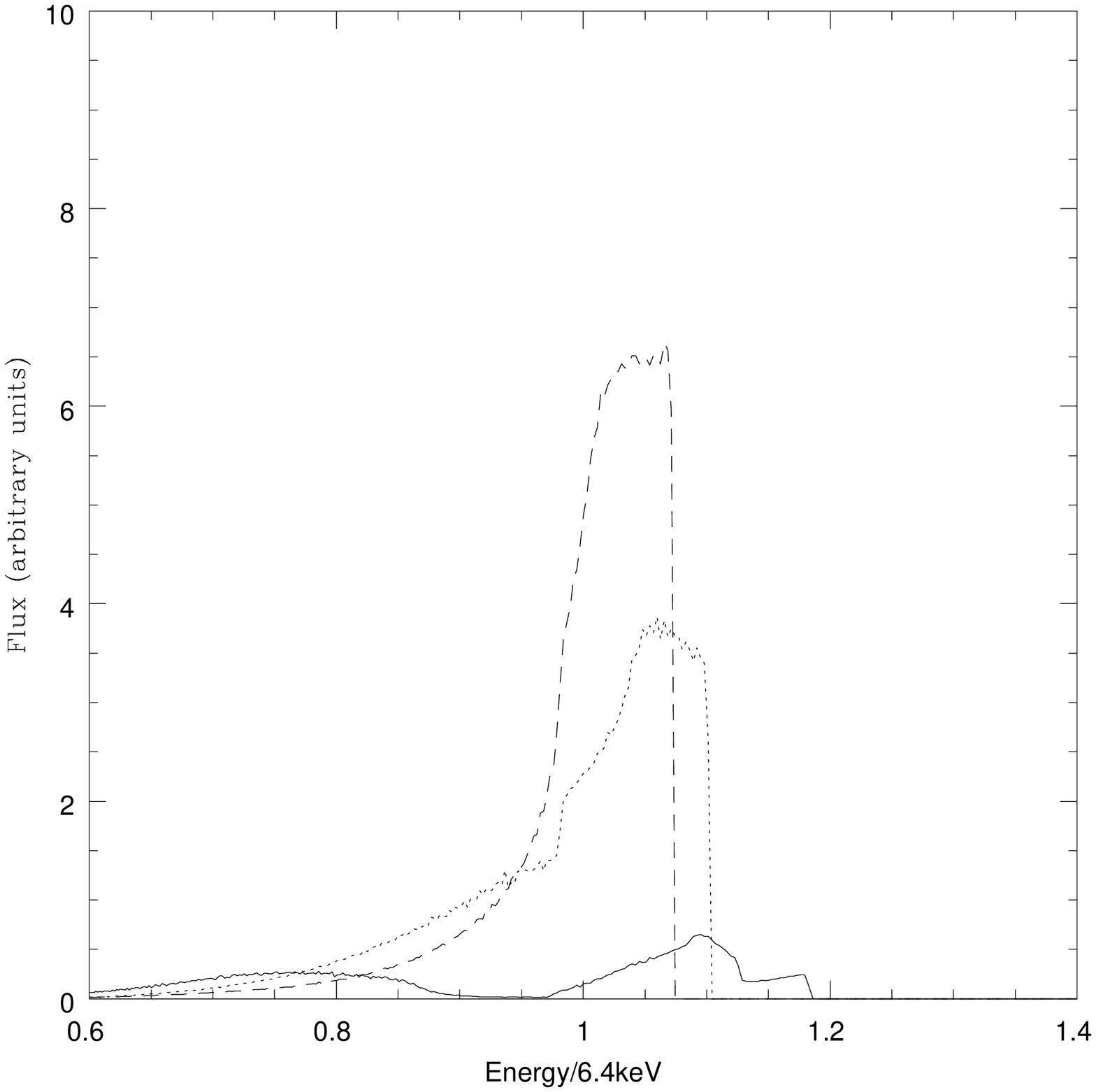,height=2in}
(h)\epsfig{file=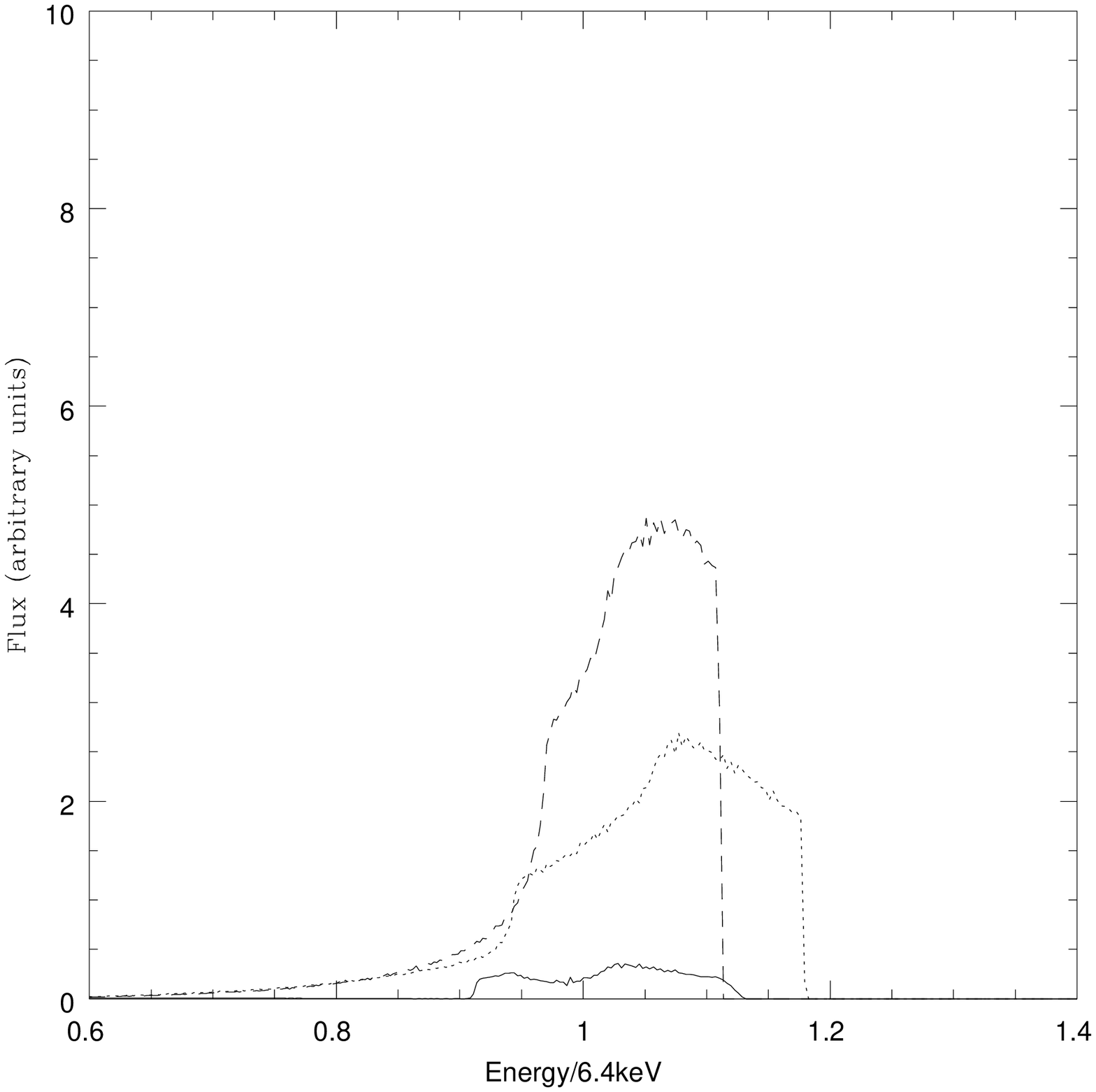,height=2in}
\hspace{0.5cm}(flat)\epsfig{file=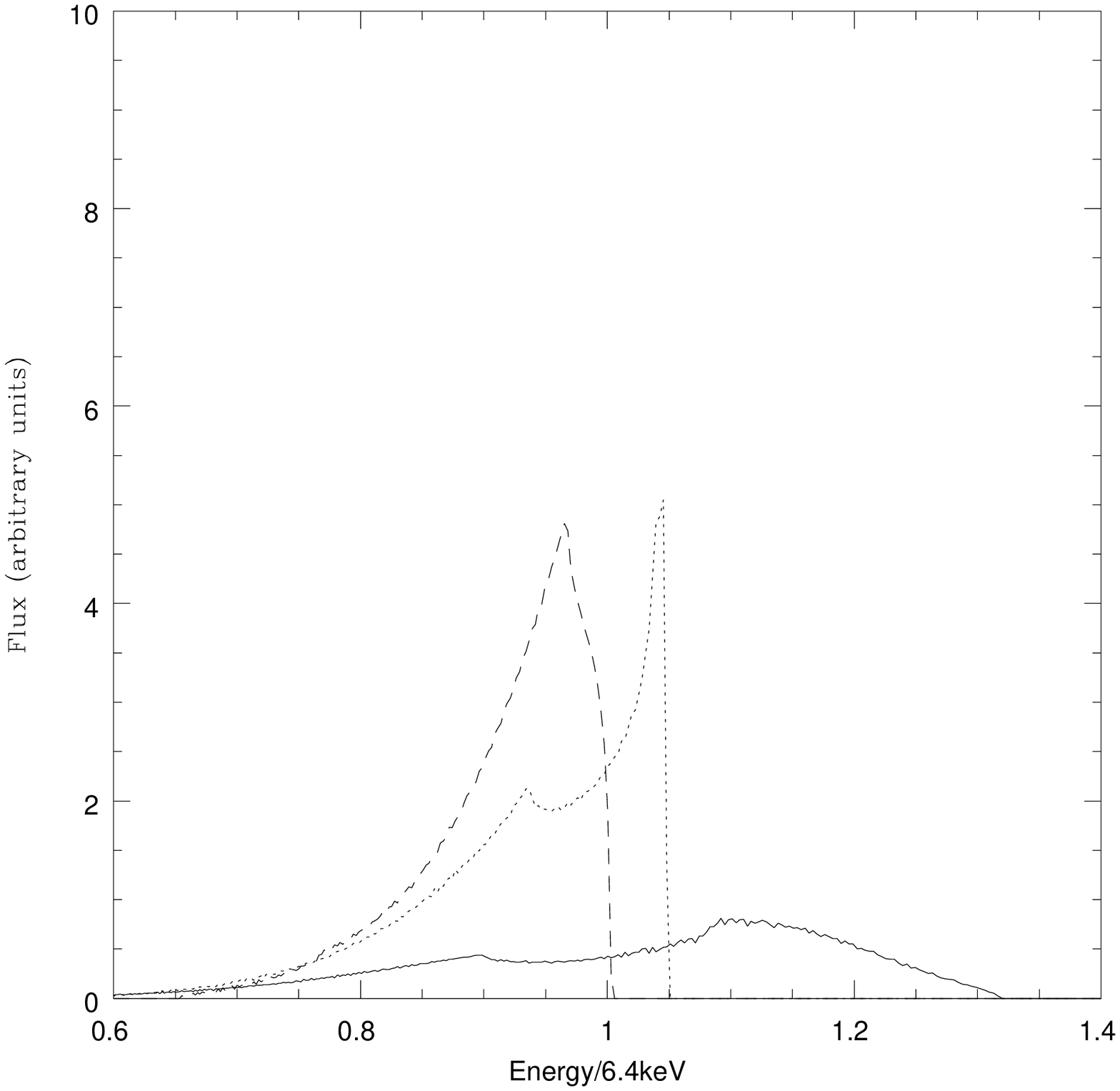,height=2in}

\noindent\textbf{figure 9:} (a)-(h) Twisted disc with $r_{out}=10^{2}$, $a_{2}=1$. (flat) is a flat disc with $r_{out}=10^{2}$ for comparison. 
The dashed, dotted and solid lines in each box 
represent inclination  viewing angles of $10^o$, $30^o$, and $70^o$ respectively from the x-y plane. 
However the inner part of the twisted discs are tilted
with respect to this plane, so $0^o$ does not correspond to the direction
of the X-ray source from the center.  
Note that the y-axis range on 
the central box, e, is 30 and is 15 on the boxes immediately on each side to include the top of the sharp peaks.
Each box represents viewing from a different azimuthal angle 
proceeding from $0$ to $2\pi$, 
where $0$ represents looking from the side of the disc containing the maximum
height, and from the plane containing this maximum.
The observer moves clockwise around the disc from (a)-(h) (or
alternatively, the disc moves anti-clockwise with respect to a fixed observer).

\end{figure}
\begin{figure}
(a)\epsfig{file=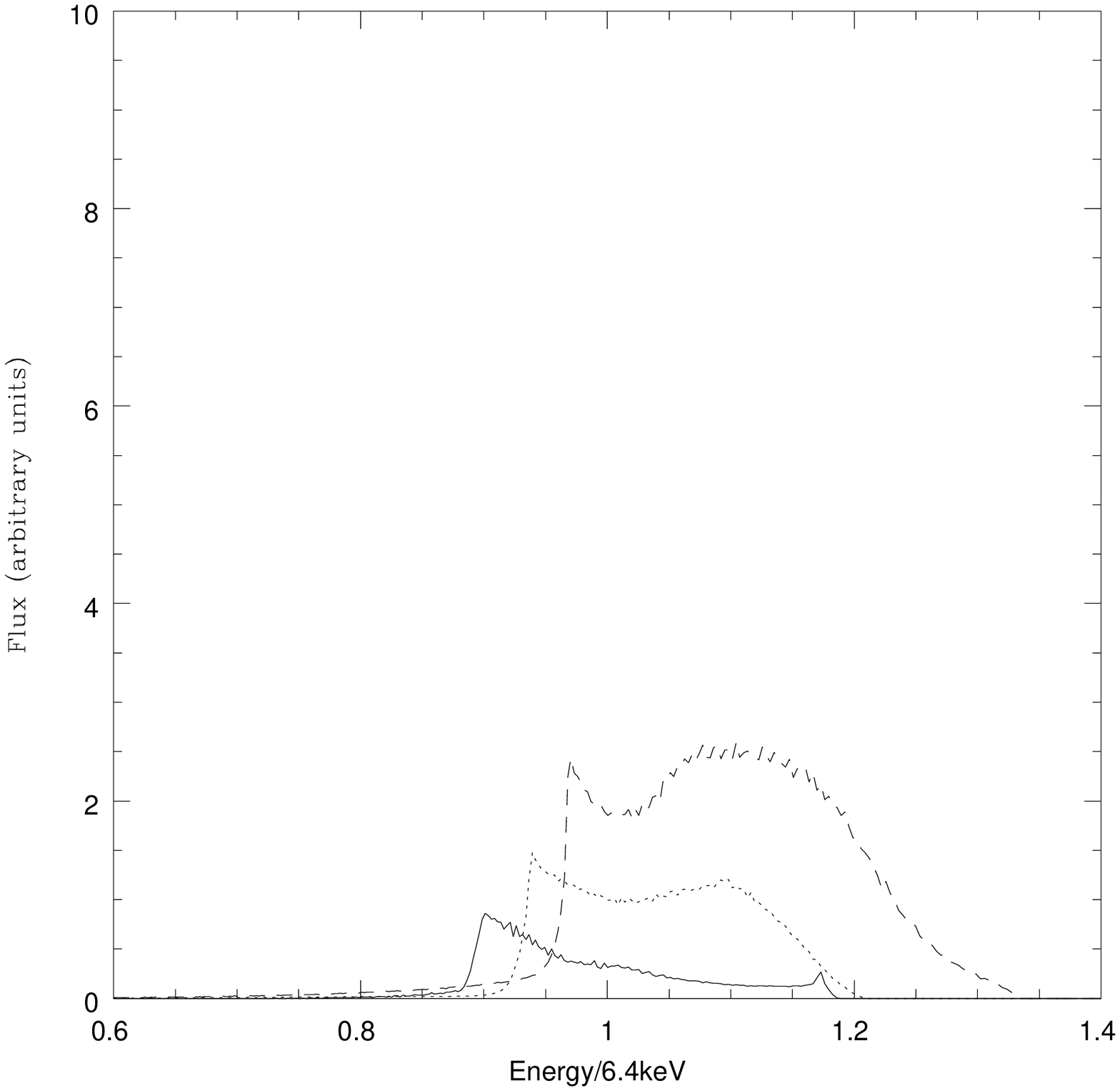,height=2in}
(b)\epsfig{file=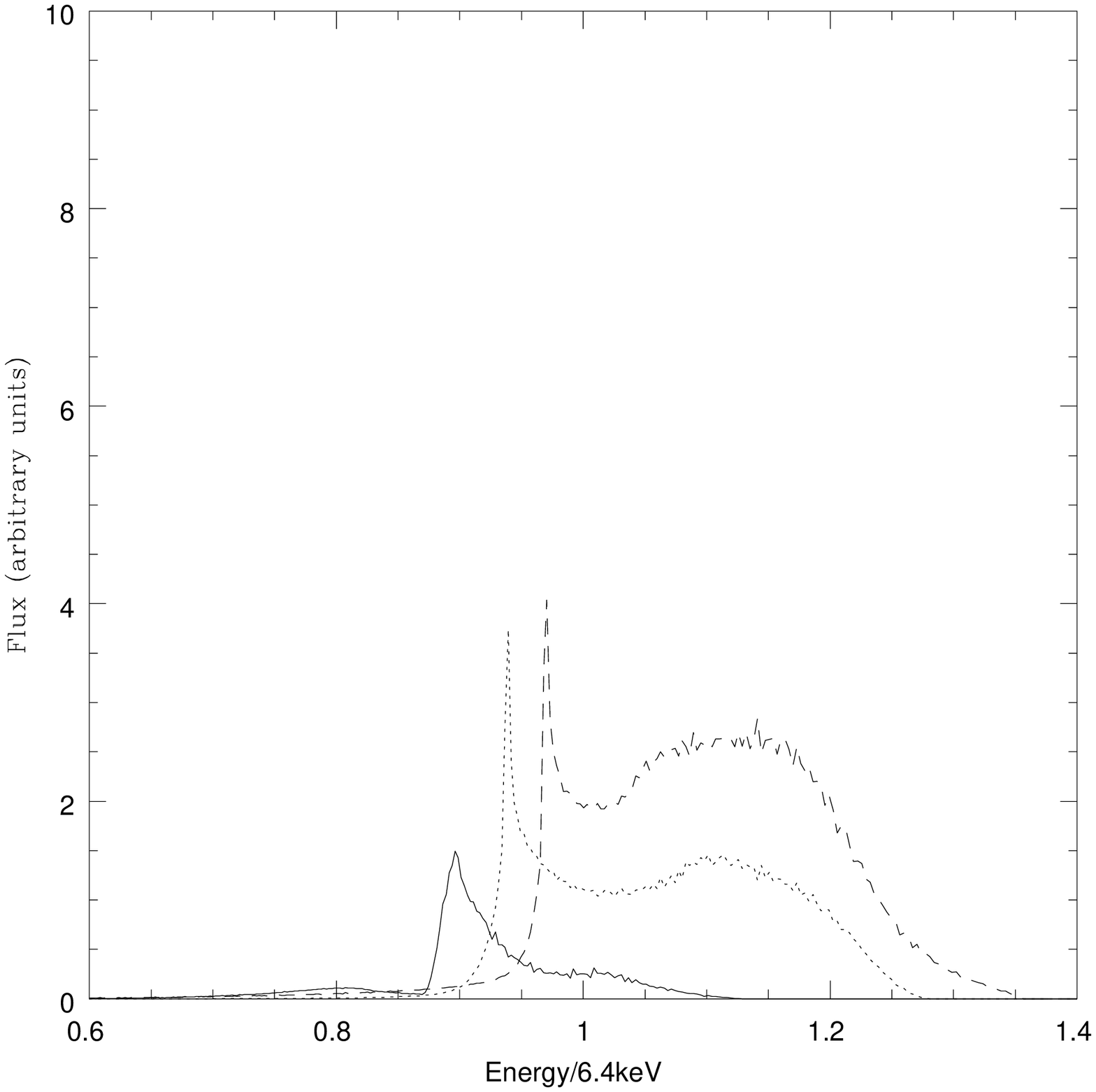,height=2in}
(c)\epsfig{file=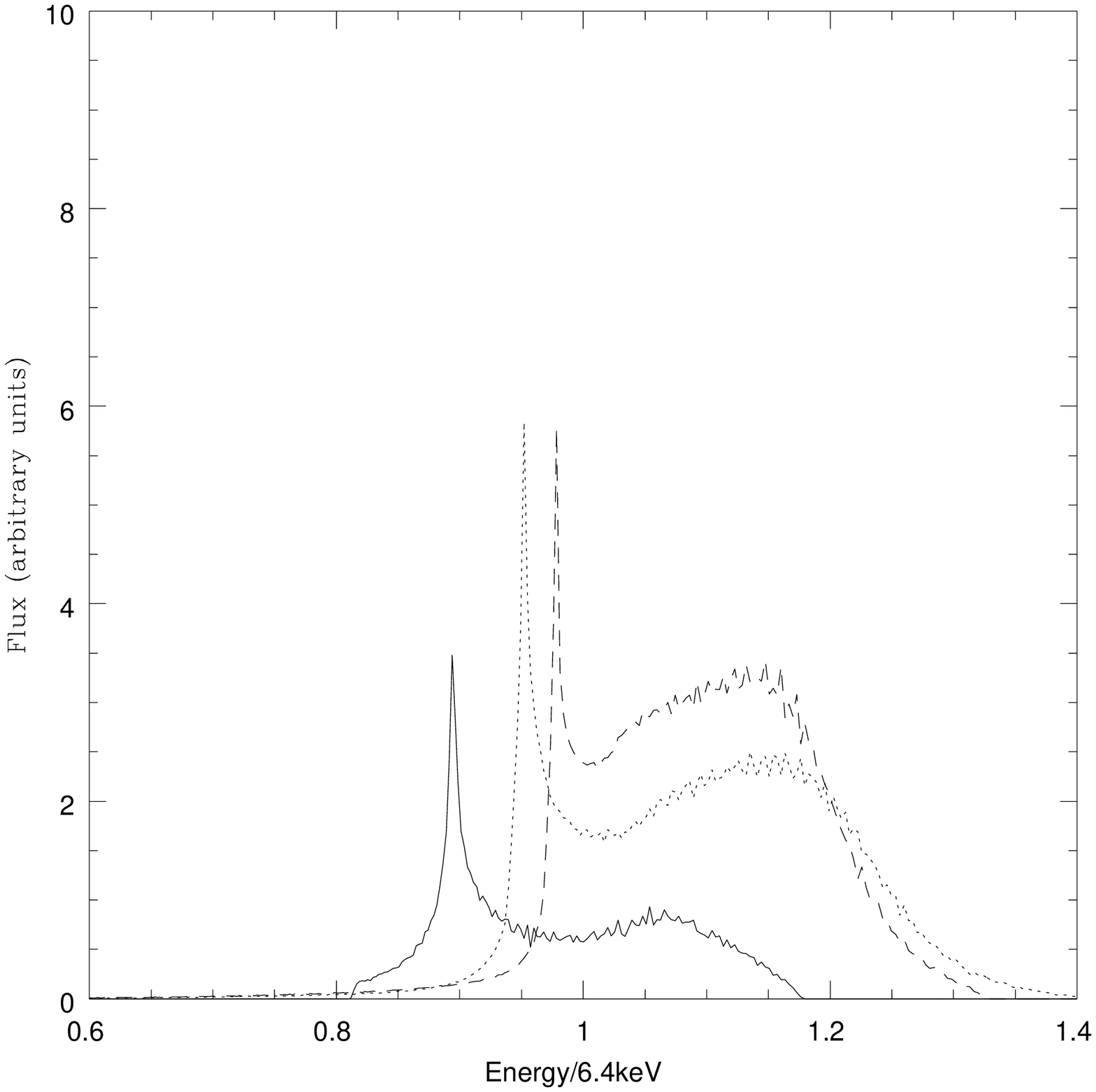,height=2in}

(d)\epsfig{file=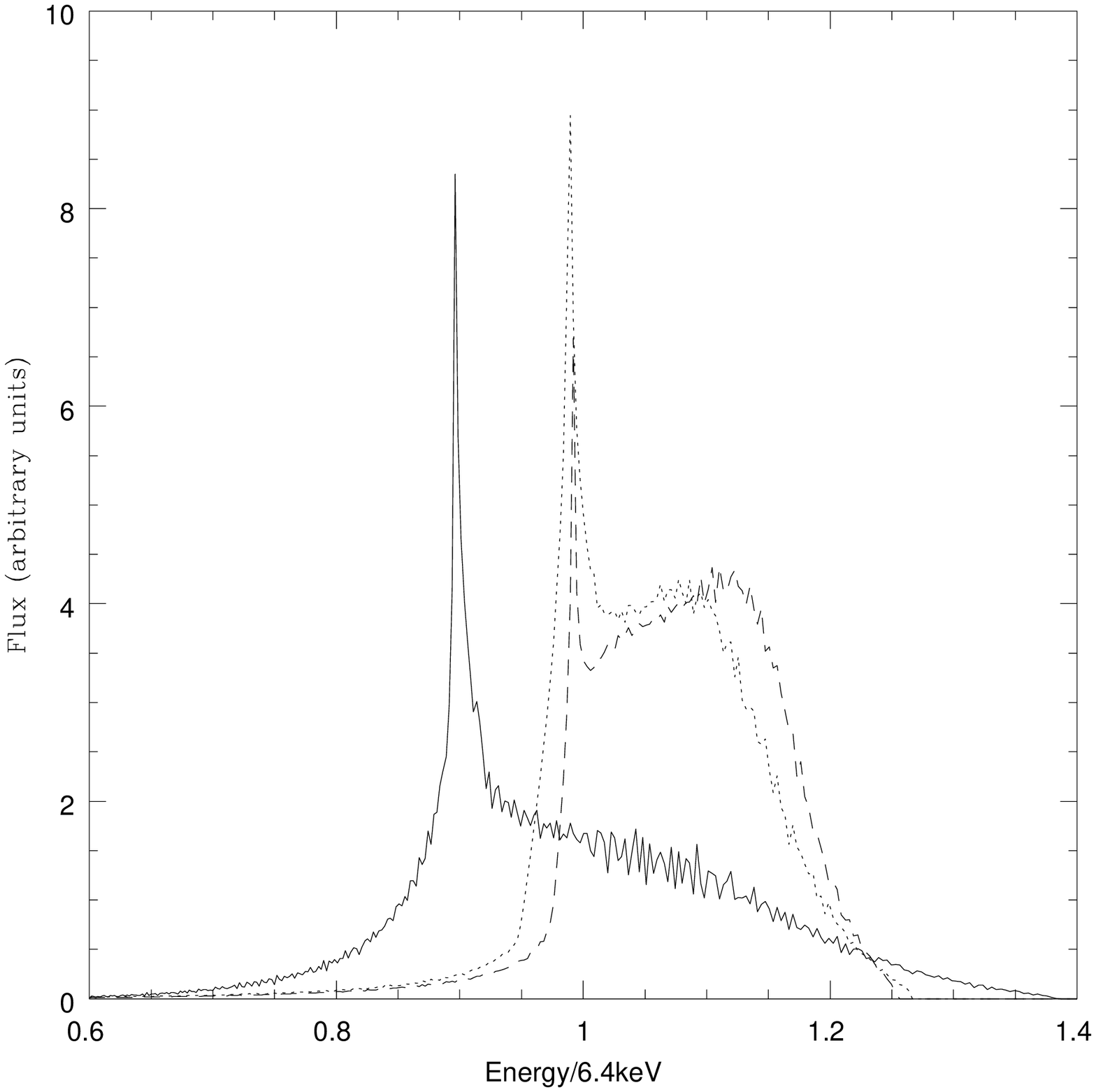,height=2in}
(e)\epsfig{file=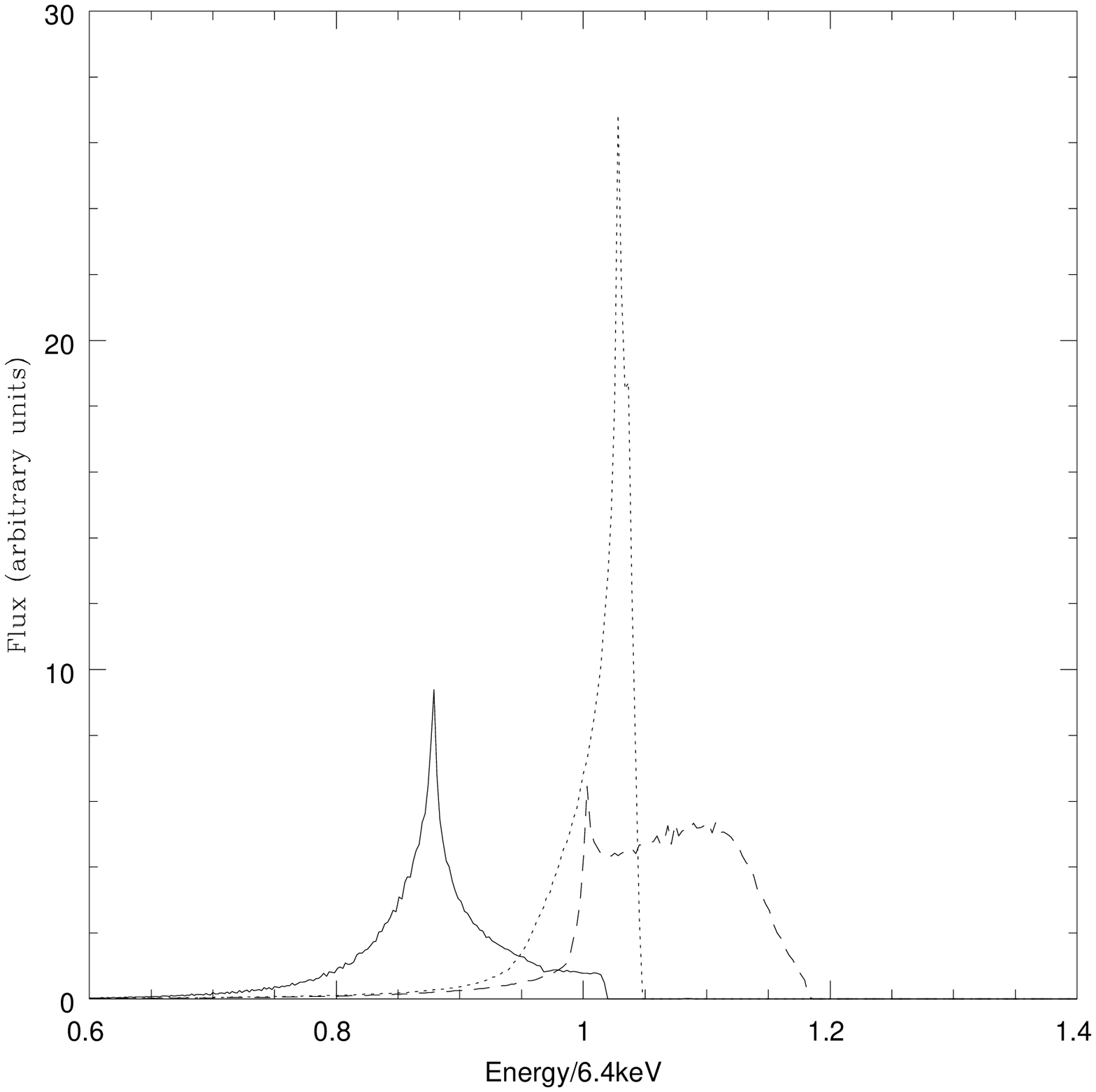,height=2in}
(f)\epsfig{file=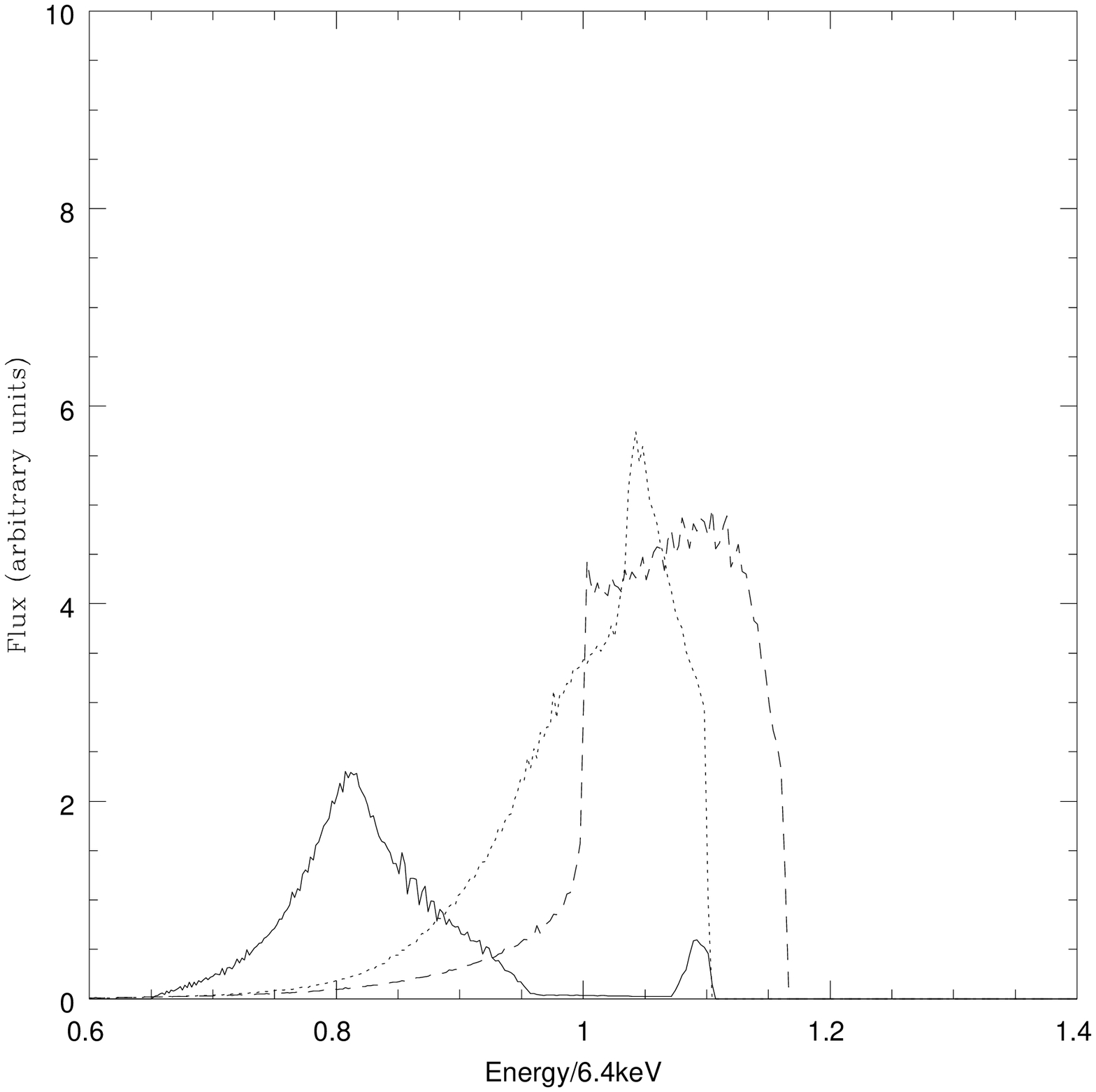,height=2in}

(g)\epsfig{file=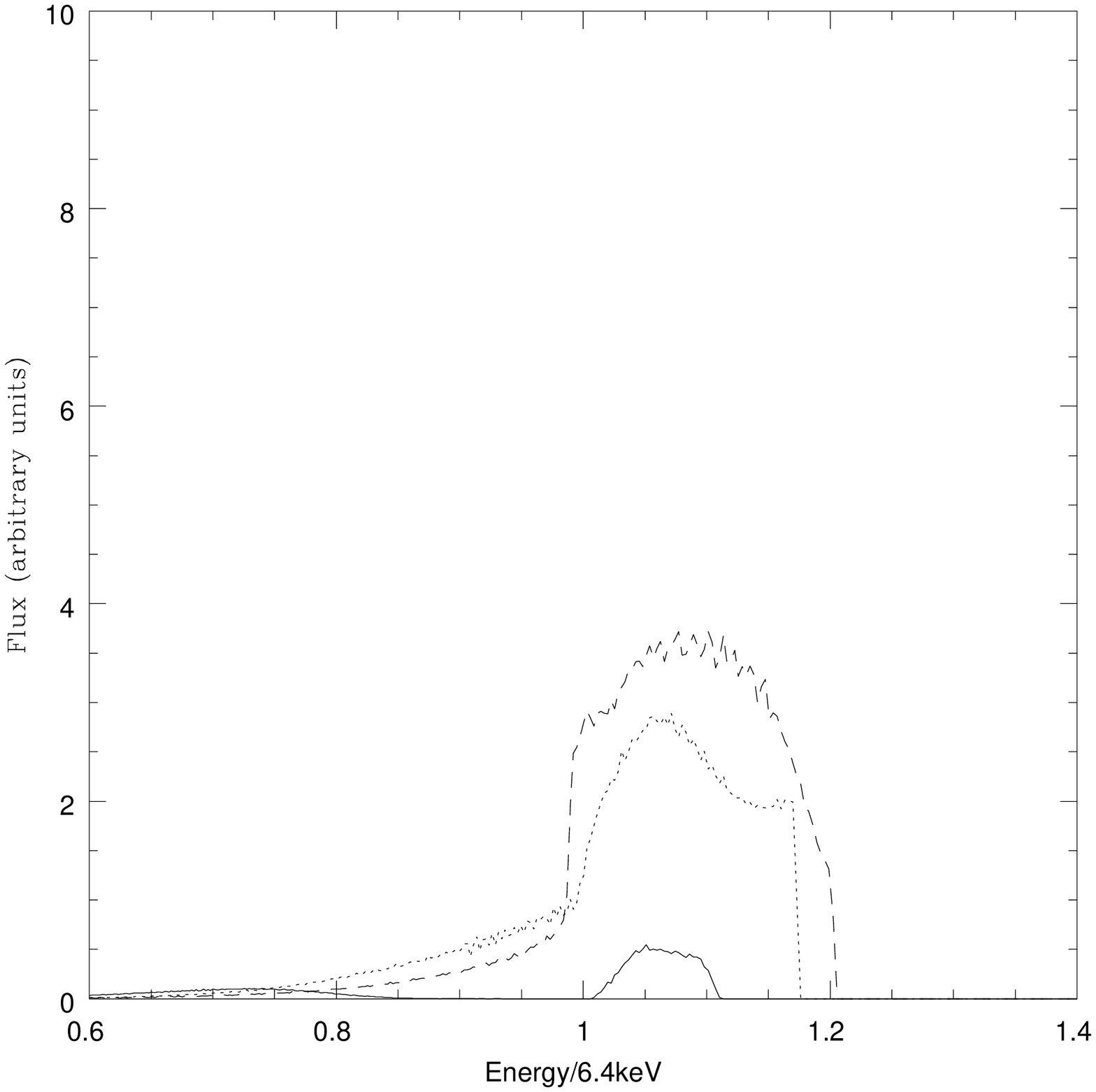,height=2in}
(h)\epsfig{file=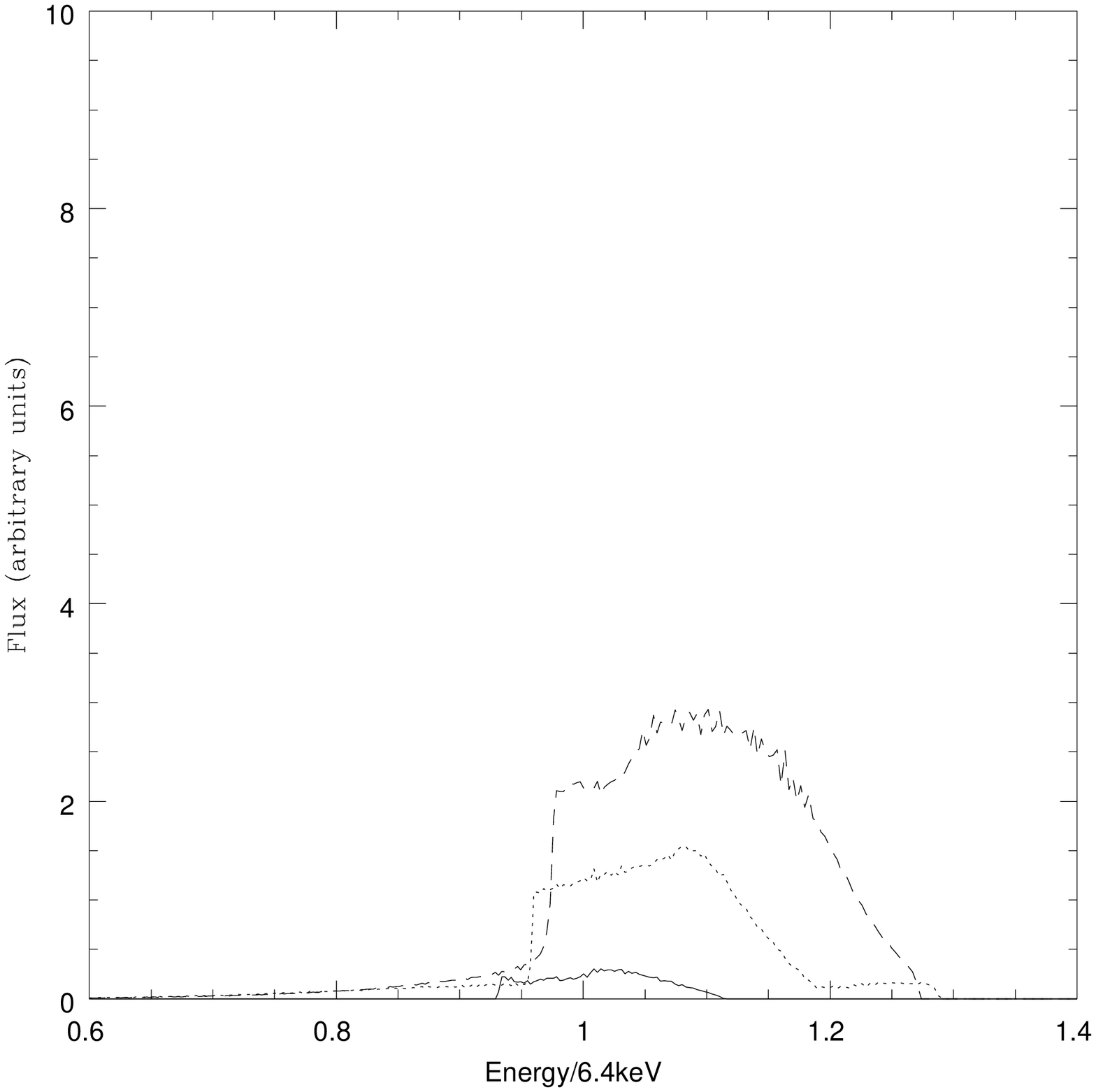,height=2in}
\hspace{0.5cm}(flat)\epsfig{file=flat2,height=2in}

\noindent\textbf{figure 10:} (a)-(h) Twisted disc with $r_{out}=10^{2}$, $a_{2}=2$. (flat) is a flat disc with $r_{out}=10^{2}$ for comparison.
boxes represent azimuthal progression as in Fig. 9,
and dashed, dotted, and solid lines in each box 
again represent inclination viewing angles of $10^o$, $30^o$, and $70^o$ respectively
(see Fig. 9 caption). Note the y-axis range on box e is 30.

\end{figure}

\begin{figure}
(a)\epsfig{file=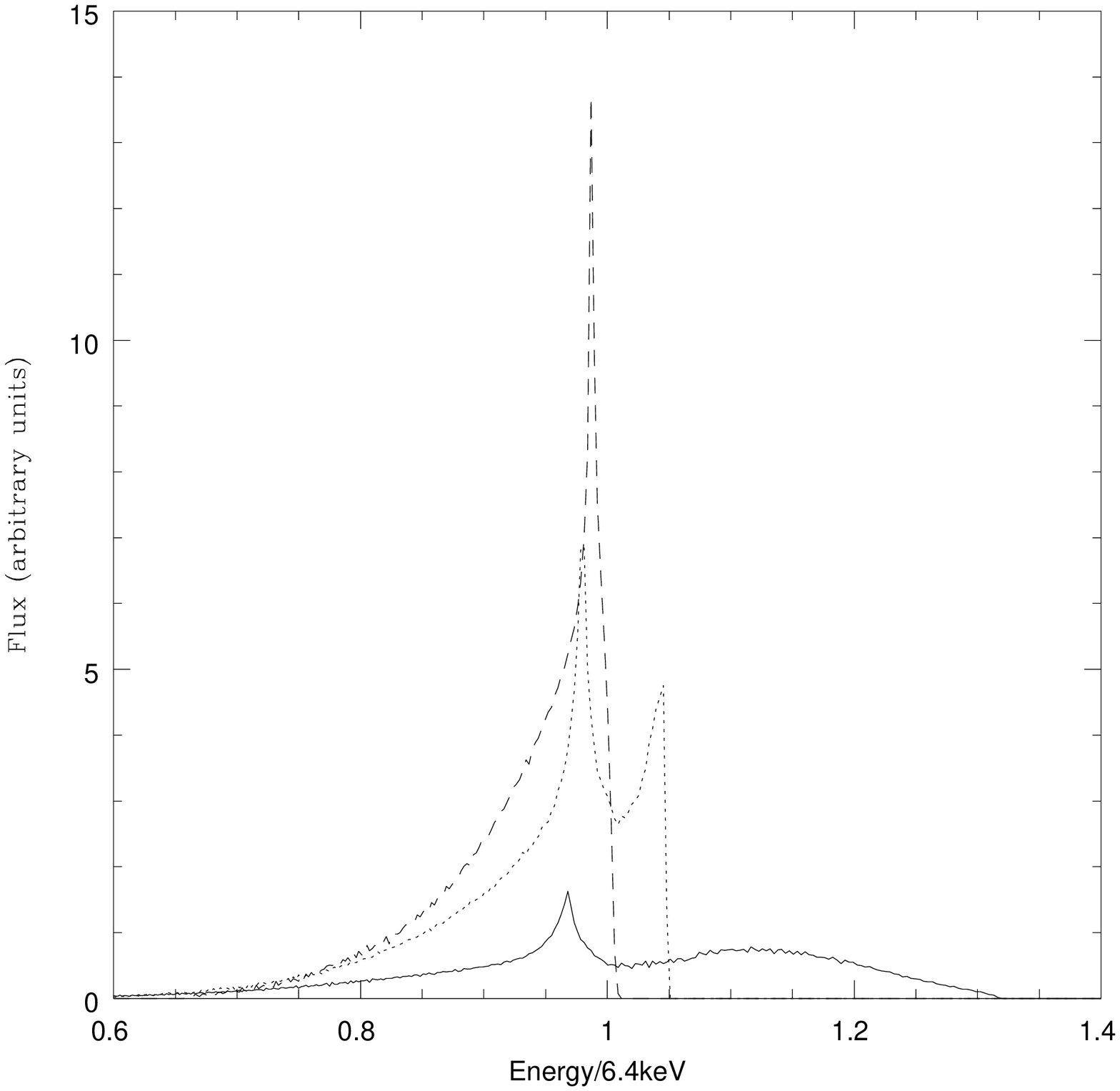,height=2in}
(b)\epsfig{file=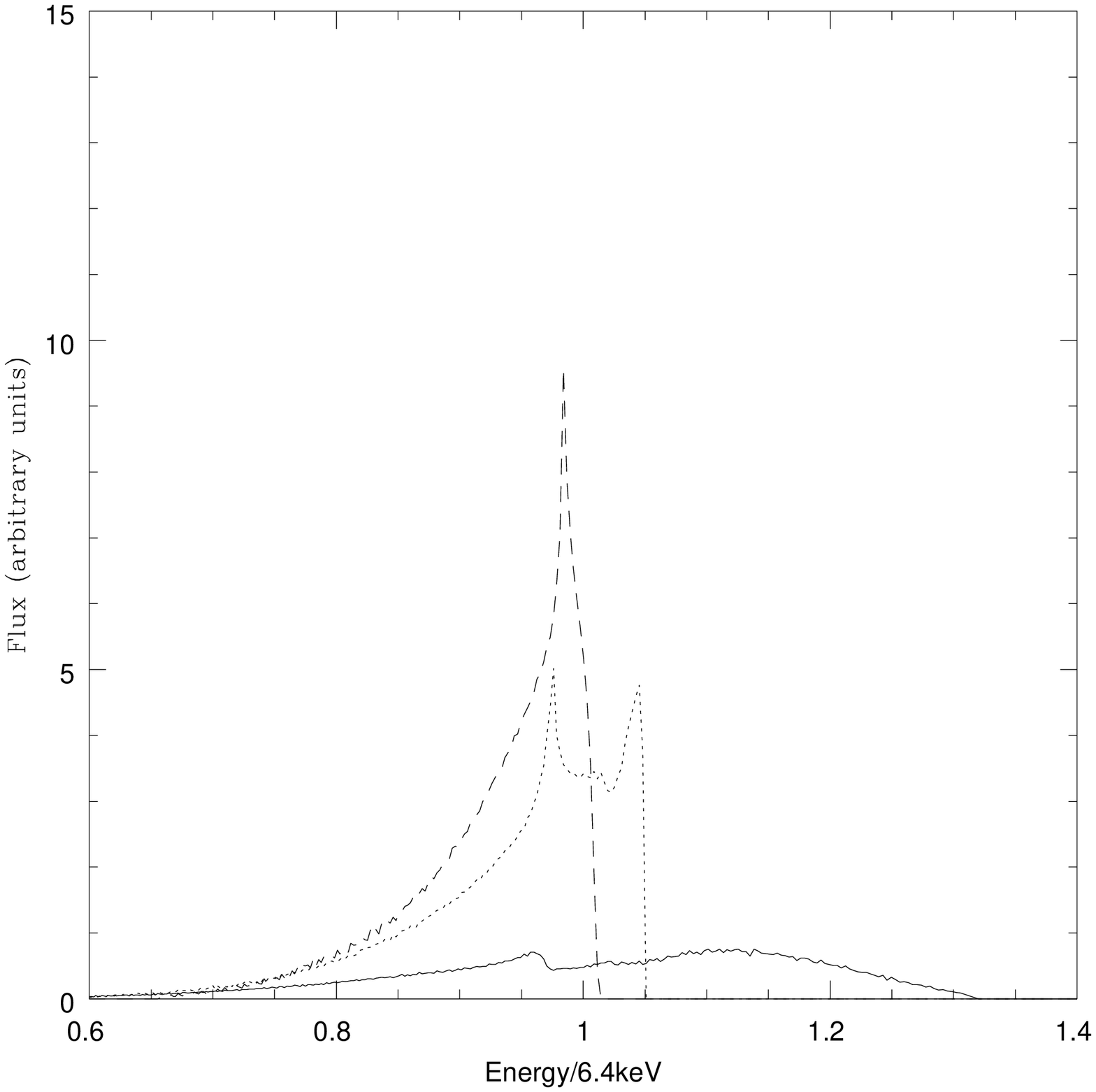,height=2in}
(c)\epsfig{file=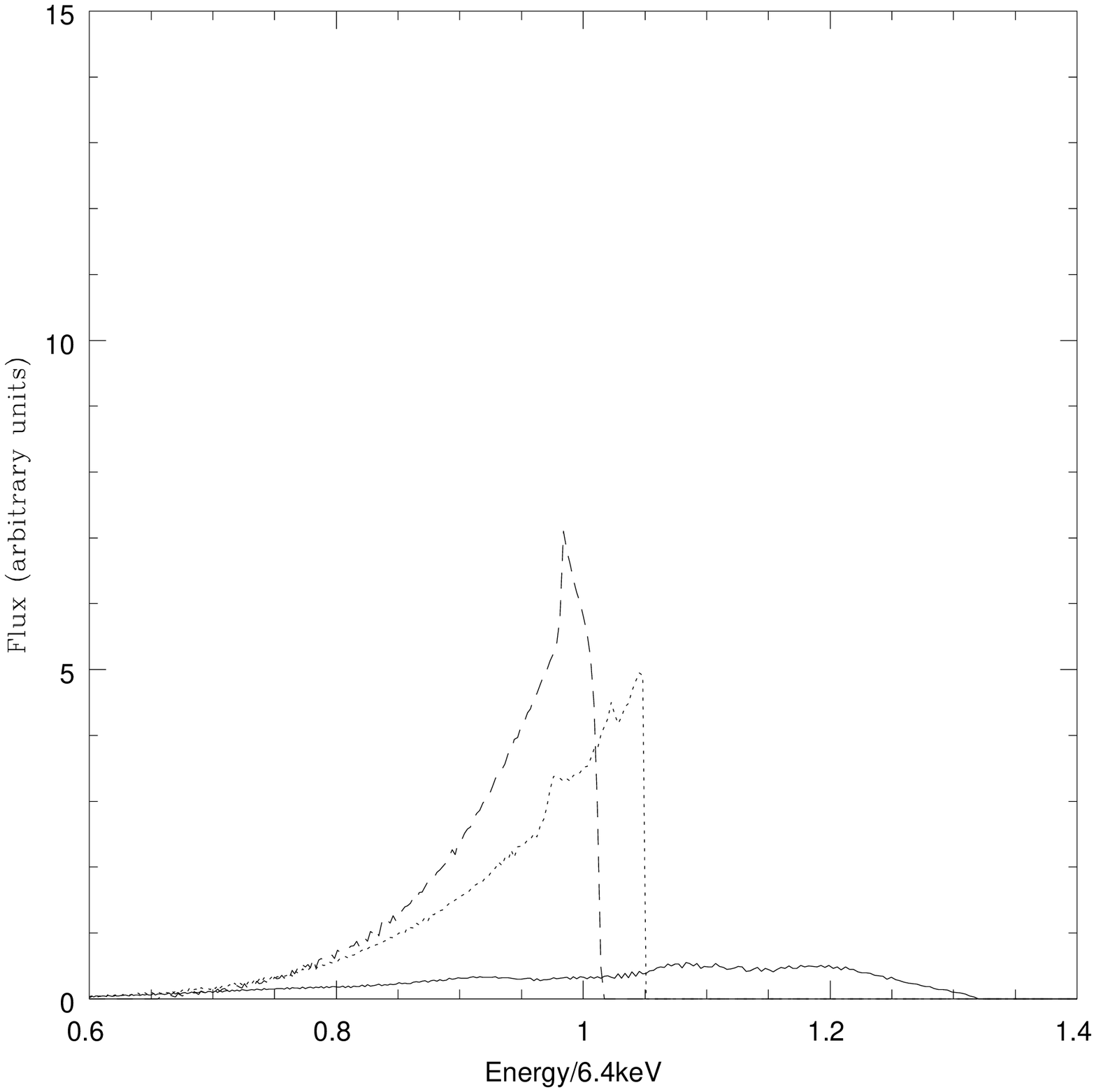,height=2in}

(d)\epsfig{file=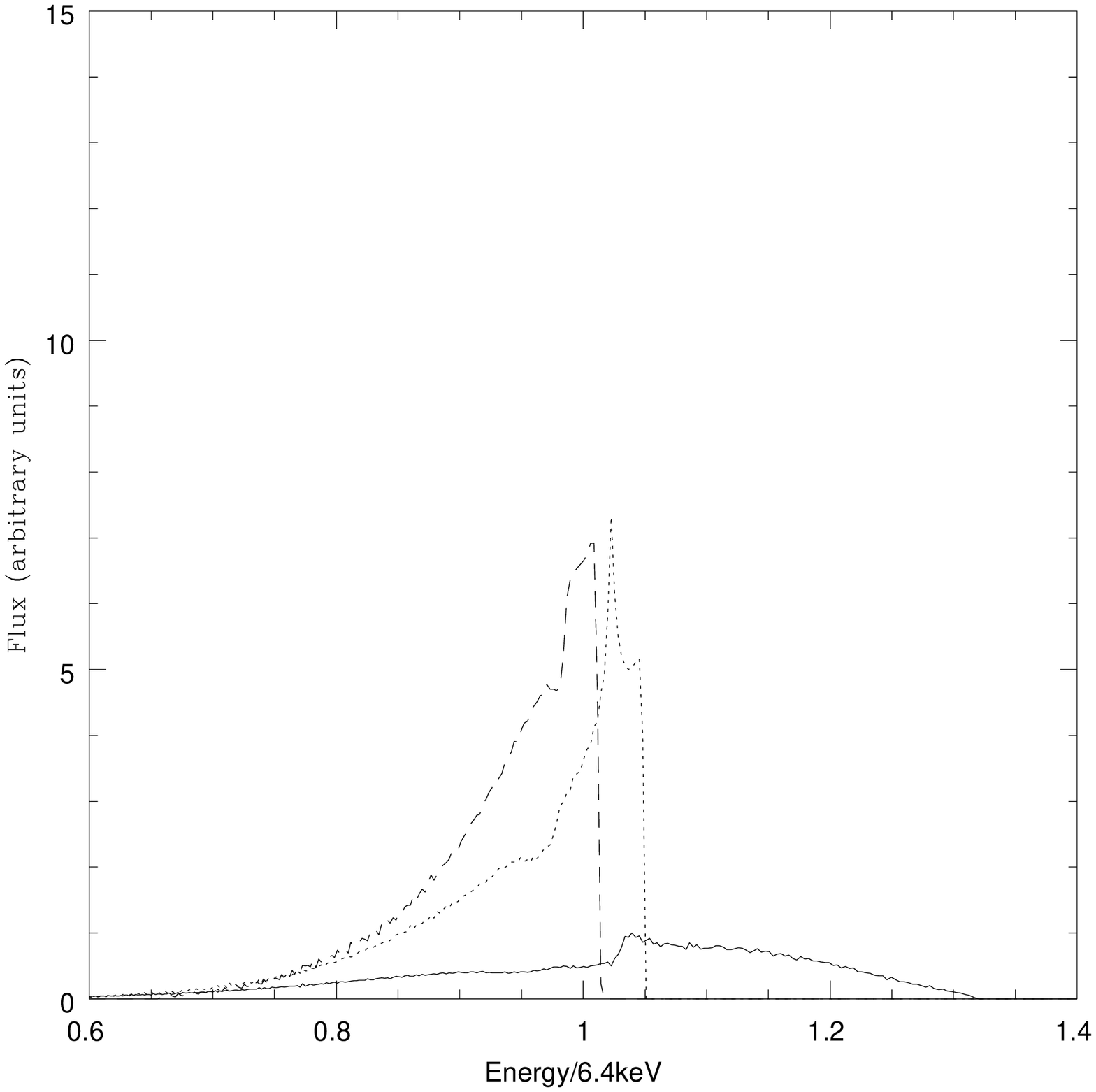,height=2in}
(e)\epsfig{file=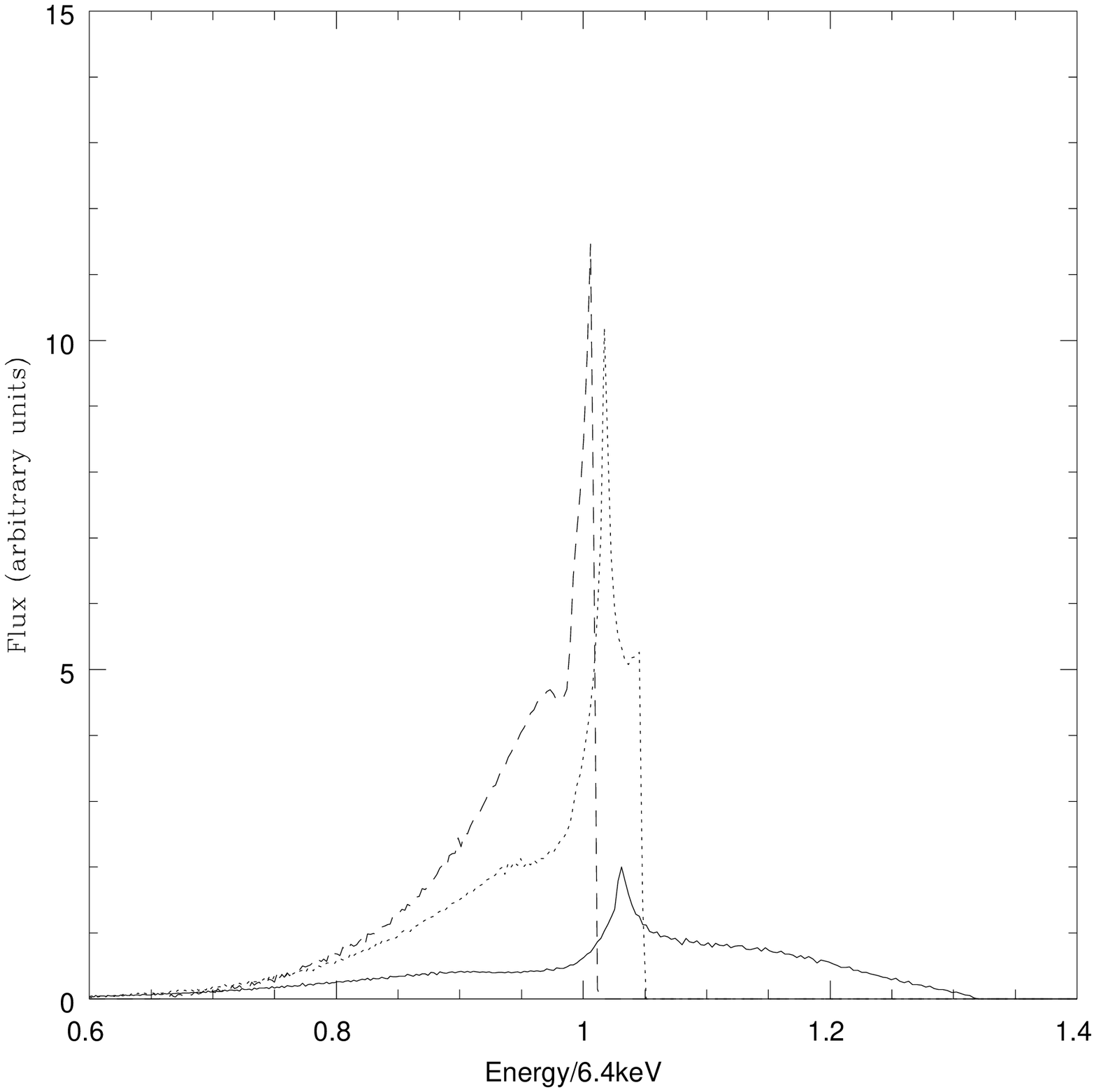,height=2in}
(f)\epsfig{file=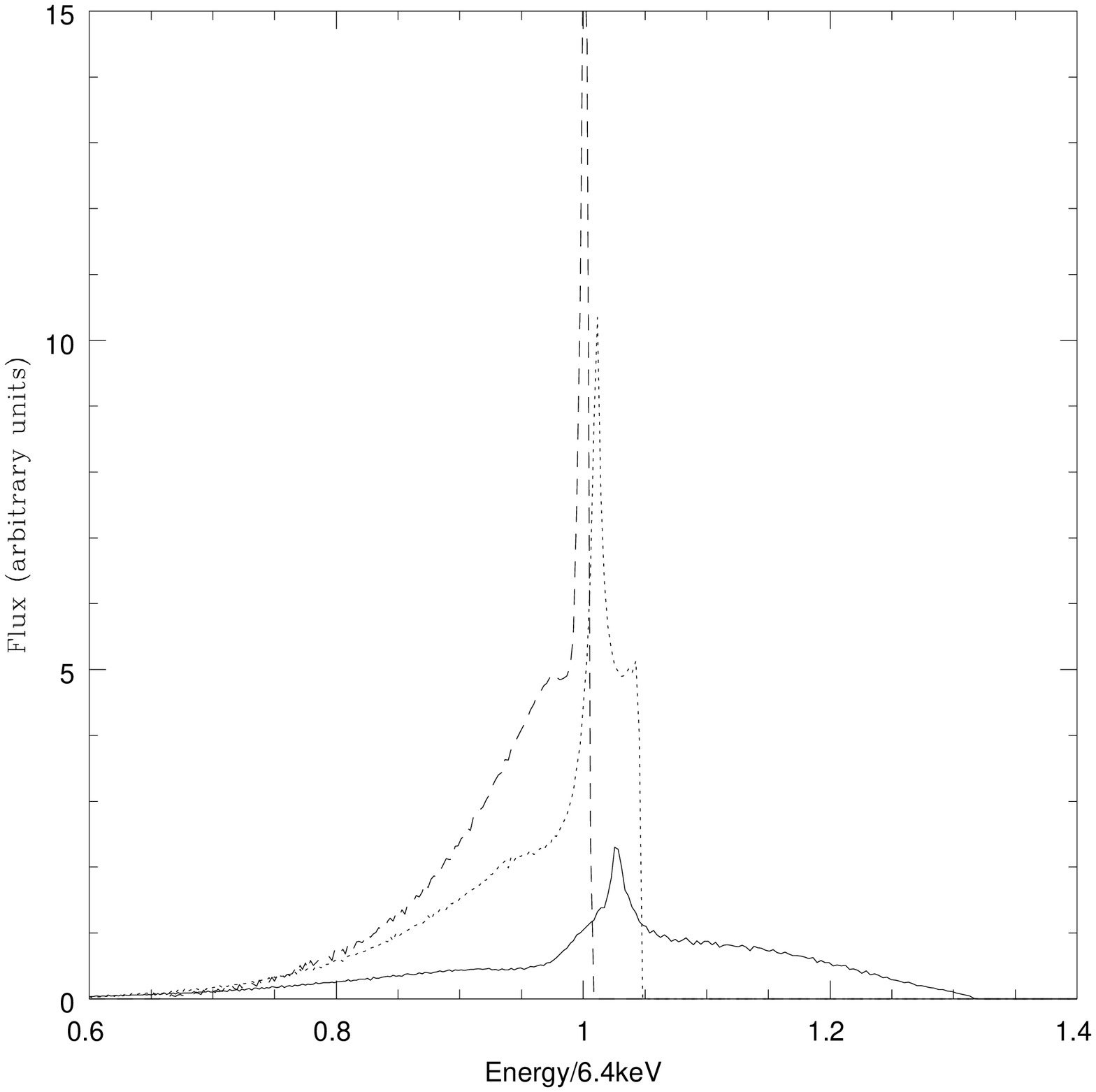,height=2in}

(g)\epsfig{file=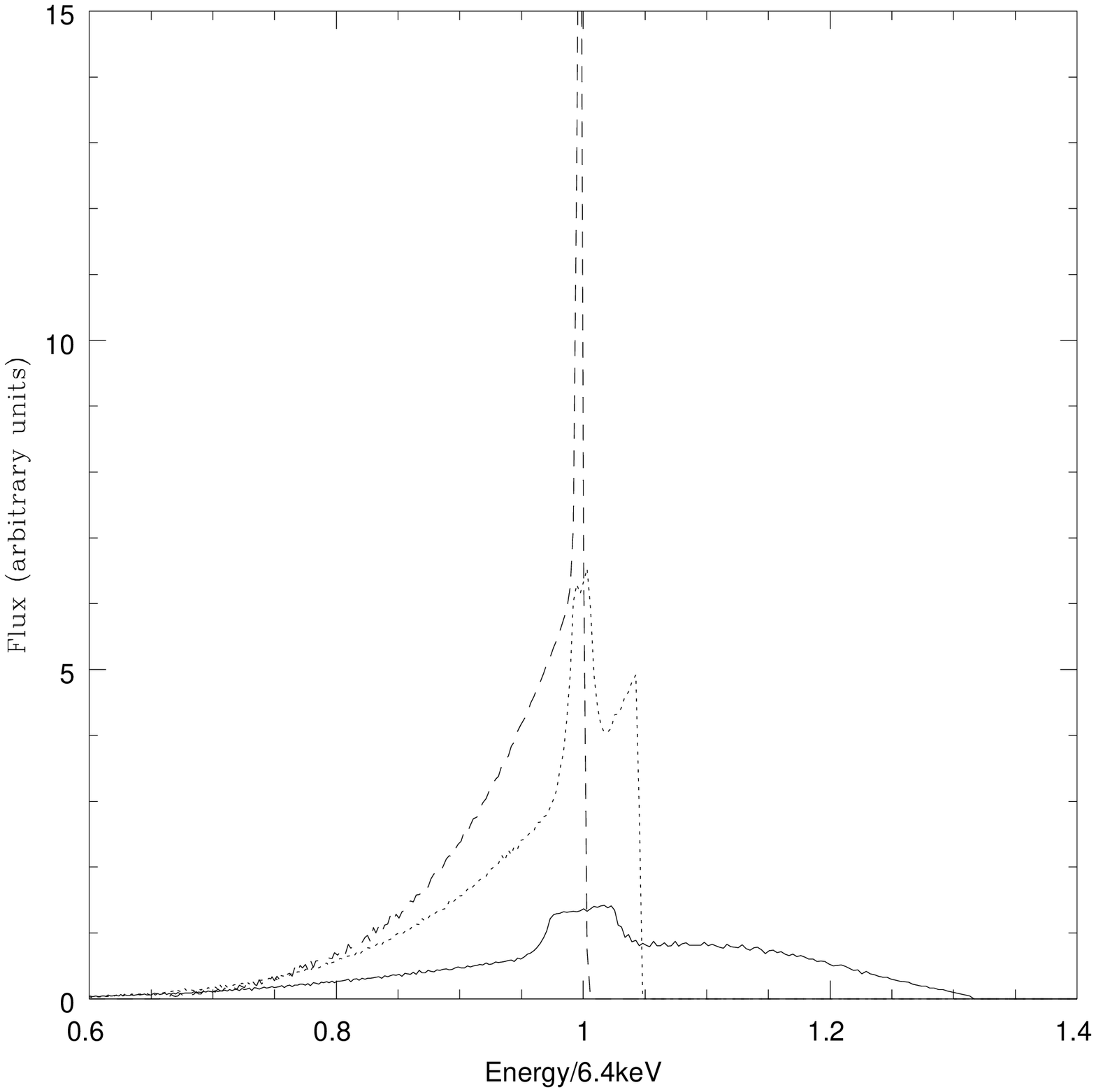,height=2in}
(h)\epsfig{file=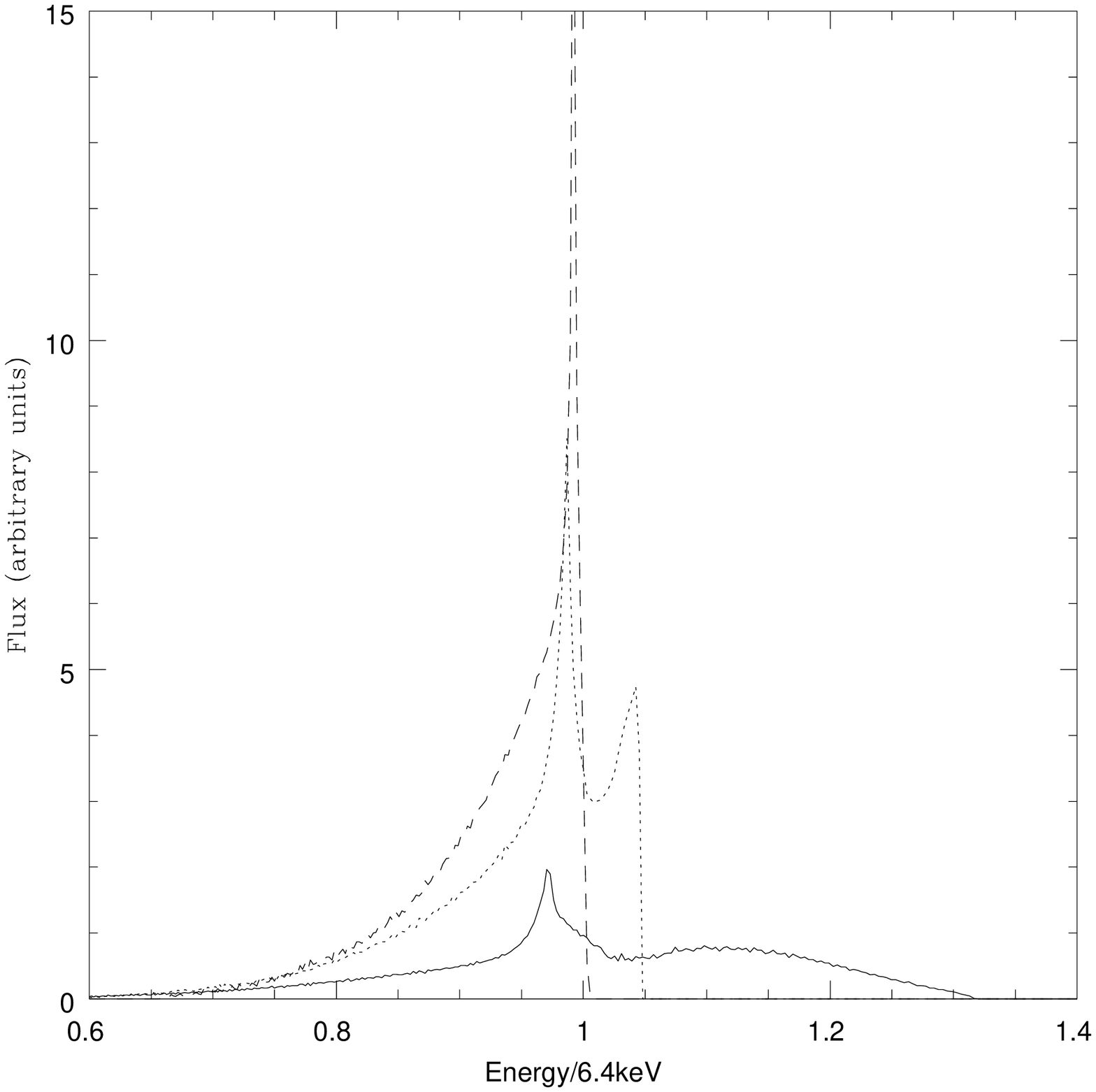,height=2in}
\hspace{0.5cm} (flat)\epsfig{file=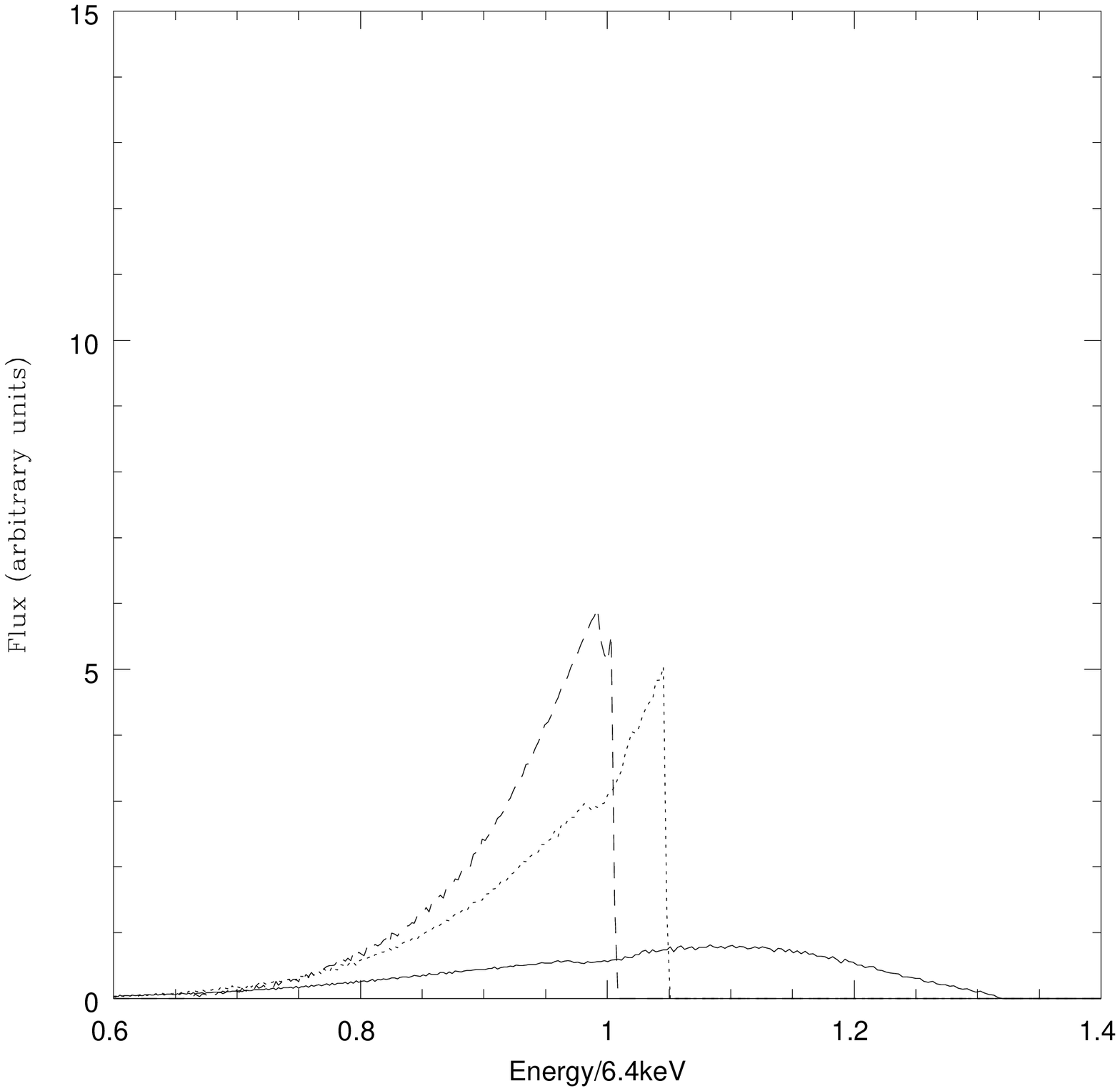,height=2in}

\noindent\textbf{figure 11:} (a)-(h) Twist-free disc with 
$r_{out}=10^{3}$, $a_{1}=0.25$ and $b=2$. (flat) is a flat disc with 
$r_{out}=10^{3}$ for comparison.
The dashed, dotted and solid lines in each box represent inclination viewing angles 
of $10^o$, $30^o$, and $70^o$ respectively from the x-y plane.
The azimuthal progression is as in Figs. 9 and 10, but here
$\pi/2$ corresponds to looking from the side of the disc containing 
the maximum height, and from the plane containing this maximum.
For twist free discs, the $0^o$ corresponds to the direction of the 
x-ray source above the disc, since the inner disc is essentially
flat and hence lies in the x-y plane.
\end{figure}
\begin{figure}
(a)\epsfig{file=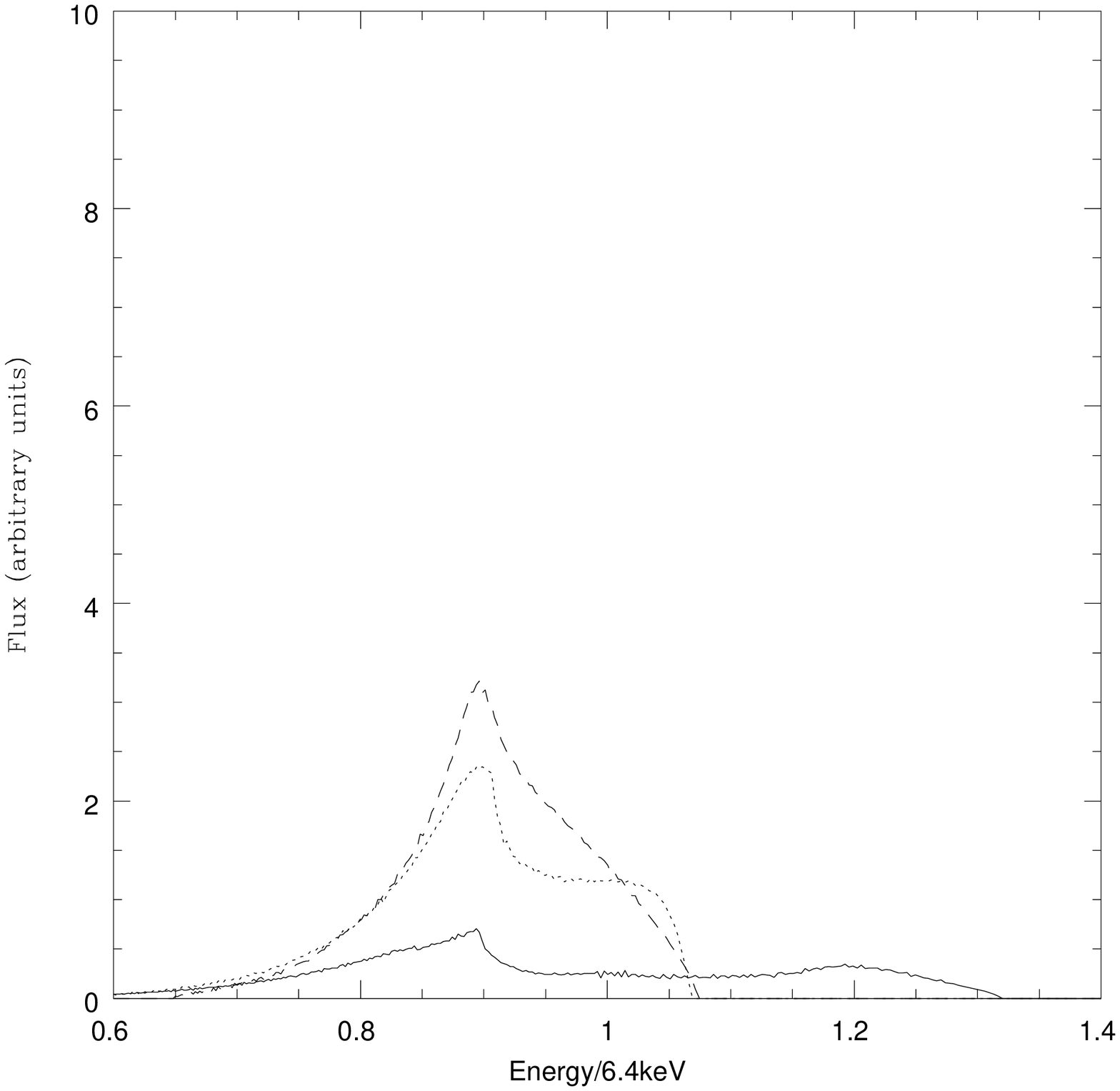,height=2in}
(b)\epsfig{file=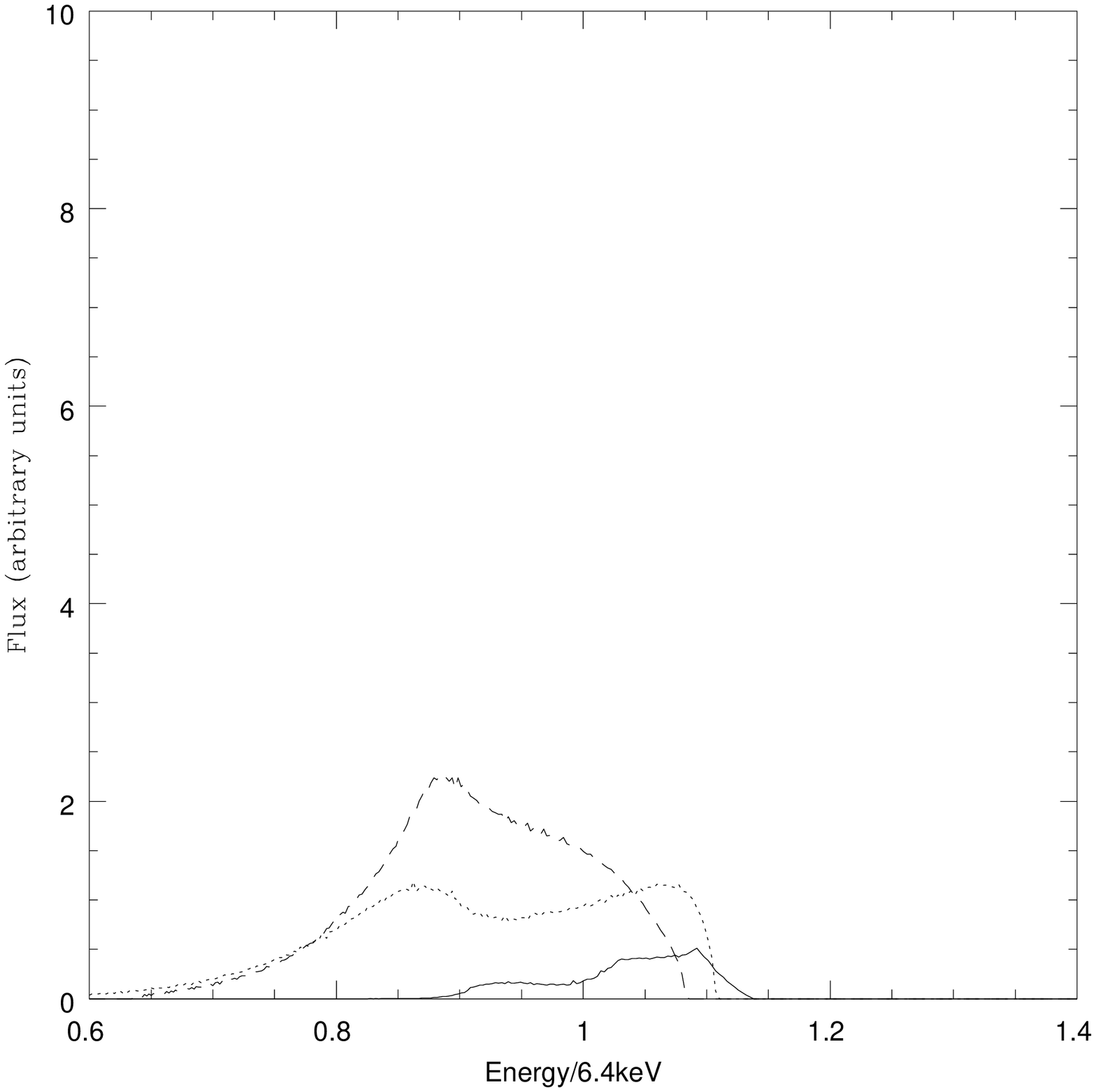,height=2in}
(c)\epsfig{file=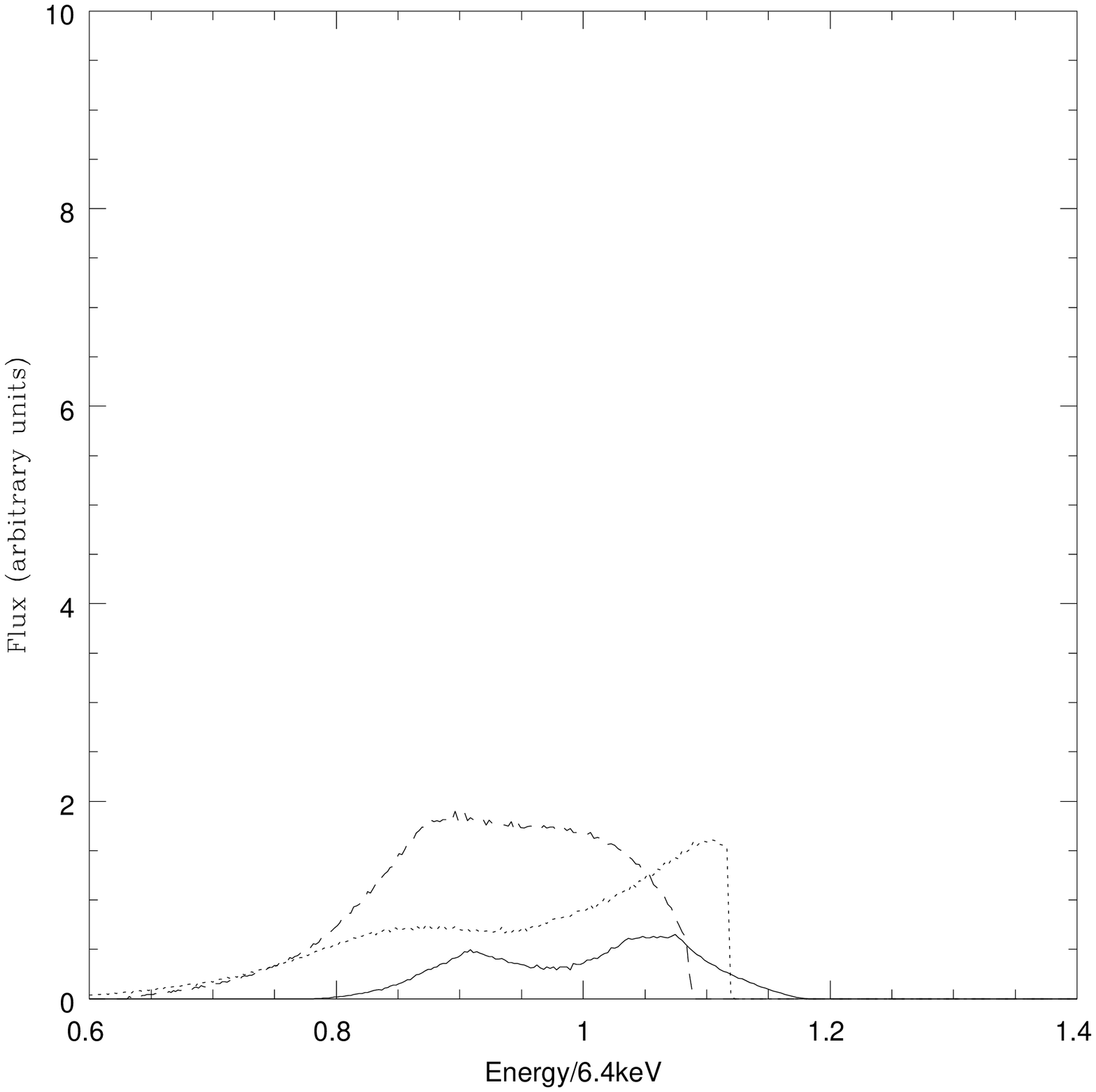,height=2in}

(d)\epsfig{file=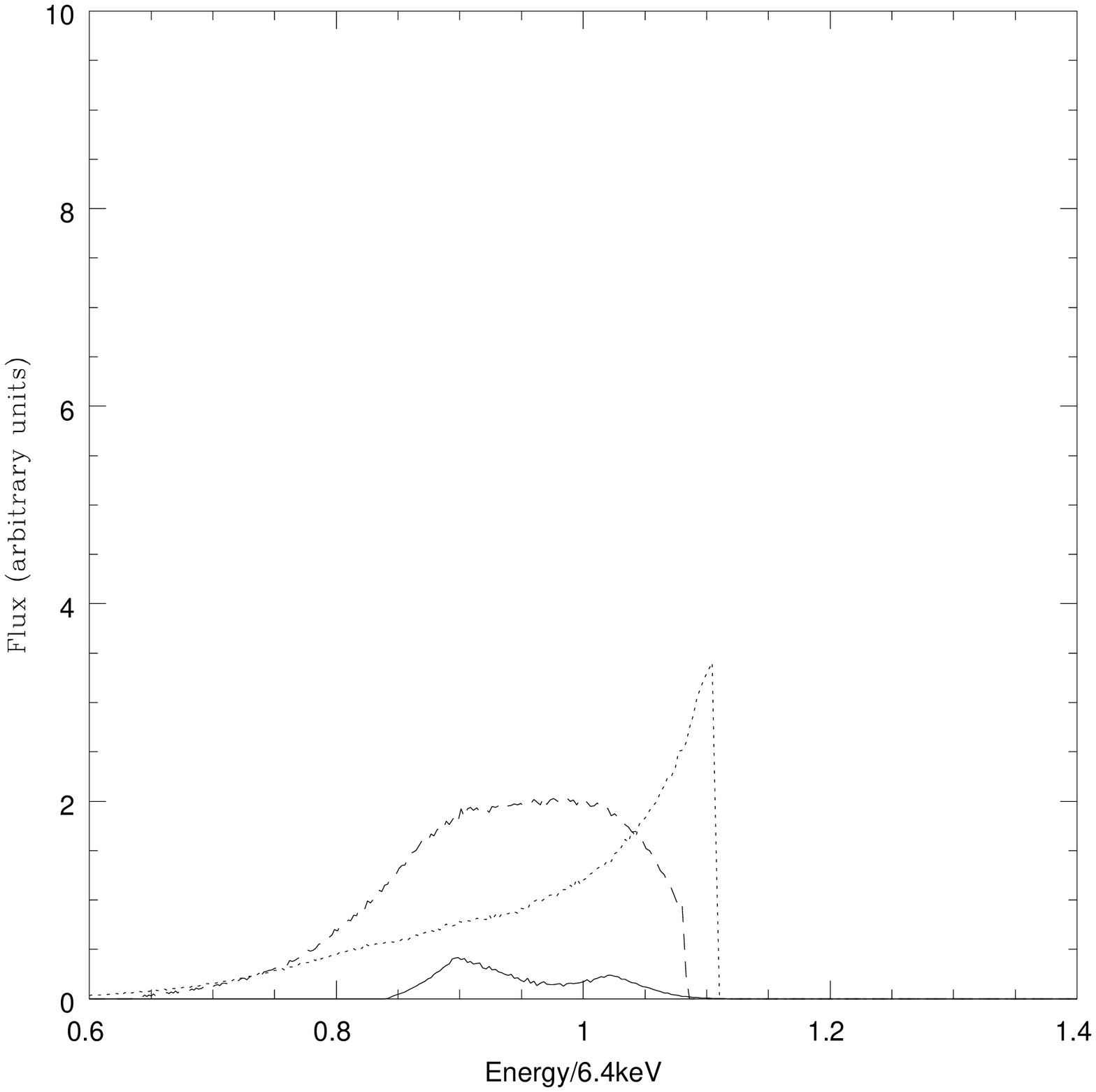,height=2in}
(e)\epsfig{file=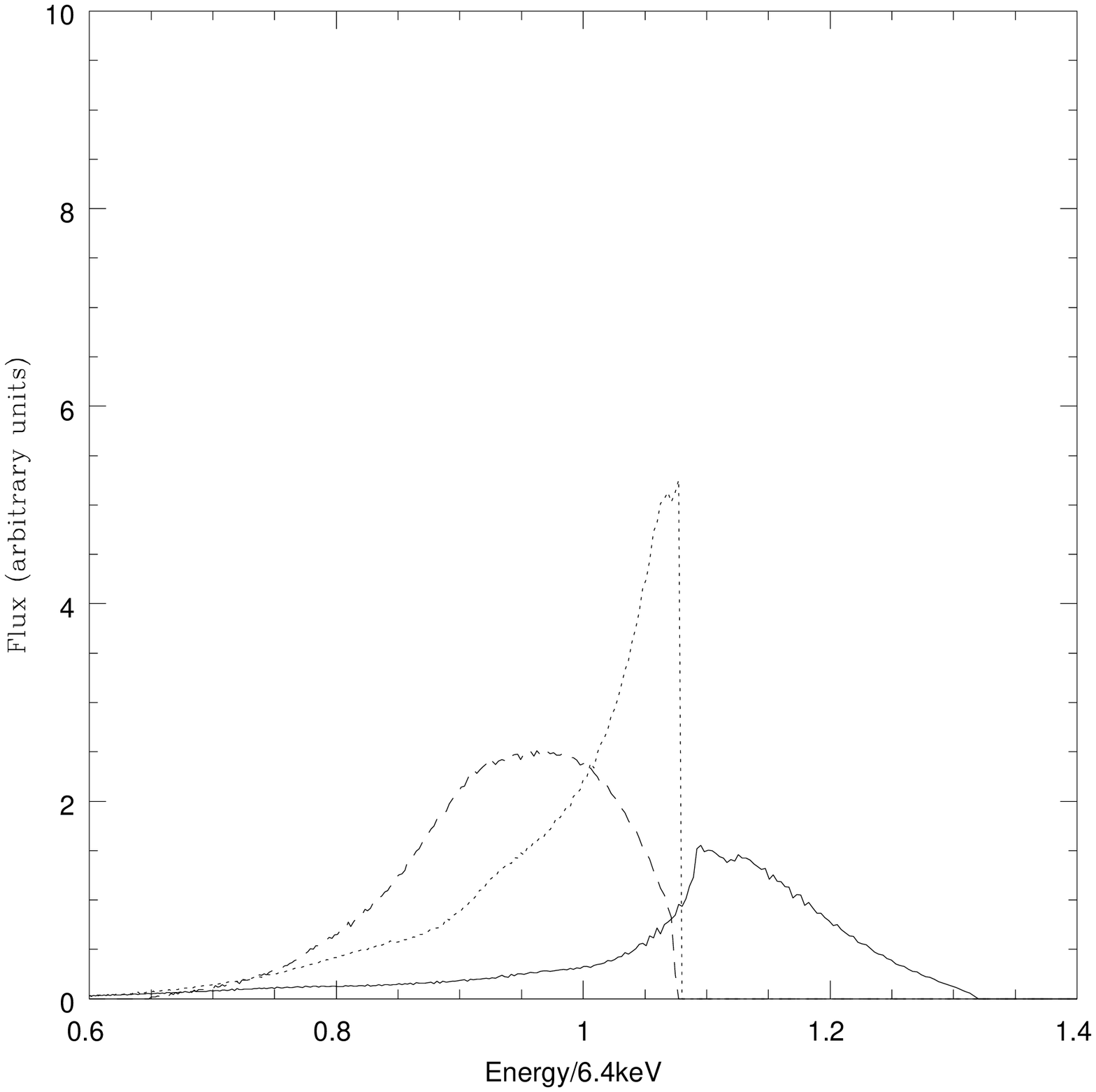,height=2in}
(f)\epsfig{file=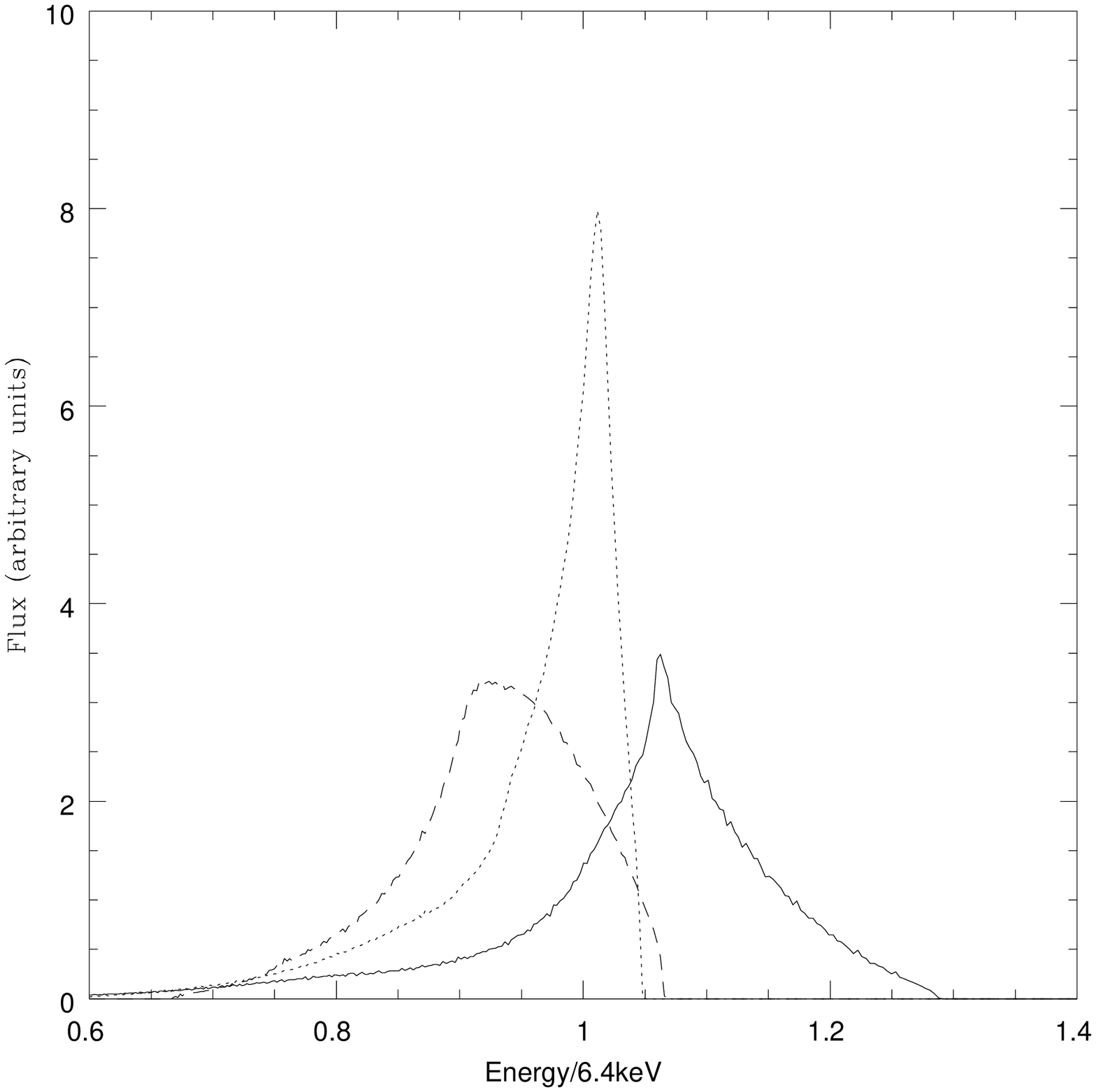,height=2in}

(g)\epsfig{file=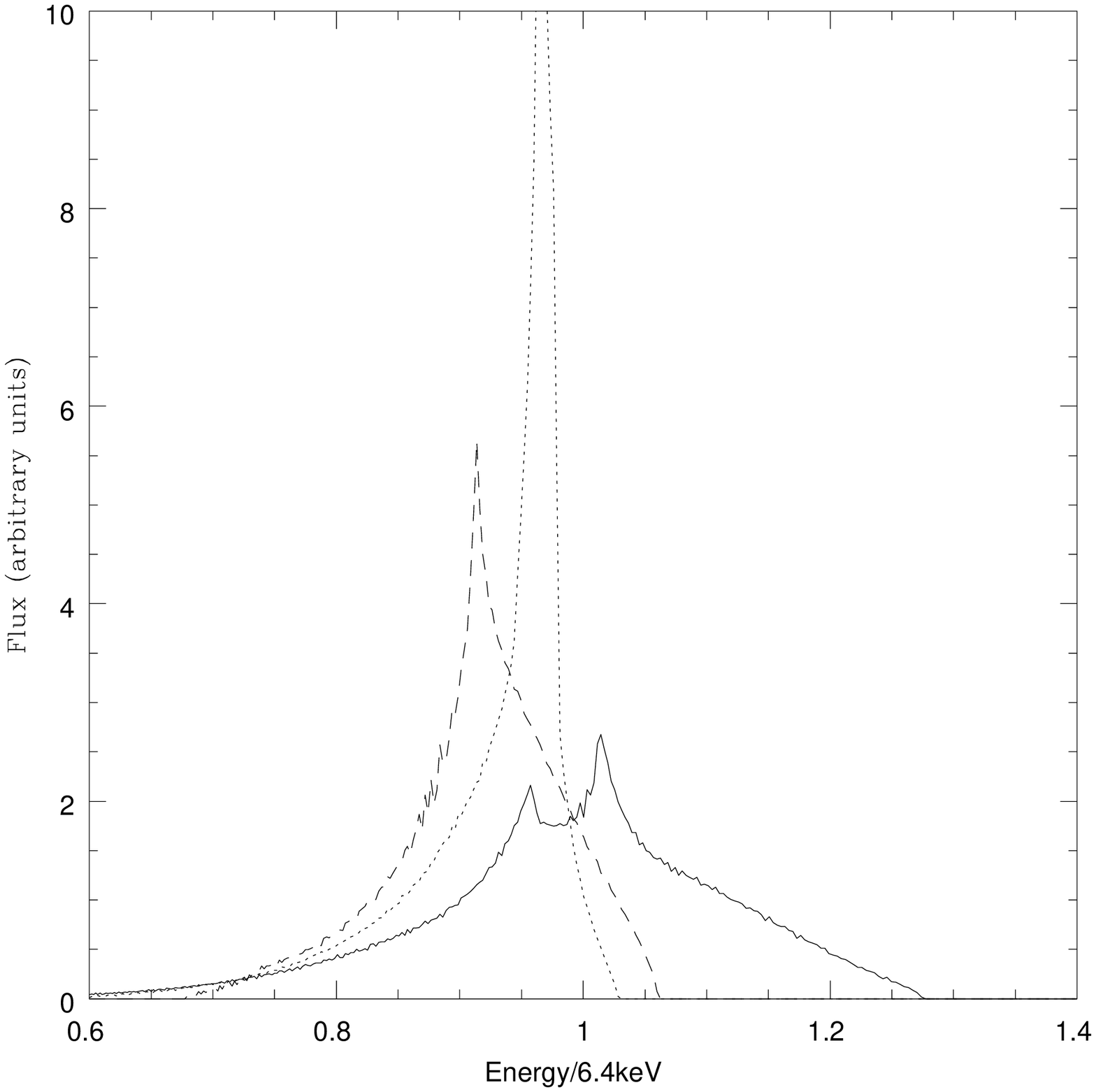,height=2in}
(h)\epsfig{file=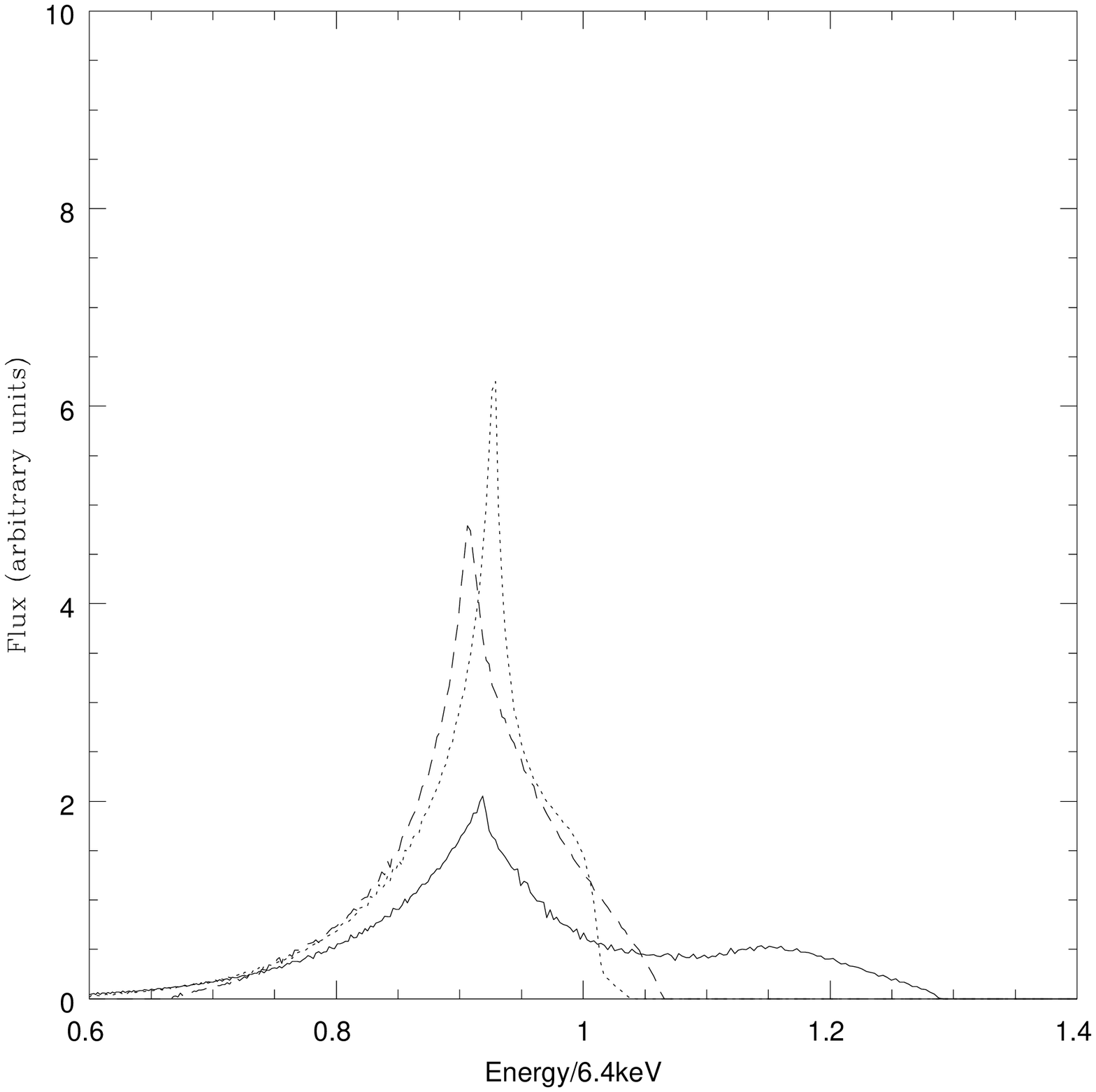,height=2in}
\hspace{0.5cm}(flat)\epsfig{file=flat2,height=2in}

\noindent\textbf{figure 12:} (a)-(h) Twist-free disc with $r_{out}=10^{2}$, $a_{1}=1$. (flat) is a flat disc with $r_{out}=10^{2}$ for comparison.
Boxes represent same azimuthal progression as in Fig. 11.
The dashed, dotted and solid lines in each box again represent inclination
viewing angles of $10^o$, $30^o$, and $70^o$ respectively from the x-y plane
(see Fig. 11 caption).

\end{figure}
\begin{figure}
(a)\epsfig{file=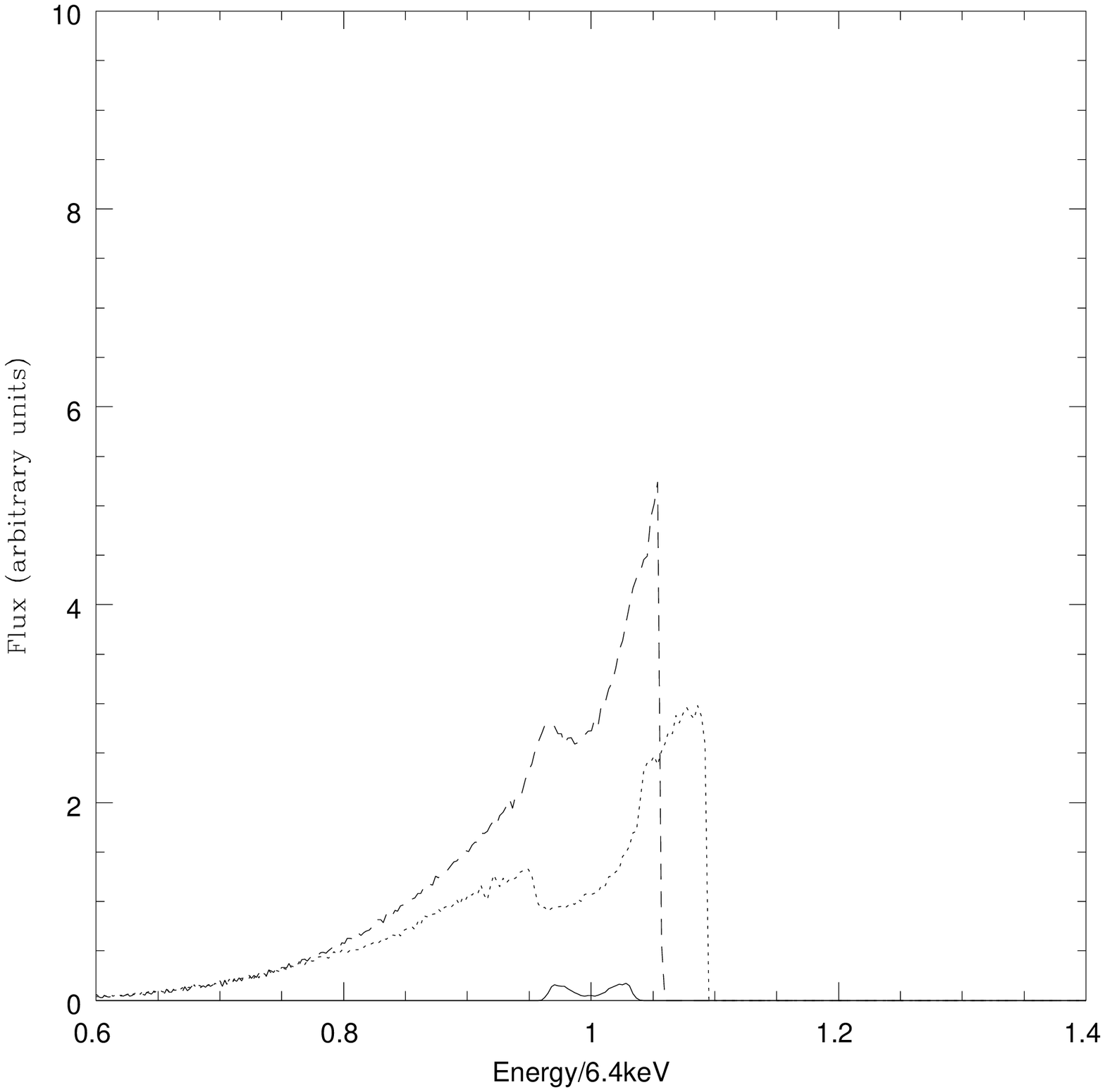,height=2in}
(b)\epsfig{file=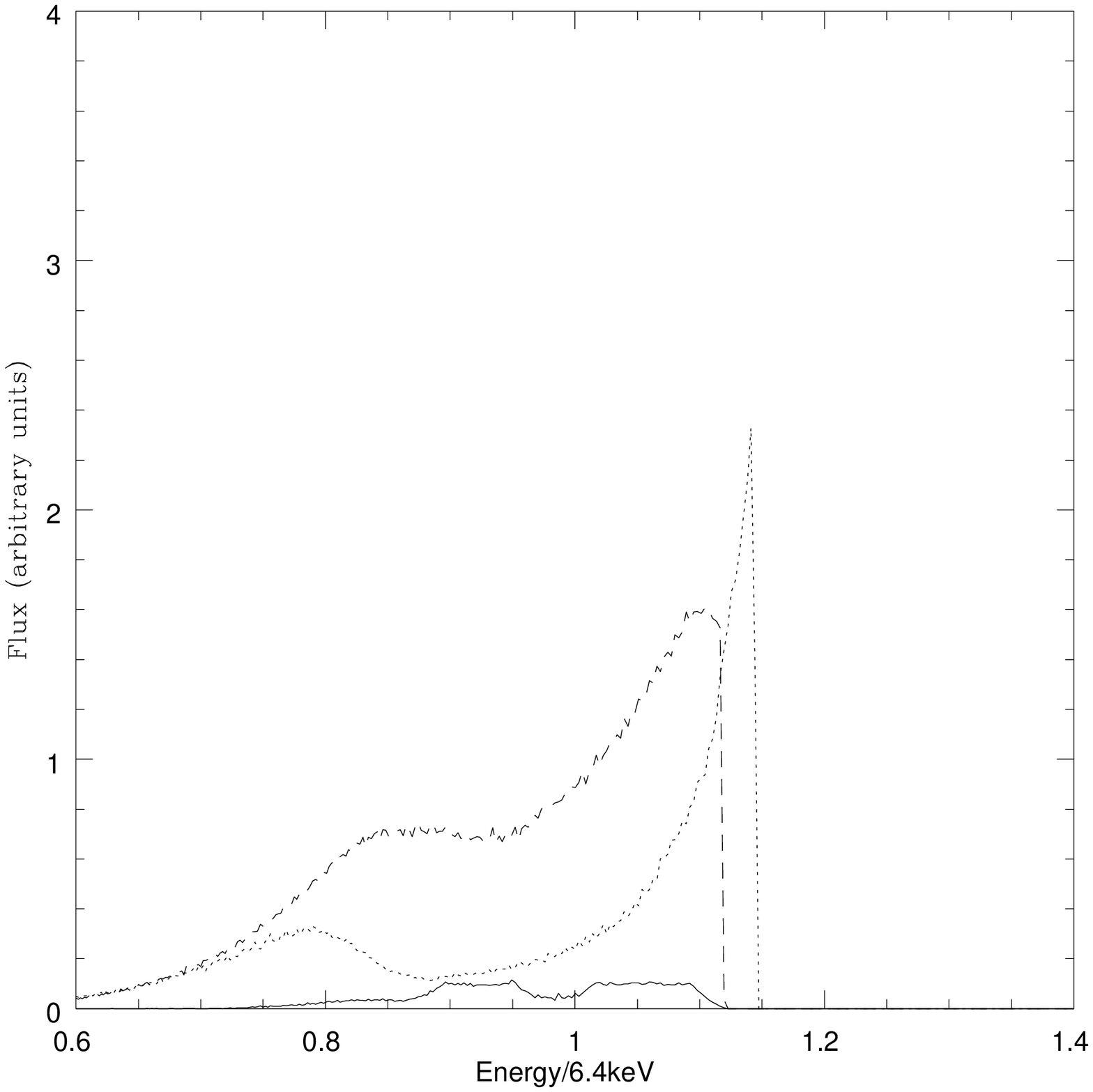,height=2in}
(c)\epsfig{file=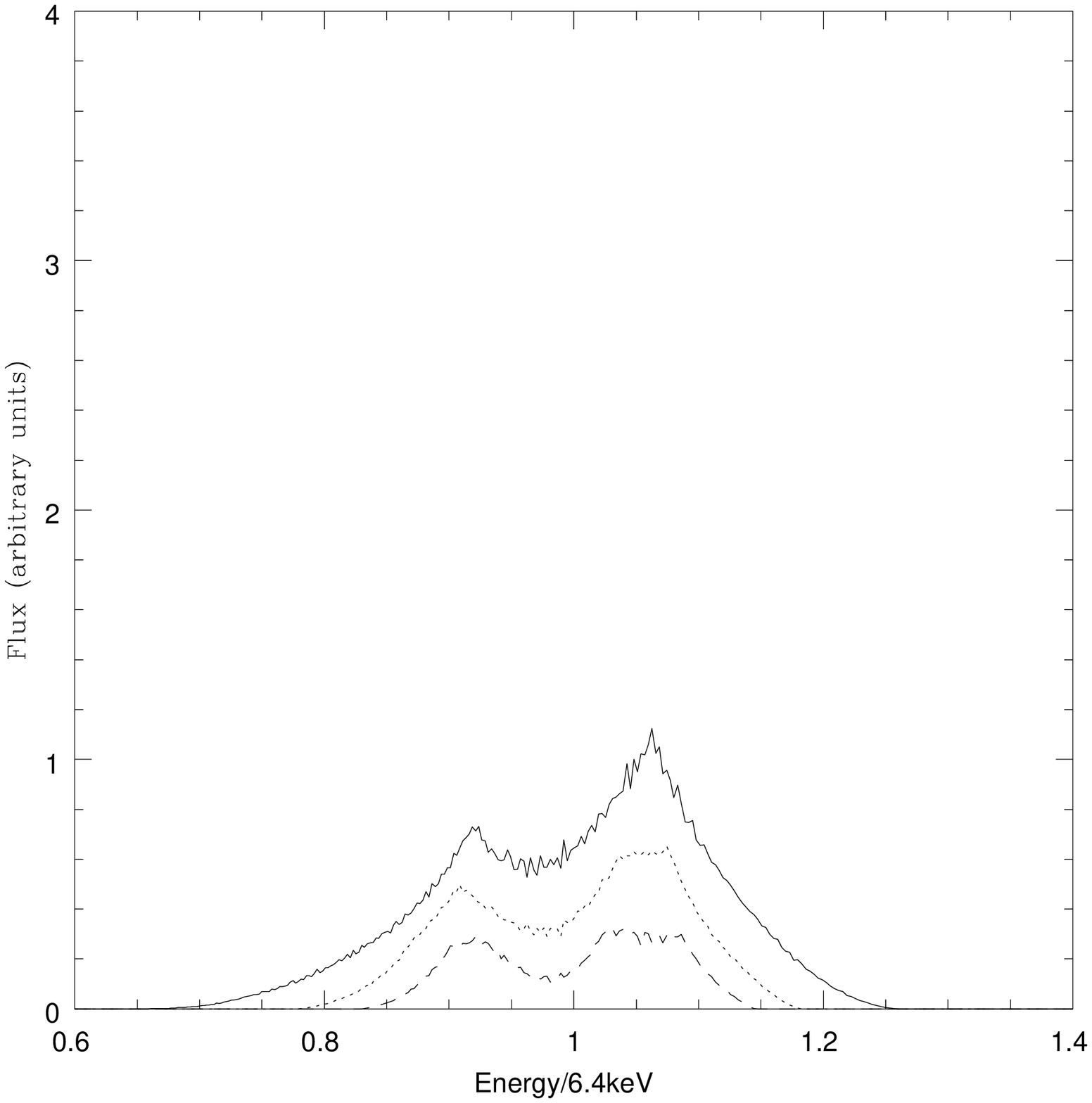,height=2in}

\noindent\textbf{figure 13:} Illustration of the effect of shadowing of the disc to the observer and of the possibilty of flux from the underside of the disc. All discs are twist-free and have $\omega t = \pi/2$, $a_{1}=1$ so the warp grows towards the observer (a) $r_{out}=10^3$, inclinations of $30^o,40^o,50^o$ from top to bottom. Shadowing to the observer 
starts at $40^o$ causing a drop in flux in the central spectral regions. At $50^o$ the disc is almost completely shadowed. (b) shows the same effect for $r_{out}=10^2$ and (c) continues from (b) with inclinations of $60^0,70^o,80^o$, the order is now reversed however, as greater inclinations give grater flux \emph{from the underside}. Note the ordinate upper bounds are different.

\end{figure}
\subsection{Line profiles}

Figs. 9,10,11,12 show a selection of line profiles.
Figs. 9 and 10 are for twisted discs and Figs. 11 and 12 
are for twist-free discs.  In each box, the line profiles are shown for 3 
different inclination angles ($a_{i}=10^{o},30^{o},70^{o}$) from the
x-y plane and each box represents one of 8 different 
azimuthal angles, $\omega t$, equally spaced between $0$ and $2 \pi$.
For the twist-free discs, 
the $wt=\pi/2$ azimuthal angle corresponds to looking at the disc from
the maximum disc height side, in the plane which intersects the maximum
(see Fig. 1). For the twisted discs, $wt=0$ corresponds to looking at the disc
from approximately the local inner disc maximum height side,
along the plane intersecting this (see Fig. 2).
Figs. a-h represent an anti-clockwise azimuthal progression 
of the disc. The last box in each figure is a flat disc for comparison.
The warp moves around the observer in the 
opposite (same) direction to the the fluid flow for positive 
(negative) $w$. 
Note that the choice of upper bound on the ordinate and abscissa
of the boxes can stretch profiles to qualitatively different shapes. 
We have adjusted only the ordinate in a few places.
This must be kept in mind when comparing line profiles 
here and elsewhere in the literature.

The inclination angles chosen in  
Figs. 9,10,11,12 do not show the full range of effects shadowing of the 
disc from the observer can play, although shadowing of the source plays
an important role in these plots. To exhibit the effects of disc shadowing
to the observer, Fig. 13 shows variation with inclination angle of several 
twist-free discs  at intermediate angles where
this secondary shadowing becomes important.

Generally speaking, there are a number of trends highlighted in
the line profiles 
which can be understood by a careful consideration of shadowing,
disc curvature, and orientation of the disc with respect to the 
observer.  

\subsubsection{Twisted disc profiles}

Consider first the twisted discs, Figs. 9, 10.
Fig. 9 has a slightly flatter warp than Fig. 10, as measured
by the different choice of $a_1$.
Line profiles from these discs are very sensitive to the
the azimuthal angle $\omega t$. This is because large
regions of the disc are shadowed from the source 
(see Fig. 6).  
The fact that only a small area remains un-shadowed by the source implies
a strong deviation in the line profile when compared to the flat disc.
In Pringle (1997), source-shadowing plays an important role 
in controlling the growth of radiative warping.

Let us now focus on Fig. 9, as the trends in Fig. 10 are similar.
For all inclinations shown, as we go through the azimuthal cycle
starting from $\omega t = 0$ (Fig. 9a), 
we see the centroid of the emission shift toward the red and then back toward 
the blue, commensurate with changes in orientation of the non-shadowed
region of the disc with respect to the observer.
 
%Notice the possibility of a significant blue hump and
%shallow blue fall-off even at $10^o$ inclination in box b.

The profiles at $10^o$ and $30^o$ are narrowest in Figs. 9d-f
because the inner region defines the extrema of the observed frequencies (i.e. the profile width) and 
is oriented more face on to the observer at these angles.
The Doppler shifts from the inner regions are therefore weaker. 
(Figs. 2 and 6 may be helpful in this regard.)
 
%innermost orbit that is the determining orbit for a disc with curvature.
%This is beacuse slightly outer orbits may actually have a larger projected 
%velocity onto the line of sight.

The tall peaks in 9c-f come from the regions between $r \approx 10$ and 
$r \approx 25$, where the curvature of the disc conspires with the incident
flux to produce many photons of coincident observed frequency.
The strong peak at $70^o$ in Fig. 10d,e is also due to this effect.
While there is some contribution from the outer regions of the
disc to these peaks, this not the dominant contribution.
If it were, the peaks in 10h,a,b would be sharper since 
the outer regions have a more favourable angle to the observer in these boxes.
It is important to distinguish this effect from 
that discussed in section 4.3, which leads to even larger peaks
near the rest frequency.

%The effect of the outer disc regions is seen 
%in the boxes of Fig. 9 for which the flux near the rest frequency is higher
%than that of the flat disc for the same
%inclination angle,  but where the blue wing extends significantly
%beyond this (e.g. Fig. 9a, top curve).  The effect 
%is not as dramatic as it might be since so much of the disc 
%area is shadowed to the source for such a twisted disc. 

Notice the presence of line profiles whose asymmetry is 
qualitatively reversed from the flat disc case.  In particular, 
the soft blue fall-off in 9b-d for the lower inclination angles is not 
obtainable from flat disc models, which have sharper blue fall-offs for low 
inclination angles. (Here particularly, 
it is important to compare plots with the same coordinate aspect ratios.)
This characteristic of flat disc models 
means that for Seyfert Is, which are thought to be selected 
at low inclinations by class, soft blue tails would  not be expected.
However, soft blue tails do seem to be present  
in some  line spectra presented in Nandra \emph{et al.} (1997). 
This effect can result in our profiles when 
the inner regions are at a higher inclination 
angle to the observer than the outer parts of the disc.
For steeper inclination angles,
soft blue fall-offs can be even more proununced 
when the combination of shadowing and inclination leads
to a significant red peak (Fig. 10d, $70^o$).

In the unified scheme of Seyferts, flat disc models do not allow
edge on views of the innermost regions for Seyfert Is or 
face on views for Seyfert IIs.  Twisted discs 
allow all comibinations, which can therefore produce a
wider variety of line profiles, as described above.

Generally speaking, twisted discs 
would either lead to an extremely variable time profile if the warp 
precession were fast enough, or an extremely wide range of different 
profiles for different objects.  We are not certain that the present data 
suggest such a wide range. Therefore, extremely twisted warps do not 
seem to be indicated by iron line profiles in Seyferts.
 
\subsubsection{Twist-free disc profiles}

Consider now the twist-free discs, Figs. 11 and 12.
Fig. 11 employs a warp magnitude  $a_1=0.25$ and an
outer radius of $r_{out}=10^3$ while
Fig. 12. employs $a_1=1$ and an
outer radius of $r_{out}=10^2$. The curvature index $b=2$
was used in both cases.
In general, these  discs show less deviation from the flat disc 
than the twisted disc does because of less severe source-shadowing. 

There are a number of other trends in the twist-free case which should 
be mentioned.  
In both Fig. 11 and 12, note the change from an initially red-dominated 
profile at $wt=0$ to progressively blue-heavy profiles and then back again.
Again, this is consistent with non-axisymmetric source-shadowing.   
The large peaks near the rest frequency correspond to seeing the outer 
regions of the disc.  Peaks in Fig. 11 are larger 
than peaks in Fig. 12 because even though Fig. 12 warps have
$a_1=1$ compared to $a_1=0.25$ of Fig. 11, the outer 
radius is 10 times larger for Fig. 11 and the contribution from the 
increased outer radii is more important.
The largest magnitude of the peak comes in Fig. 12g, 
where the concavity is at the most directly favourable angle to the observer,
maximizing the contributions from the outer regions.
One half of the disc is largely shadowed; the profiles can then be 
understood by considering the rotation of the illuminated half of the 
disc toward and away from the observer as the azimuth cycles, 
moving from boxes a-h.

The small peak in the $70^o$ discs of Fig. 11,
which results in a sharper than usual red fall-off, is also due to greater 
flux from the outer regions. 
%(in fact, this is same effect 
%considered above that was a problem for discs with $r_{out}=10^4$, it is not a%problem here because 
The high inclination angle means less observed flux
and a smaller peak, but this peak is still much larger
than allowed by a flat disc. This shape at large inclination
angle shows resemblance to 
some line profile observations of Seyfert IIs (Turner \emph{et al.} 1998), 
and the AGN unification paradigm holds that Seyfert IIs are indeed 
inclined at large angles. Such peaks are not present in 
flat disc modelling.

In Figs. 12bcd at $30^o$ and possibly also at $10^o$, we see
a form resembling the 
``deep minimum''  of  MGC-6-30-15
which has been modelled by ionization effects (Iwasawa \emph{et al.} 1996).
For our case, this shape 
would be explained by noting that the outer regions are more edge on 
to the observer than for a flat disc of the same inclination angle, 
even though the innermost regions are reasonably face on.  
Weaver \& Yaqoob (1998) suggest that the deep minimum may be related to 
geometric effects. In particular, they consider
occultation by an optically thick cloud. 
%They point out that the apparent increase in intensity of the line during
%the deep minimum phase may not be fully explained by ionization effects
%and material in a Kerr orbit.  
%They also worry that 
%Kerr effects only manifest themselves in the deep minimum phase.
How this will be resolved remains to be seen, but
we note that a natural geometric shadowing due to warping could play  
some role based on the profiles we find.

Note that the large required fraction of reprocesed X-rays 
required by observations of ultra-soft narrow-line Seyfert 
may aslo be indicative of disc concave-curvature effects,
as mentioned in section 1.

\subsubsection{Effect of warp precession}

Using the the standard Shakura-Sunaeyev accretion disc viscosity formula 
$\nu=\alpha c_s r_g H$ where, H is the disc thickness in units of $r_g$, $c_s=(H/R)v_{k}$ is the sound speed, and $v_k= c R^{-1/2}$ is the Keplerian speed, we can rewrite the 
radiatively driven disc precession frequency given by Maloney \emph{et al.}
 (1996) as
\begin{equation}
\Gamma_{0}/2=
1.8\left(\frac{H_0R_0^{-1}}{20}\right)^2\left(\frac{\epsilon}{0.1}\right)\left(\frac{\alpha}{0.01}\right)\left(\frac{R_0}{10^{12}
{\rm cm}}\right)^{-1}{\ \rm yr^{-1}},
\end{equation}
where $\alpha$ is the viscosity parameter at the arbitraily chosen
fiducial radius $R_0$, $\epsilon$ is the accretion efficiency, 
$H_0$ is the disc height at $R_0$.
There is significant uncertainty in this value because
of the choice of $H_0$ and $R_0$. The main purpose of presenting this
formula here is to note that significant variation between epochs
of observation could have something to do with non-axisymmetric
disc effects in addition to whatever may be happening 
with the state of source activity/flaring (c.f. Iwasawa \emph{et al.} 1999).

For X-ray binaries, depending on the choice of $R_0$,
such variations could be as short as $\le 1$sec or much longer (days-month) 
again depending on $R_0$. The possibility of long periodicities
from radiatively driven warps in X-ray binaries was
studied by Wijers \emph{et al.} (1999).
X-ray iron lines measured in Cyg X-1 (Ebisawa \emph{et al.} 1993) 
show significant variation in both the line profile
and the equivalent width with no binary orbital phase
dependence found. 
Perhaps warped precession could account for a separate periodicity.
Also, the observed lines do not seem to show 
the strong redshifted tail.
Whilst this may result from the inner regions being more
highly ionized, a 
combination of disc curvature and shadowing may be important; 
note e.g. Fig. 9a-d 
in this regard, for the inclination angles of $10^o$ and $30^o$.

\section{Conclusion}

We have developed a method for calculating relativistic 
line profiles that result from reprocessing of emission 
by an arbitrarily warped accretion disc orbiting a Schwarzchild black hole,
and subject to various shadowing effects associated with disc warping.
We have applied the method to calculating a selection of reprocessed 
X-ray iron line profiles from a point source located above
and below the disc.  The essential generalisation from
flat disc models is the non-axisymmetry in quantities
such as the normal to the disc and the orbit plane of points on the disc,
 which result from warping.
We have also included shadowing effects both {\it by the disc of the source},
and {\it by the disc of the disc} toward the observer.

We considered two classes of disc warps,
twisted and twist-free, distinguished by whether the line of nodes
twists or does not twist. The purpose of our study was two-fold:
First, as warped discs are observed on a variety of scales in astrophysics,
it is important to develop diagnostics for their presence and predictions
for specific warp models.  The iron line profile predictions serve
as such a diagnostic for their presence in the central regions of 
black hole accretion engines.   Second, the combination 
of shadowing effects and non-axisymmetry considerably extends the variety
of line profiles which can result from purely geometric and orientation
effects, which are potentially important.

For a specific warped disc, there are two angles which 
determine the observed profile, the inclination angle and
the azimuthal viewing angle.  For a flat disc, all azimuthal viewing
angles are equivalent.  
Time variability of the line profile is expected on the time scale
of the warp precession around the disc.  This time scale is determined
by the dynamics of particular warp models. 
Shadowing also changes the equivalent width
of the line as a function of azimuthal viewing angle.

Aside from the generally increased variety of profiles compared to a 
flat disc, we note some general trends:
(i) The relative height of the red and blue parts of the line
changes for different azimuthal angles for a fixed
inclination angle. (ii) Red and blue fall-offs can vary significantly
for a fixed inclination angle, depending on azimuth.
The shape can either a sharper red cutoff than blue, 
which is not possibile for a flat disc case,  or a sharp blue cutoff 
as for a flat disc.  
(iii) There can be sharper peaks near the rest 
frequency compared to a flat disc. 
(iv) Twisted warped discs show a larger variability in profiles than 
twist-free discs due to more substantial   
source shadowing and also show more extreme deviation from the flat disc
profiles.  

It seems that although observations do show
a wide variety of profiles, the twisted warp profiles 
seem to be more varied than the observations. 
Whether this is evidence against twisted warps requires more analysis. 
Some observed features that we have been able to reproduce 
qualitatively with warped discs and which 
may (or may not) present some difficulty to standard flat disc models include 
(i) red peaks or steep red fall-offs in some Seyfert II profiles, 
(ii) soft blue fall-offs seen in some Seyfert I's (Nandra \emph{et al.} 1997), 
(iii) line profile time variations if the warp precesses at an appropriate
frequency,
(iv) `deep minima' states (Iwasawa \emph{et al.} 1996). 
In this context, Weaver \& Yaqoob (1998) suggested that
occultation (shadowing) may play a role in line profiles.
The warp disc shadowing described herein could perhaps play 
the role of their postulated obscuration. Finally,
(v) misalignmnet of central disc plane with the obscuring 
torus: in the unified paradigm of Seyferts, warped discs would
relax the constraints on the predicted profiles that would
be otherwise expected from flat
disc models. This means that although Seyfert Is represent edge on 
views through the torus, the inner disc could actually be oriented
face on, with corresponding line profile effects.

As the data improve, it will be important to distentangle the
effects of plauisble non-planar disc geometries with other effects
such as ionization fraction changes.  
Exploring a well motivated range of 
geometric possibilities is important for furthering our understanding 
of accretion engines.  

\section*{Acknowledgements}

Many thanks to Ruben Krasnopolsky for help and suggestions with computing 
aspects of the work. S.A.H. acknowledges the hospitality of the California Institute of Technology, and we acknowledge funding from the SURF programme 
and the Theoretical Astrophysics group.

\end{document}